\newif\ifsingle
\singletrue 

\ifsingle
	\documentclass[12pt,onecolumn,journal]{IEEEtran}
\else	
	\documentclass[10pt,twocolumn,journal]{IEEEtran}
\fi

\newif\iftcom

\usepackage{graphicx}
\usepackage{rotating}
\usepackage{tikz}
\usepackage{amssymb}
\usepackage{epsfig}
\usepackage{epstopdf}
\usepackage{algorithm}
\usepackage{algorithmic}
\usepackage{amsmath}
\usepackage{amsfonts}
\usepackage{dsfont}
\usepackage{url}
\usepackage{chemarr}
\usepackage{chemarrow}
\usepackage{bbold}
\usepackage[version=3]{mhchem}
\usepackage{mathabx}
\usepackage{enumitem}
\usepackage{subcaption}
\usepackage{multirow}
\usepackage{lipsum}
\newif\ifusetables
\usetablesfalse 
\ifCLASSINFOpdf
\else
\fi
\newtheorem{example}{Example}
\newtheorem{remark}{Remark}

\begin{document}

\title{Using spatial partitioning to reduce the bit error rate of diffusion-based molecular communications}

\author{Muhammad Usman Riaz, Hamdan Awan, Chun Tung Chou,~\IEEEmembership{Member IEEE}
        
\thanks{M. Usman Riaz and C.T. Chou are with the School of Computer Science and Engineering, The University of New South Wales, Sydney, New South Wales 2052, Australia. Hamdan Awan is with Department of Electrical Engineering and Computer Science, York University, Toronto, Canada.
Emails: usmanriaz@cse.unsw.edu.au, c.t.chou@unsw.edu.au, 
hawan@eecs.yorku.ca}}


\maketitle
\begin{abstract}
This work builds on our earlier work on designing demodulators for diffusion-based molecular communications using a Markovian approach. The demodulation filters take the form of an ordinary differential equation (ODE) which computes the log-posteriori probability of observing a transmission symbol given the continuous history of receptor activities. A limitation of our earlier work is that the receiver is assumed to be a small cubic volume called a voxel. In this work, we extend the maximum a-posteriori demodulation to the case where the receiver may consist of multiple voxels and derive the ODE for log-posteriori probability calculation. This extension allows us to study receiver behaviour of different volumes and shapes. In particular, it also allows us to consider spatially partitioned receivers where the chemicals in the receiver are not allowed to mix. The key result of this paper is that spatial partitioning can be used to reduce bit-error rate in diffusion-based molecular communications.
\end{abstract}


\IEEEpeerreviewmaketitle



\section{Introduction}
\label{sec:SECTION_1_Introduction}
\IEEEPARstart{M}{olecular} communication is an emerging field which focuses on realizing communication between nano-scale devices \cite{Akyildiz:2008vt,Nakano:2014fq} and especially the internet of bio-nano things \cite{akyildiz2015internet}. A key characteristic of molecular communication is the use of molecules as the information carrier. This paper considers diffusion-based molecular communications \cite{farsad2016comprehensive}\cite{Pierobon:2010kz}.

The receiver is an important component in any communications system. We can divide the techniques for improving receiver performance of diffusion-based molecular communications into two categories. The first category of work uses signal processing techniques. Some examples in this category are: the paper \cite{kilinc2013receiver} designs a receiver based on minimum mean square error method; \cite{mahfuz2015comprehensive} uses multiple samples per symbol and maximum likelihood method to design a receiver; \cite{jamali2017design} designs a matched filter to maximise the signal-to-interference-plus-noise ratio; \cite{tiwari2017estimate} studies the design of estimate-and-forward relay nodes; and, \cite{fang2017convex} uses data fusion of decisions from multiple receivers. 


The second category of work uses physical or chemical properties to improve communications performance. Some examples are: \cite{noel2014improving} uses enzyme in the medium to reduce interference; \cite{noel2014optimal} uses flow to improve the performance of the weighted sum detector at the receiver; \cite{farahnak2019medium} uses chemical reactions in the transmission medium to reduce interference.

This paper proposes to use the physical mechanism of spatial partitioning to improve communication performance. Our work is inspired by the fact that receptors on the cell membrane are organized into spatially separated clusters of receptors \cite{mugler2013spatial}. In this paper, we  assume that the receiver uses receptors which can be activated by the signalling molecules. We propose to segregate these receptors into a number of clusters separated spatially. We consider two configurations: partitioned and mixed. In the partitioned configurations, spatial isolation is perfect and receptors cannot move between clusters; however, in mixed configurations, which can be considered as imperfect isolation, receptors can diffuse between the clusters. For both configurations, we derive the maximum-a-posteriori (MAP) demodulator by leveraging our earlier work on designing demodulators for diffusion-based molecular communications using a Markovian approach \cite{chou2015markovian}\cite{awan2017generalized}. 

This paper also removes a limitation of our earlier work in \cite{chou2015markovian}\cite{awan2017generalized} where the receiver is assumed to be a small cubic volume called a voxel. In this paper, we assume that the receiver consists of multiple voxels and the union of these voxels define the shape of the receiver. This method of modelling a 3-dimensional volume is similar to that in finite difference method. In this paper, we extend the method of demodulator design in \cite{chou2015markovian}\cite{awan2017generalized} to the multi-voxel case. 

The contributions of this paper are: 
\begin{itemize} 
\item We propose to use spatial partitioning to reduce the bit error rate (BER) of diffusion-based molecular communications. 
\item We derive the MAP demodulator for both the partitioned and mixed configurations. This derivation of the MAP demodulator applies to a receiver with multiple voxels. 
\item We derive approximations of the derived MAP demodulators. We show that the approximate demodulator for the partitioned case has a lower BER than that of the mixed configuration. We also show that our approximate demodulator for the mixed configuration can offer gradual degradation in BER from the partitioned configuration which means that it can deal with imperfect partitioning. 
\item We derive a method to analytically estimate the BER of the approximate demodulator. 
\end{itemize}





The rest of the paper is organized as follows. Section \ref{sec:SECTION_2_relatedwork} discusses the related work. In Section \ref{sec:SECTION_3_Back_ground} we present the background and a summary of our previous work in \cite{chou2015markovian}\cite{awan2017generalized}. Section \ref{sec:SECTION_5_map_demodulator} presents the MAP demodulator for the partitioned and mixed configurations. 
Section \ref{sec:LNA} presents an analytical method to calculate the BER of the approximate demodulators. 
Section
\ref{sec:SECTION_6_Numerical_Examples} presents numerical studies on comparing the performance of the partitioned and mixed configurations. Finally, Section
\ref{sec:SECTINO_7_Conclusions} concludes the paper.


\section{Related work} 
\label{sec:SECTION_2_relatedwork}

A number of insightful surveys have been written on the topic of molecular communications, see \cite{Akyildiz:2008vt,Nakano:2014fq,Hiyama:2010jf,Nakano:2012dv,farsad2016comprehensive}.  

There are three main components of a molecular communications system, namely transmitter, propagation medium and receiver. For molecular communications transmitters, many different modulation methods have been proposed. These include Molecule Shift Keying, Frequency Shift Keying, on-off keying and Concentration Shift Keying  \cite{ShahMohammadian:2012iu,Kuran:2011tg,Mahfuz:2011te}. In this paper, our transmitter uses Reaction Shift Keying (RSK) which is proposed in our earlier work \cite{awan2017generalized}, \cite{chou2015markovian}, \cite{awan2017improving} \cite{Awan:2015:IRM:2800795.2800798}. In RSK, the transmission symbols are generated by a set of chemical reactions and are characterised by the time-varying concentration of the signalling molecules. 
    
For the modelling of the propagation medium, various models have been proposed in literature. A common model is to assume that space is continuous, see e.g. \cite{Mahfuz:2014vs,Pierobon:2011ve,Pierobon:2011vr,noel2014improving,farahnak2019medium}. In this paper, we assume that the medium is divided into voxels \cite{Chou:rdmex_tnb,awan2016reducing,Chou:gc}. The voxel based setting allows us to model the molecular communication networks, which consist of both diffusion and chemical reactions, by using the reaction-diffusion master equation (RDME) \cite{Gardiner}.



In Section \ref{sec:SECTION_1_Introduction}, we discussed two categories of work on improving receiver performance. Another method to classify different receivers that have been used in molecular communication is: those that use chemical reactions at the receiver, e.g. \cite{thomas2016capacity,aminian2015capacity,Chou:2014jca,kuscu2018maximum,Chou:hf}, and those that do not, e.g. \cite{Atakan:2010bj,Einolghozati:2011cj,Pierobon:2013cl,Mahfuz:2011te}. In this work, we assume that the receiver uses chemical reactions at the receiver. 

The demodulation problem considered in this paper can be considered as using data fusion to combine the measurements coming from multiple voxels in order to carry out demodulation. Data fusion has previously been considered in molecular communications in \cite{fang2017convex}. The paper \cite{fang2017convex} assumes that there are multiple receivers; each receiver makes a hard decision for ON-OFF keying and a fusion centre makes an overall decision based on the decisions from the receivers. This paper uses a different method for data fusion. For our proposed demodulator, each voxel computes an approximate log-posteriori probability and we add these probabilities up to obtain the log-posteriori probability of the receiver.




In this paper, we consider using spatial partitioning to reduce the BER in molecular communications. Spatial partitioning has been studied in biophysics, e.g. the authors of \cite{mugler2013spatial} study how spatial partitioning can reduce noise, and the paper \cite{aquino2011optimal} studies the impact of receptor clustering on the noise in cell signalling. In the area of chemical based computation, the authors of \cite{chatterjee2017spatially} use partitioning to improve the performance of a molecular computing system. 

In our previous paper \cite{riaz2018using} we show that the partitioning can be used to reduce the signal variance in the receiver. However, no receiver mechanisms have been proposed. This paper derives the MAP demodulators for both the partitioned and mixed configurations.

\begin{figure}
\centering
\begin{minipage}{.5\textwidth}
  \centering    
    \includegraphics[scale=0.35]{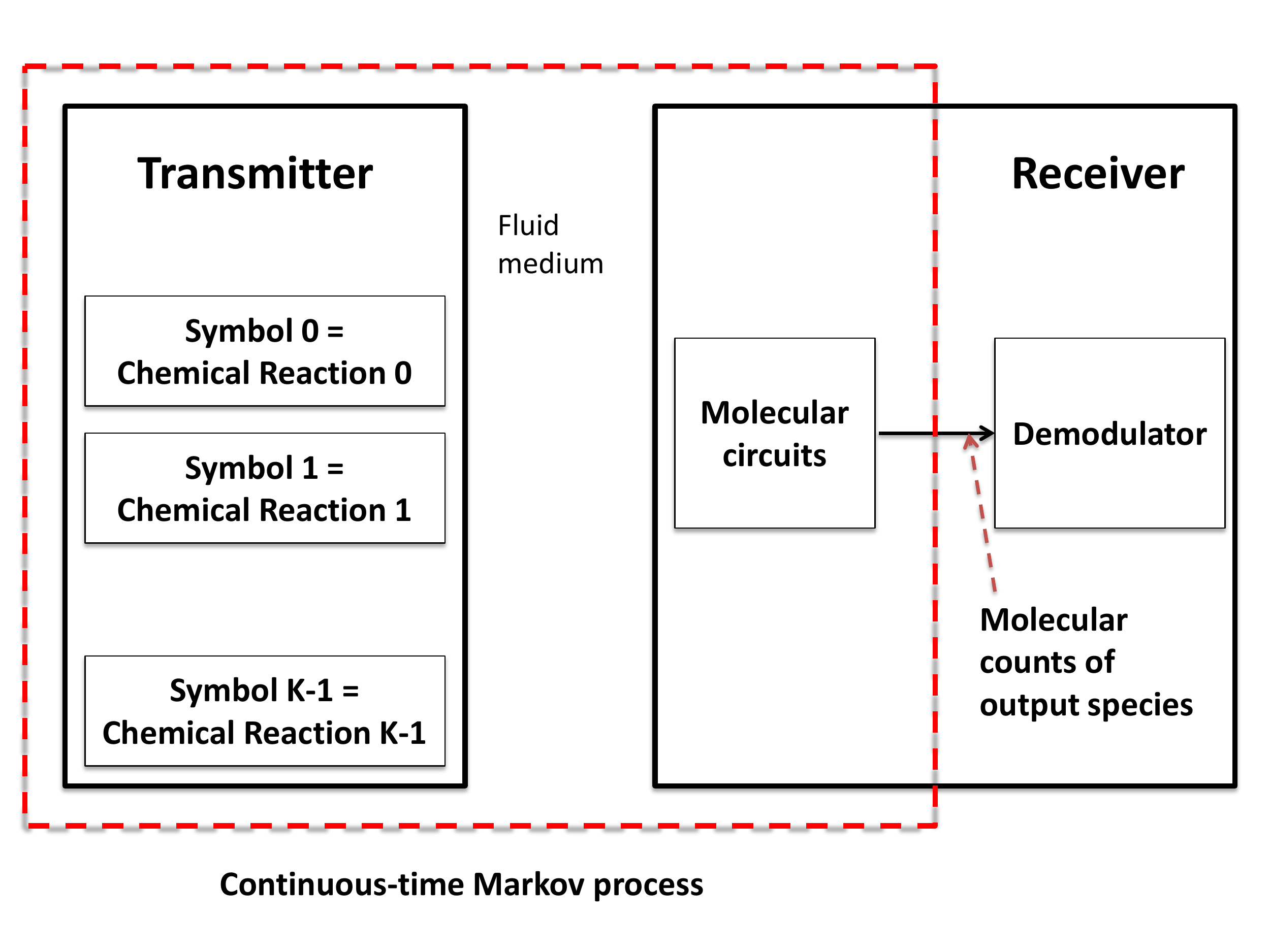}
    \caption{System Overview. }
    \label{fig:FIGURE_1_System_Overview}
    \label{fig:overall}
\end{minipage}%
\begin{minipage}{.5\textwidth}
  \centering
    \centering
    \includegraphics[scale=0.35]{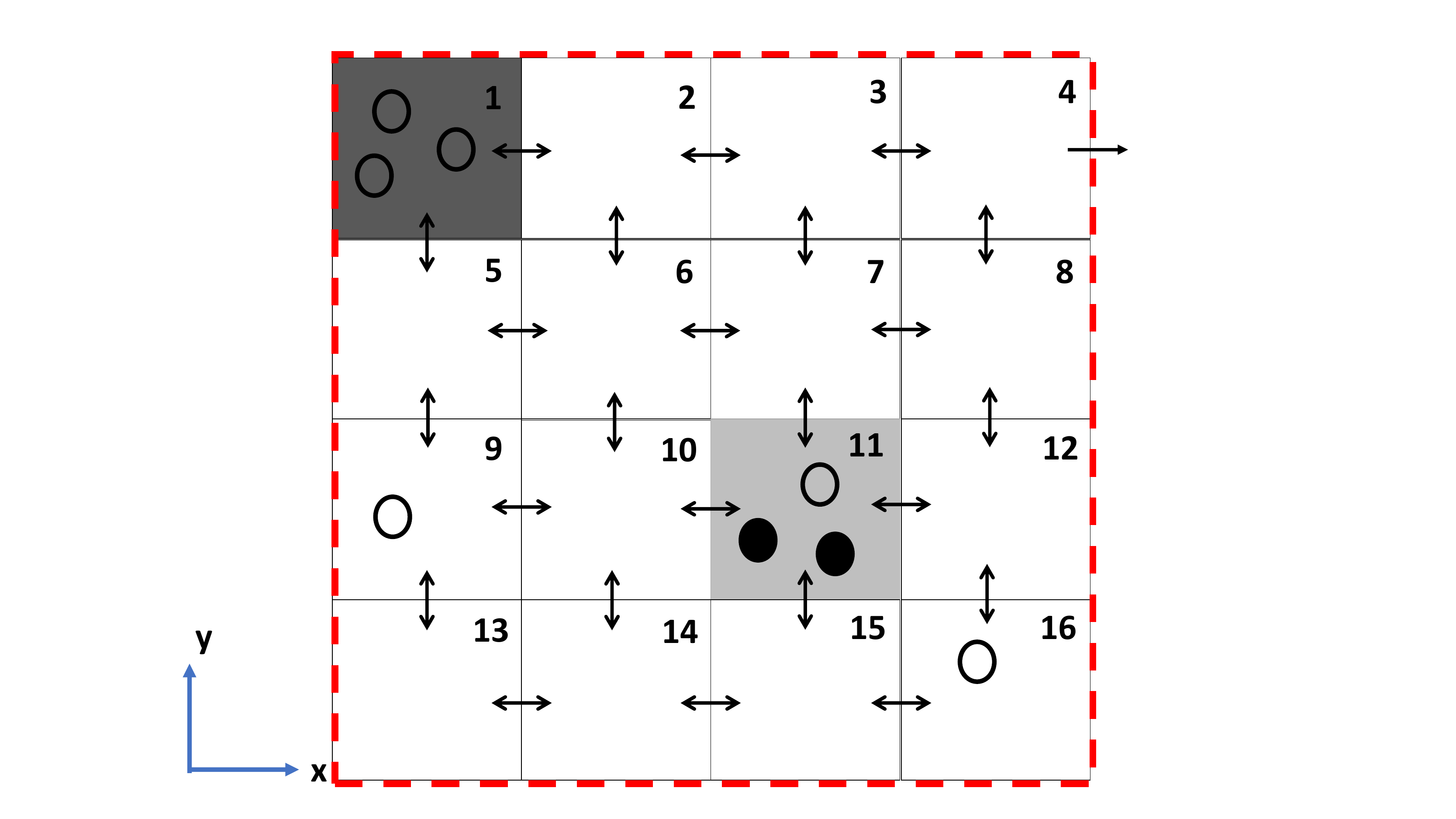}
    \caption{The propagation medium is divided into voxels.}
    \label{fig:FIGURE_2_A_Model_of_Molecular_Communication_Network}
\end{minipage}    
\end{figure}


\section{Background and summary of previous Work} 
\label{sec:SECTION_3_Back_ground} 
This section provides a summary of our previous work in \cite{chou2015markovian,awan2017generalized} on using a Markovian approach to derive a MAP demodulator. This section is divided into two subsections. Section \ref{sec:bg:model} presents the modelling framework while Section \ref{sec:bg:demod} presents the MAP demodulator. 

\subsection{Modelling framework}
\label{sec:bg:model}
In \cite{chou2015markovian,awan2017generalized}, we consider a molecular communication system with a transmitter and a receiver inside a fluid propagation medium. Fig.~\ref{fig:overall} shows an overview of the system. We assume that the transmitter and the receiver communicate with one type of signalling molecules denoted by {\cee S}. We will now discuss the modelling of the system components in further details. 

\subsubsection{Propagation medium} 
\label{sec:prop_medium}
Signalling molecules diffuse from transmitter to receiver through a propagation medium. We model the propagation medium as a rectangular prism. We divide the medium into voxels. This is the same as applying a finite difference spatial discretization to a volume. Fig.~\ref{fig:FIGURE_2_A_Model_of_Molecular_Communication_Network} shows a 2-dimensional projection of a medium consisting of 4-by-4-by-1 voxels. In order to facilitate the description later on, we index each voxel in this example by using integers 1,..,16 which are shown in the top-right corner of each voxel. 
We model the diffusion of the signalling molecules by using spatially discrete jumps between neighbouring voxels. For example, a signalling molecule in Voxel 1 in Fig.~\ref{fig:FIGURE_2_A_Model_of_Molecular_Communication_Network} can diffuse to any of its neighbouring voxels, which are Voxels 2 and 5. The double headed arrows in Fig.~\ref{fig:FIGURE_2_A_Model_of_Molecular_Communication_Network} show the direction of diffusion. We assume the diffusion of the signalling molecules are independent each other. 

We assume the medium is homogeneous with a constant diffusion coefficient $D$ for the signalling molecules. By applying a finite difference discretization to the 3-dimensional diffusion equation \cite{Gardiner}, it can be shown that within an infinitesimal time $\Delta t$, the probability that a signalling molecule will jump from a voxel to a neighbouring voxel is $\frac{D}{w^2} \; \Delta t$ where $w$ denotes the length a voxel edge. 

Lastly, our model can be used to model two types of boundary conditions: reflecting boundary condition where signalling molecules are not allowed to leave the medium; or, absorbing boundary condition where signalling molecules may leave the medium forever, e.g. the single headed arrow in Voxel 4 in Fig.~\ref{fig:FIGURE_2_A_Model_of_Molecular_Communication_Network} shows that signalling molecules may leave the medium.

\subsubsection{Transmitter} 
The transmitter is assumed to occupy only one voxel. The role of the transmitter is to produce the signalling molecules for the transmission symbols. We assume that the transmitter uses RSK. This means that each symbol corresponds to a time-varying concentration profile of signalling molecules produced by a set of chemical reactions, Fig.~\ref{fig:overall} depicts a transmitter which can produce $K$ different transmission symbols. Once the signalling molecules have been produced by the transmitter, they are free to diffuse in the medium.

\subsubsection{Receiver} 
In \cite{chou2015markovian,awan2017generalized}, we assume that the receiver occupies one voxel, which is an assumption which we will change in Section \ref{sec:SECTION_5_map_demodulator} of this paper. In this section, we follow the assumption in \cite{chou2015markovian,awan2017generalized}.  

We divide the operation of the receiver into two blocks, which we will refer to as the front-end and back-end blocks, see Fig.~\ref{fig:overall}.  The front-end block consists of a molecular circuit, which is another name for a set of chemical reactions. In \cite{chou2015markovian}, the front-end block is assumed to be a reversible ligand-receptor binding while in \cite{awan2017generalized}, the front-end block can be any molecular circuit. For example, the following molecular circuit is considered in \cite{chou2018designing}: 
\begin{subequations}
\begin{align}
\cee{
S + X & ->[$g_+$] S + X^* \label{cr:on} \\ 
X^* &  ->[$g_-$] X \label{cr:off}}
\end{align}
\label{cr:all} 
\end{subequations}
where $g_+$ and $g_-$ are propensity function constants; and \cee{X} and \cee{X^*} are, respectively, the inactive and active forms of a species. Reaction \eqref{cr:on} is an activation reaction where the signalling molecule \cee{S} turns inactive \cee{X} into active \cee{X^*}, while Reaction \eqref{cr:off} is a deactivation reaction. 

Note that Table \ref{table:1} contains a list of commonly used constants and chemical symbols for easy referral. 

One or more chemical species in the front-end molecular circuit are chosen as the output species. The molecular counts of the output species over time are the output signals of the front-end, which are fed into the back-end as the input signals, see Fig.~\ref{fig:overall}. The back-end of the receiver is the demodulator whose aim is to infer the symbol that the transmitter has sent by using the input signals to the demodulator, i.e. the molecular counts of the output species of the front-end molecular circuit.


\subsection{MAP demodulator}
\label{sec:bg:demod} 
This section summarises the steps of deriving the MAP demodulator in our Markovian framework \cite{chou2015markovian,awan2017generalized}. For illustration, we assume that the front-end molecular circuit of the receiver is given by Reactions \eqref{cr:all}. We further assume that the molecules \cee{X} and \cee{X^*} can only be found in the receiver voxels, and they are uniformly distributed in the voxel. We also assume that the total number of \cee{X} and \cee{X^*} molecules is a constant $M$. We designate \cee{X^*} as the output species. Let $X(t)$ and $X^*(t)$ denote, respectively, the {\sl number} of $X$ and $X^*$ molecules at time $t$. Note that both $X(t)$ and $X^*(t)$ are piece-wise constant because they are molecular counts. 

We model the transmitter, medium and receiver front-end by using a RDME\footnote{Note that RDME assumes that space is discretised into voxels and time is continuous, therefore RDME is compatible with the discretisation of the medium discussed in Section \ref{sec:prop_medium}. There is another rationale why we choose RDME. For the design of molecular communications systems, we need a stochastic model that can model systems with both diffusion and chemical reactions. There are three main classes of such models: Smoluchowski equation \cite{smoluchowski1918versuch}, RDME and the Langevin equation \cite{Erban:2007we}. The Smoluchowski equation is based on particle dynamics. It is a fine grained model but hard to work with analytically. Both RDME and Langevin are easier to work with analytically but master equation has a finer scale and granularity compared to the Langevin equation \cite{DelVecchio:book}. Therefore we choose to use RDME which allows us to use the Markovian theory for analysis and is at the same time a finer grained model. Note that there is some recent work in combining the voxel-based approach (also known as the mesoscopic approach) and the Smoluchowski equation (also known as the microscopic approach) in simulating systems with both reactions and diffusion, see \cite{noel2017simulating} and \cite{hellander2017mesoscopic}.}. Note that RDME is a specific type of continuous-time Markov process (CTMP). This means the signal $X^*(t)$ is a realization of a CTMP. The RDME assumes that the species in each voxel is well-mixed. For RDME, the position of each molecule is only known to the spatial scale of a voxel; in other words, the exact position (say, in terms of $x$, $y$, $z$ co-ordinates) of each molecules is not known. This means that for RDME, the state of a molecular system is characterised by using the count of each possible type of molecules in each voxel. 

Since \cee{X^*} has been designated as the output species of the front-end molecular circuit, the signal $X^*(t)$ is available to the demodulator. For the derivation of the demodulation filter, we assume that at time $t$, the data available to the demodulation filter is  $X^*(\tau)$ for all $\tau \in [0,t]$. We use ${\cal X}^*(t)$ to denote the continuous-time history of $X^*(t)$ up to time $t$. 
We use the Bayesian framework for demodulation. Let ${\mathbf P}[k | {\cal X^*}(t)]$ denote the posteriori probability that symbol $k$ has been sent given the history ${\cal X^*}(t)$. 
Instead of working with ${\mathbf P}[k | {\cal X^*}(t)]$, we will work with its logarithm. Let $L_k(t) = \log ({\mathbf P}[k | {\cal X^*}(t)])$. Note that the log-posteriori probability diverges in continuous time, but we are able to compute a shifted version of it. For ease of reference, we will simply refer to the shifted version as the log-posteriori probability, We can use the method in \cite{awan2017generalized} to show that we can compute $L_k(t)$ by using the following ordinary differential equation (ODE):
\begin{align}
\frac{dL_k(t)}{dt} =& \left[ \frac{dX^*(t)}{dt} \right]_+ \log( {\mathbf E}[N_R(t) | k, {\cal X^*}(t)] ) - 
g_+ (M - X^*(t)) {\mathbf E}[N_R(t) | k, {\cal X^*}(t)]
\label{eqn:logpp_dd} 
\end{align}
where $[\xi]_+ = \max(\xi,0)$ and ${\mathbf E}[N_R(t) | k, {\cal X^*}(t)]$ is the estimation of the mean number of signalling molecules in the receiver voxel. Recall that $X^*(t)$ is the count of \cee{X^*} molecules at time $t$. It is therefore a piecewise constant signal and the derivative $\left[ \frac{dX^*(t)}{dt} \right]_+$ is a series of Dirac deltas situated at the times at which the count of \cee{X^*} changes.

\begin{table}[]
 \centering
 \caption{Notation and Chemical Symbols}
 \begin{tabular}{|c|l|}
 \hline
 \multicolumn{1}{|c|}{Symbols}	&	\multicolumn{1}{|c|}{Notation and Value}	\\ \hline
 $w$ &       Dimension of one side of a voxel
 \\ \hline
 $D$ &      diffusion coefficient of signalling molecules  
 \\ \hline
 $d$ &     Inter-voxel diffusion constant for signalling molecules 
 \\ \hline
 $d_r$ &   Inter-voxel diffusion constant for \cee{X^*}    
 \\ \hline
 $g_+$ &    Propensity function constant for the activation reaction \eqref{cr:on}
 \\ \hline
 $g_-$ &      Propensity function  constant for deactivation reaction \eqref{cr:off}
 \\ \hline
  $P$      &     Number of receiver voxels \\ \hline 
\cee{S}, \cee{X}, \cee{X^*}&   Chemical symbols for signalling molecule, inactivate receptor and active receptor 
 \\ \hline
$X_p(t)$ ($X^*_p(t)$) &  Number of active (inactive) receptors in the $p$-th receiver voxel at time $t$    \\ \hline
$N_{R,p}(t)$  &  Number of signalling molecules in the $p$-th receiver voxel at time $t$ 
 \\ \hline
 \end{tabular}
 \label{table:1}
 \end{table}

The computation of ${\mathbf E}[N_R(t) | k, {\cal X^*}(t)]$ requires extensive computation because it requires an optimal Bayesian filtering problem to be solved\footnote{ In order to solve the optimal Bayesian filtering problem, we need to track the probability of each possible system state. For our problem, the system state is the count of each possible type of molecule in each voxel. The total number of possible states grows rapidly due to combinatorial explosion. This explanation also applies to subsequent problems considered in this paper.}. In \cite{chou2015markovian}, we propose to replace the hard-to-compute ${\mathbf E}[N_R(t) | k, {\cal X^*}(t)]$ by ${\mathbf E}[N_R(t) | k]$, which is the mean number of signalling molecules if Symbol $k$ is transmitted. Let $\sigma_k(t)$ denote ${\mathbf E}[N_R(t) | k]$. With the proposed replacement, Eq.~\eqref{eqn:logpp_dd} becomes:
\begin{align}
\frac{dZ_k(t)}{dt} =  \left[ \frac{dX^*(t)}{dt} \right]_+ \log(\sigma_k(t)) - g_+ (M - X^*(t)) \sigma_k(t)
\label{eqn:logmap_s}
\end{align}
where $Z_k(t)$ is an approximation of $L_k(t)$. We can interpret Eq.~\eqref{eqn:logmap_s} as using $\sigma_k(t)$ as an internal model or prior knowledge for demodulation. The use of internal model is fairly common in signal processing and communications, e.g. in matched filtering. 

Note that Eq.~\eqref{eqn:logpp_dd} is the optimal solution for the demodulation problem. The replacement of ${\mathbf E}[N_R(t) | k, {\cal X^*}(t)]$ by ${\mathbf E}[N_R(t) | k]$ means that Eq.~\eqref{eqn:logmap_s} is a approximate solution. We show in \cite{chou2015markovian} that $Z_k(t)$ from Eq.~\eqref{eqn:logmap_s} approximates $L_k(t)$ in Eq.~\eqref{eqn:logpp_dd} well. 

\begin{figure}
\centering
\begin{minipage}{.5\textwidth}
  \centering    
    \includegraphics[scale=0.38]{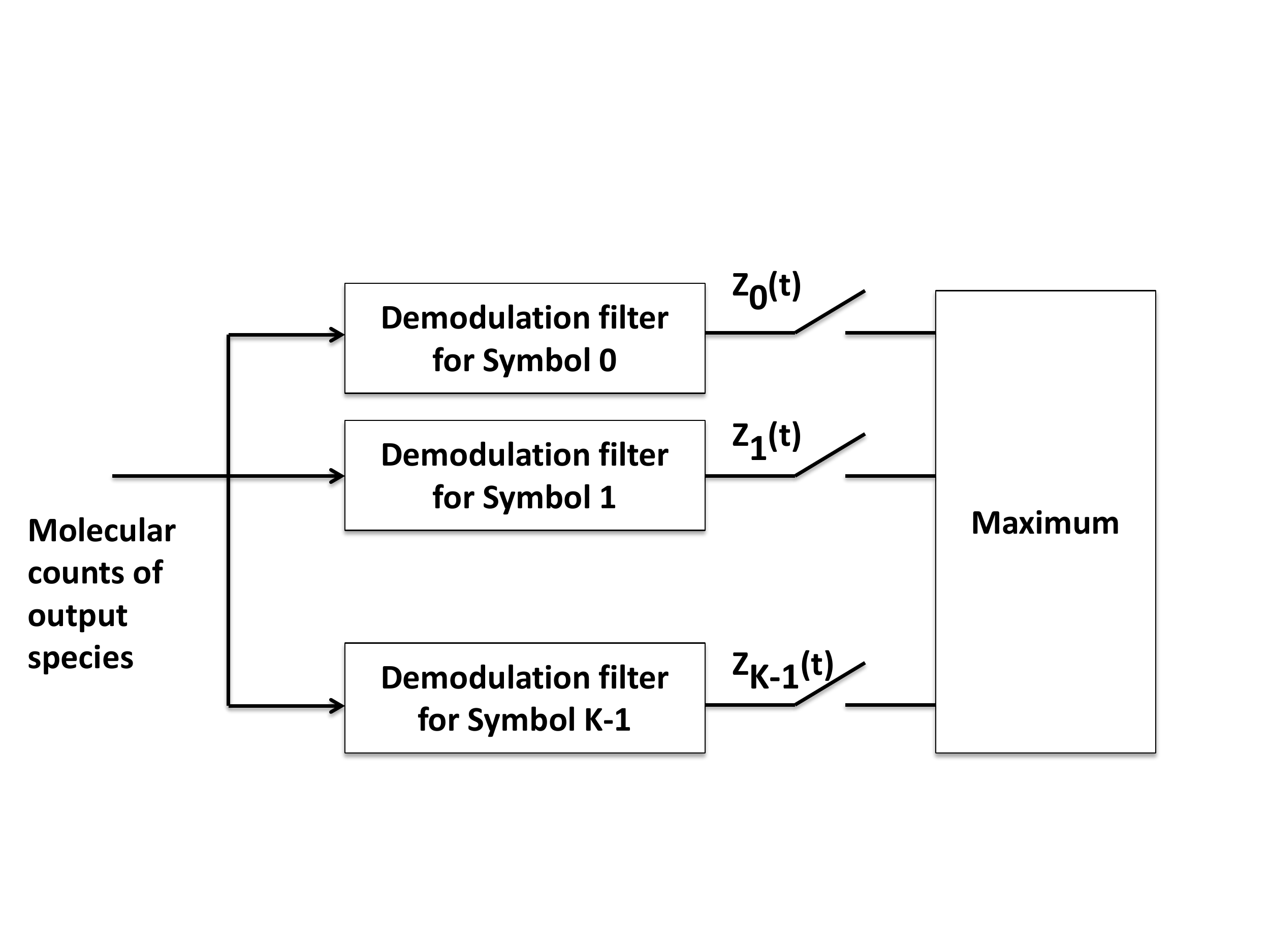}
     \caption{The structure of the demodulator.}
 \label{FIGURE_4}
\end{minipage}%
\begin{minipage}{.5\textwidth}
  \centering
    \centering
    \includegraphics[scale=0.2]{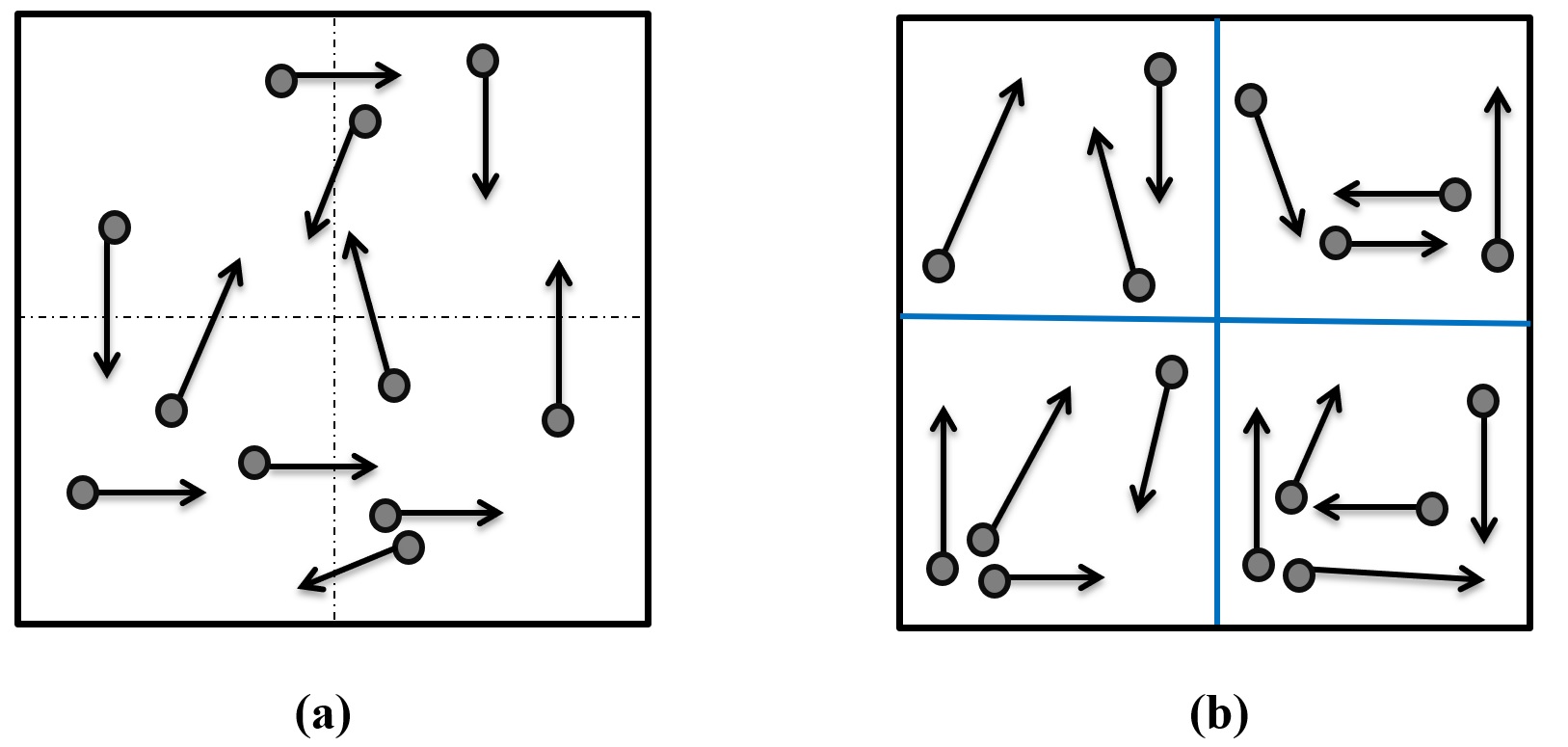}
    \caption{(a) The mixed configuration. (b) The partitioned configuration.}
\label{fig:FIGURE_3_partVSmix}
\end{minipage}    
\end{figure}

To make the decision at time $t$, the demodulator decide symbol $\hat{k}$ is transmitted if $\hat{k} = {\arg\max}_{k = 0, ..., K-1} Z_k(t)$. Also, $Z_k(0)$ is initialized to the logarithm of the prior probability that the transmitter sends Symbol $k$. The structure of the demodulator is depicted in Fig.~\ref{FIGURE_4}. The demodulator consists of $K$ filters computing $Z_k(t)$ and a maximum block to determine the transmission symbol. 

\section{Receiver with multiple voxels}
\label{sec:SECTION_5_map_demodulator}
\label{sec:multi_voxel}

This section derives the demodulator when the receiver consists of multiple voxels. We will consider both the partitioned and mixed configurations.  

\subsection{Multi-voxel receiver demodulation problem}
We assume the receiver consists of $P$ voxels where $P > 1$. Without loss of generality, we assume these $P$ voxels form a connected volume. We will use $p$ to index the voxels in the receiver, where $p = 1, \ldots, P$. We will explain the demodulation problem using an example. 

\begin{example}
\label{ex:demod}
In this example, we assume the molecular circuit \eqref{cr:all} is present in all the $P$ receiver voxels. We will continue to use the RDME to model the system. The RDME framework requires us to distinguish between the \cee{S}, \cee{X} and \cee{X^*} in different voxels. We will use \cee{S_p}, \cee{X_p} and \cee{X^{*}_{P}} to denote the \cee{S}, \cee{X} and \cee{X^*} in the $p$-th receiver voxel where $p = 1, ..., P$. The $P$ molecular circuits in the receiver voxels are:
\begin{subequations}
\begin{align}
\cee{
S_p + X_p &  ->[$g_+$] S_p + X^{*}_{P} \label{cr:general_on} \\
X^{*}_{P} &  ->[$g_-$] X_p \label{cr:general_off}}
\end{align}
\label{cr:general_all} 
\end{subequations}
We designate \cee{X^{*}_{P}}, for $p = 1,\ldots,P$, as the output species of the receiver. Let $X_p(t)$ and $X^{*}_{p}(t)$ denote respectively the number of \cee{X_p} and \cee{X^{*}_{P}} molecules at time $t$. 

Let also ${\cal X}^{*}_{P}(t)$ be the continuous history of the species \cee{X^{*}_{P}} up to and including time $t$. Our goal is to determine the posteriori probability ${\bf P}[k | {\cal X}^{*}_{1}(t), \ldots, {\cal X}^{*}_{P}(t)]$ that the $k$-th transmitter symbol has been sent given the histories ${\cal X}^{*}_{1}(t)$, $\ldots$, ${\cal X}^{*}_{P}(t)$. \hfill $\Box$
\end{example}

Note that it is possible to generalise the above examples in a few different ways: we can use molecular circuits other than \eqref{cr:all}; we can use different molecular circuits in different receiver voxels; we can use different choices of output species. We note that the methodology that we have developed to derive the demodulator can deal with all these generalisations. 

We make the assumption that the species in the front-end molecular circuits (e.g. \cee{X} and \cee{X^*} in \eqref{cr:all}) can only be found in the receiver. Also these species are confined within the receiver and cannot get into the propagation environment. In the text below, we will to refer to them as the receiver species. 

\subsection{Mixed and partitioned configurations}
\label{Mixed_and_partitioned configurations}
Inspired by the study \cite{mugler2013spatial} which shows that spatial partitioning of the receptors on cell membrane can be used to reduce noise, we will consider two different configurations for the receiver. For illustration, we assume the receiver consists of 4 voxels arranged in 2-by-2 configuration as shown in Fig.~\ref{fig:FIGURE_3_partVSmix}. 

In the first configuration, which is illustrated in Fig.~\ref{fig:FIGURE_3_partVSmix}a, a receiver species is allowed to diffuse from a receiver voxel to another receiver voxel. We will refer to this configuration as mixed as the receiver species is allowed to mix among the receiver voxels. 

In the second configuration, which is illustrated in Fig.~\ref{fig:FIGURE_3_partVSmix}b, a receiver species cannot move from a receiver voxel to another. In this case, we can assume the receiver voxel wall is a selective membrane which prevent certain species from moving between voxels. We will refer to this configuration as partitioned. 

We note that the voxel framework is well suited to model the partitioned and mixed configurations. This is because, for the voxel framework, it is possible to choose the value of the diffusion coefficient of a species between any two voxels. If a species is not allowed to pass between the interface of two voxels, then its diffusion coefficient for that interface is zero.

We now discuss how the two configurations can be realised. We assume that the cross-sectional area of the signalling molecules is small while that of the receiver species is large. We assume that the boundaries between receiver voxels are made of membrane with pores on it. If we choose the size of the pores to be large enough for the signalling molecules to pass through unhindered but small enough to prevent the receiver species from passing through, then we obtain the partitioned configuration. If we increase the size of the pores to allow the receiver species to pass through them with a non-zero probability, then we get the mixed configuration. A possible physical implementation is based on structural DNA nanotechnology \cite{Ke:2018jq,Xavier:2018gv} which can create 3-dimensional wireframes with pores of different shapes and dimension, e.g. \cite{Han:2013ju} demonstrated a design with pores down to approximately 6nm by 6nm. In nature, many signalling molecules and receptors are protein. The cross-sectional area of large protein molecules can be 10nm or more, while that of small protein molecules is a few nanometres \cite{Milo:BioNumbers}.

We will now derive the MAP demodulator for the partitioned configuration.  

\subsection{MAP demodulator for the partitioned configuration} 
\label{sec:demod:part}
In this section, we will derive the MAP demodulator for the demodulation problem illustrated in Example \ref{ex:demod}. In particular, in this section, we consider the partitioned configuration where the species \cee{X} and \cee{X^*} are not allowed to move among the receiver voxels. This means the total number of \cee{X} and \cee{X^*} molecules in a receiver voxel remains constant. We will use $M_p$ to denote the total number of \cee{X} and \cee{X^*} molecules in the $p$-th receiver voxel. 

The aim of the demodulation problem is to determine the posteriori probability ${\bf P}[k | {\cal X}^{*}_{1}(t), \ldots, {\cal X}^{*}_{P}(t)]$ that the $k$-th transmitter symbol has been sent given the histories ${\cal X}^{*}_{1}(t)$, $\ldots$, ${\cal X}^{*}_{P}(t)$. A key step in the derivation is to compute the probability ${\bf P}[X^{*}_{1}(t+\Delta t), X^{*}_{2}(t+\Delta t), \ldots, X^{*}_{P}(t+\Delta t) | k, {\cal X}^{*}_{1}(t), \ldots, {\cal X}^{*}_{P}(t)]$ which predicts the counts of the output species based on their histories. This is an optimal Bayesian filtering problem. The structure of this filtering problem is identical to the one considered in \cite{awan2017generalized}. In \cite{awan2017generalized}, we consider a filtering problem where the state vector evolves according to a CTMP. In that problem, only some elements of the state vector can be observed and the aim is to predict the future values of the observable elements from their past histories. We can therefore apply the method in \cite{awan2017generalized} to the demodulation problem considered in this paper. We remark that although the presentation in \cite{awan2017generalized} states that the output species come from one voxel, it is the structure of the filtering problem that really matters; in other words, it is not important whether the output species come from one or many voxels. 

In order to simplify the notation, we will use the shorthand ${\cal X}^{*}_{R}(t)$ to denote the histories ${\cal X}^{*}_{1}(t)$, $\ldots$, ${\cal X}^{*}_{P}(t)$ from all the receiver voxels; note that the subscript R in ${\cal X}^{*}_{R}(t)$ is short for receiver. Let $L_k(t) = \log({\bf P}[k | {\cal X}^{*}_{R}(t)]$ be the log-posteriori probability that the $k$-th symbol has been sent given the histories. \iftcom
In Appendix A of our technical report \cite{riaz2019using} , we show that $L_k(t)$ evolves according to:
\else
In Appendix \ref{app:part}, we show that $L_k(t)$ evolves according to:
\fi
\begin{align}
\frac{dL_k(t)}{dt} =& 
\sum_{p=1}^{P} ( \left[ \frac{dX^{*}_{p}(t)}{dt} \right]_+  \log( {\mathbf E}[N_{R,p}(t) | k, {\cal X}^{*}_{R}(t)] ) - 
g_+  (M_p - X^{*}_{p}(t)){\mathbf E}[N_{R,p}(t) | k, {\cal X}^{*}_{R}(t)] ) \label{MAP_generalized_partitioned} 
\end{align}
where ${\mathbf E}[N_{R,p}(t) | k, {\cal X}^{*}_{R}(t)]$ is the estimated number of signalling molecules in the $p$-th receiver voxel given $k$ and the histories ${\cal X}^{*}_{R}(t)$.

Although the term ${\mathbf E}[N_{R,p}(t) | k, {\cal X}^{*}_{R}(t)]$ can be computed by solving an optimal Bayesian filtering problem, its computational complexity is high. Another issue, which is found in the multi-voxel receiver case but not the single-voxel receiver case, is that the estimation of the mean number of signalling molecule in the $p$-th receiver voxel via ${\mathbf E}[N_{R,p}(t) | k, {\cal X}^{*}_{R}(t)]$ requires the history in the $p$-th receiver voxel as well as other receiver voxels. In other words, the computation of ${\mathbf E}[N_{R,p}(t) | k, {\cal X}^{*}_{R}(t)]$ requires communications between receiver voxels, which adds burden to the receiver. We follow our earlier work \cite{chou2015markovian}\cite{awan2017generalized} and replace ${\mathbf E}[N_{R,p}(t) | k, {\cal X}^{*}_{R}(t)]$ with the prior knowledge ${\mathbf E}[N_{R,p}(t) | k]$, which will be denoted by $\alpha_{p}(t)$. 
We will use a numerical example in Section \ref{sec:num:sub:part} to demonstrate that this approximation is accurate.
With this replacement, an approximate demodulator is: 
\begin{align}
\frac{Z_k(t)}{dt} =& \sum_{p=1}^{P} Z_{k,p}(t) 
\label{eqn:generalized_partitioned} \\
Z_{k,p}(t) =& 
\left[ \frac{dX^{*}_{p}(t)}{dt} \right]_+ \log(\alpha_{p}(t)) - g_+ (M_p - X^{*}_{p}(t)) \alpha_{p}(t)
\label{eqn:part:zkp} 
\end{align}
We can interpret $Z_{k,p}(t)$ in Eq.~\eqref{eqn:part:zkp} as the approximate log-posteriori probability computed using the measurements $X^{*}_{p}(t)$ from the $p$-the voxel. These approximate log-probabilities are then added up or fused in Eq.~\eqref{eqn:generalized_partitioned} to obtain the approximate log-posteriori probability of the receiver. 

\begin{figure}
\begin{center}
\begin{minipage}{.6\textwidth}
  \centering    
    \includegraphics[width=9cm,height=8cm]{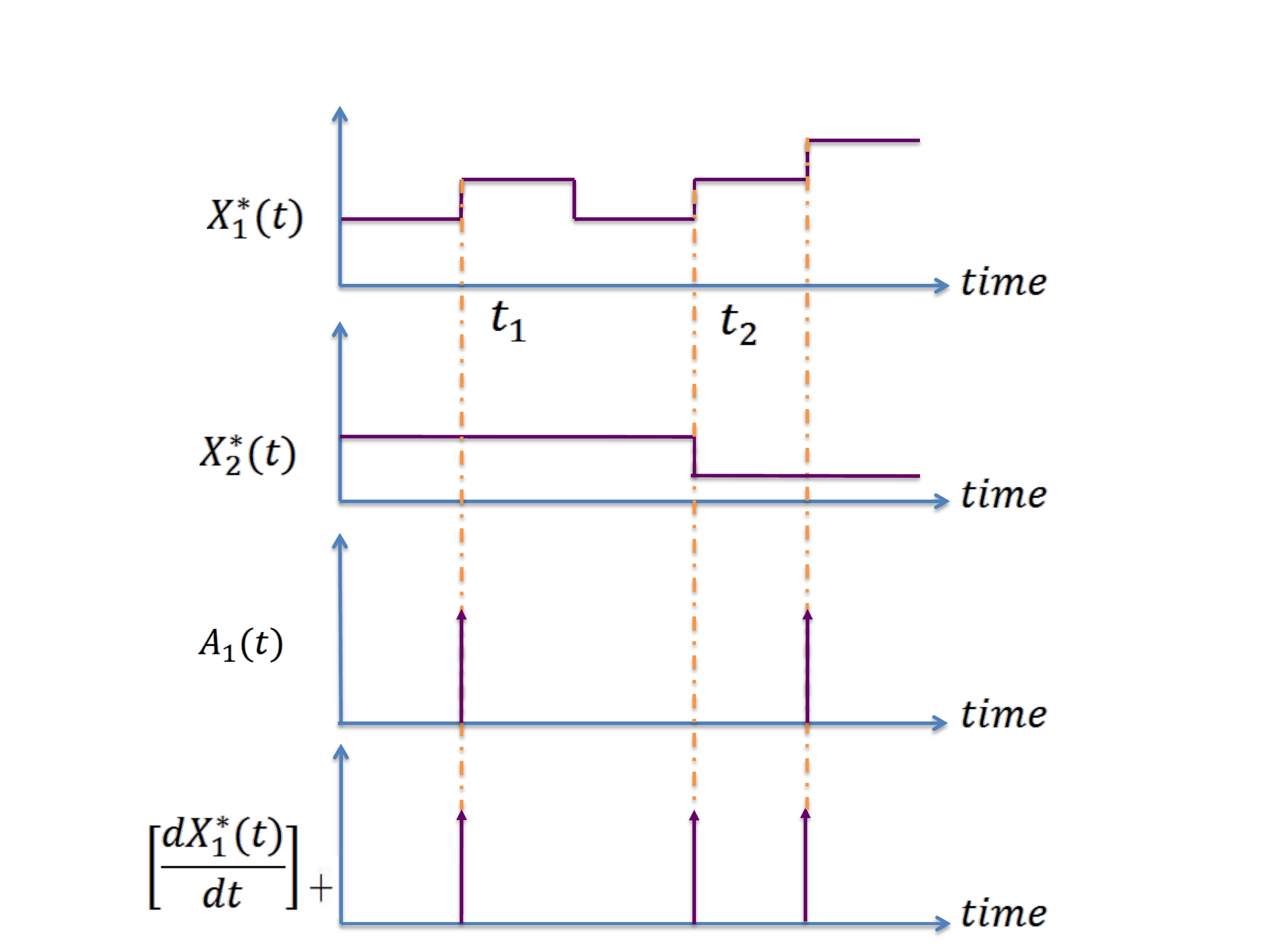}
    \caption{Illustrating the difference between $A_p(t)$ and $\left[ \frac{dX^{*}_{p}(t)}{dt} \right]_+ $.}
    \label{fig:ap}
\end{minipage} 
\end{center}
\end{figure}

\subsection{MAP demodulator for the mixed configuration}
\label{2222_mixed}
This section considers the demodulation for the mixed configuration. For the mixed configuration, we assume that the \cee{X} and \cee{X^*} molecules are allowed to diffuse between the receiver voxels but they are not allowed to leave the receiver. Let $D_r$ be the diffusion coefficient of \cee{X^*} in the receiver voxels. Note that we do not specify the diffusion coefficient of \cee{X} because it does not enter the ODE for computing log-posteriori probability.  

If \cee{X} and \cee{X^*} are not allowed to diffuse, i.e. for the partitioned case as discussed in Section \ref{Mixed_and_partitioned configurations}, then $X_p(t) + X^{*}_{P}(t)$ is a constant for all receiver voxels $p = 1, \ldots, P$. However, this condition no longer holds for the mixed configuration. This means that $M_p$ are not parameters of the mixed configuration. 

The aim of the demodulation problem is to determine the posteriori probability ${\bf P}[k | {\cal X}^{*}_{1}(t), \ldots, {\cal X}^{*}_{P}(t)]$ that the $k$-th transmitter symbol has been sent given the histories ${\cal X}^{*}_{1}(t)$, $\ldots$, ${\cal X}^{*}_{P}(t)$. We again use ${\cal X}^{*}_{R}(t)$ to denote the histories. 

\iftcom
In Appendix B of our technical report \cite{riaz2019using}  we derive the optimal demodulation filter:
\else
In Appendix \ref{app:mixed}, we derive the optimal demodulation filter:
\fi
\begin{align}
\frac{dL_k(t)}{dt} =& 
\sum_{p = 1}^P (
A_{p}(t)  \log( {\mathbf E}[X_p(t) N_{R,p}(t) | k, {\cal X}^{*}_{R}(t))] ) - 
g_+  {\mathbf E}[X_p(t) N_{R,p}(t) | k,  {\cal X}^{*}_{R}(t)] ) 
\label{eqn:demod:mixed:opt} 
\end{align}
where $A_p(t)$ is a sequence of Dirac deltas situated at the times where activation reactions \eqref{cr:on} occur in voxel $p$ according to the given histories and ${\mathbf E}[X_p(t) N_{R,p}(t) | k, {\cal X}^{*}_{R}(t))]$ is the estimated mean of the product of $X_p(t)$ and $N_{R,p}(t)$ by using the histories. 

 We now consider how the terms in \eqref{eqn:demod:mixed:opt} can be approximated. We first use an example to point out that $A_p(t)$ is different from $\left[ \frac{dX^{*}_{p}(t)}{dt} \right]_+ $. We illustrate this by assuming that the receiver consists of two neighbouring receiver voxels, which we will refer to as RV1 and RV2. Fig.~\ref{fig:ap} shows two sample trajectories for $X^*_1(t)$ and $X^*_2(t)$, which are of the number of \cee{X^*} molecules in RV1 and RV2. The figure also depicts the signals $A_1(t)$ and $\left[ \frac{dX^*_1(t)}{dt} \right]_+$ for RV1. At time $t_1$, an activation reaction occurs in RV1 and $X^*_1(t)$ is increased by 1 and therefore a Dirac delta is situated at time $t_1$ in both $A_p(t)$ and $\left[ \frac{dX^{*}_{P}(t)}{dt} \right]_+ $. We remark that $X^*_2(t)$ remains unchanged at $t_1$, which is a fact that we will use later. At time $t_2$, an \cee{X^*} molecule moves from RV2 to RV1. This means $X^*_1(t)$ is increased by 1 while $X^*_2(t)$ is decreased by 1. We note that at time $t_2$, there is a Dirac delta in $ \left[ \frac{dX^{*}_{1}(t)}{dt} \right]_+ $ but not $A_1(t)$. 

The above example shows that Dirac deltas in $A_1(t)$ occur at the times at which $X^*_1(t)$ is increased by 1 while $X^*_2(t)$ remains the same. This means the computation of $A_1(t)$ requires both the trajectories of $X^*_1(t)$ and $X^*_2(t)$. In general, the computation of $A_p(t)$ requires $X^*_p(t)$ as well as the histories of \cee{X^*} in the neighbouring voxels. This implies that the computation $A_p(t)$ requires communications between neighbouring voxels, which is not realistic. An alternative is to change the activation reaction \eqref{cr:on} so that every time when an activation reaction occurs, an extra molecule is produced to indicate its occurrence. However, there are a number of open issues with this approach: how to ensure this extra molecule is local both in time and space, and to make sure the continual supply of this extra molecule. In this paper, we will approximate $A_p(t)$ by $\left[ \frac{dX^{*}_{P}(t)}{dt} \right]_+ $. We are conscious that $\left[ \frac{dX^{*}_{P}(t)}{dt} \right]_+ $ has more Dirac deltas than $A_p(t)$ and these extra Dirac deltas are due to the movement of \cee{X^*} into receiver voxel $p$ from its neighbouring receiver voxels. 
We will demonstrate in Section \ref{sec:SECTION_6_Numerical_Examples} that if \cee{X^*} diffuses slowly, then the error is small. 

In order to avoid solving an optimal Bayesian filtering problem, we will replace
${\mathbf E}[X_p(t) N_{R,p}(t) | k, {\cal X}^{*}_{R}(t))]$ in \eqref{eqn:demod:mixed:opt} by ${\mathbf E}[X_p(t) N_{R,p}(t) | k]$, which is denoted by $\beta_{k,p}(t)$. This replacement also removes the need for the voxels to communicate with each other. We will show that this approximation is accurate in Section \ref{sec:num:sub:mixed}. With these approximations, Eq.~\eqref{eqn:demod:mixed:opt} becomes:

\begin{align}
\frac{Z_k(t)}{dt} \approx& 
\sum_{p = 1}^P \left( 
\left[ \frac{dX^{*}_{p}(t)}{dt} \right]_+  \log(\beta_{k,p}(t)) - 
 g_+ \beta_{k,p}(t) \right) 
\label{eqn:demod:mixed}
\end{align}
This will be the approximation demodulation filter for the $k$-th symbol for the mixed configuration. 




\begin{remark}
Although we have presented the results assuming that the molecular circuit is given by Reactions \eqref{cr:all}. Our methodology can be used for any molecular circuit. The key to deriving the demodulator is to compute the counterpart of the probability ${\bf P}[X^{*}_{1}(t+\Delta t), X^{*}_{2}(t+\Delta t), \ldots, X^{*}_{P}(t+\Delta t) | k, {\cal X}^{*}_{1}(t), \ldots, {\cal X}^{*}_{P}(t)]$ when different molecular circuits are used. In words, this probability can be stated as: $\mathbf{P}[$ counts of all the output species from all receiver voxels at time  $t + \Delta t| k$, histories of the counts of all the output species from all receiver voxels up to time  t$]$. We can use the method in \cite{awan2017generalized} to determine this probability. $\Box$
\end{remark}

\begin{remark}
It is possible for a receiver to have both mixed and partitioned configurations. The derivation of the filters for this case is straightforward. We use the summand of the RHS of Eq.~\eqref{MAP_generalized_partitioned} for those voxels that are partitioned and that of Eq.~\eqref{eqn:demod:mixed:opt} for those voxels that are mixed. After that we sum up the log-posteriori probability contributed by each voxel. 

We mentioned earlier that there is no loss of generality to consider all $P$ voxels being connected. If the receiver does not form one connected component, then we can compute the log-posteriori probability for each connected component and sum them up. \hfill $\Box$
\end{remark}

\begin{remark}
We would like to discuss how the approximate demodulation filters may be implemented. For the partitioned case, we first point out that \eqref{MAP_generalized_partitioned} has the same form as \eqref{eqn:logmap_s}. We have recently derived a molecular circuit implementation of \eqref{eqn:logmap_s} for the case where the transmission symbols are concentration-modulated signals in \cite{chou2018designing}. By leveraging these results, we have recently presented a molecular circuit implementation for the partitioned case in \cite{riaz2019nanocom}. A work-in-progress is to implement the approximate demodulation filters for the mixed configuration. 

We have so far assumed that the channel is stationary. An interesting open research problem is to consider a time-varying communication channel. This would require us to adapt the molecular circuit implementation of the demodulation filters to the channel. This appears to be a challenging problem but we want to point out that molecular circuit implementation of some relevant tools do exist, e.g. least-squares estimation \cite{Zechner:2016is} and feedback control for tracking \cite{Briat:2016ha}.

$\Box$
\end{remark}

\section{Analytical approximation of BER}
\label{sec:LNA} 

This section presents an analytical method to compute the BER of the sub-optimal demodulators for the partitioned and mixed configurations. In order to analytically compute the BER, we need to derive the joint probability distribution of $Z_0(t)$, $Z_1(t)$, ... , $Z_{K-1}(t)$, where $Z_k(t)$ is the approximate log-posteriori probability given in \eqref{eqn:part:zkp}  or \eqref{eqn:demod:mixed}. This in turn requires us to compute the probability distribution of the molecular counts of the species in the system, whose exact solution is given by the solution of the RDME. However, for systems with nonlinear reaction rates, which is the case for our system, it is not possible to solve the RDME analytically because of the moment closure problem \cite{Munsky:2006es} which refers to the issue that the computation of the lower order moments requires the values of the higher order moments. An alternative method is the Linear Noise Approximation (LNA) \cite{Gardiner,Cardelli:2016bp} which uses Gaussian distribution to approximate the distribution of the molecular counts when the system is in the steady state. The LNA is known to be a good approximation when the number of the molecules in the system is large. We will use LNA in this section. 

In this section, we use a system which consists of three voxels, to illustrate the method. The voxels are arranged in a line and are numbered as 1, 2 and 3. The transmitter voxel is Voxel 3 and the two receiver voxels are in Voxels 1 and 2. Receiver voxel 1 (resp. 2) is in Voxel 1 (2). Note that LNA is based on concentration rather than molecular count. Earlier, we have used the notation $X_p(t)$, $X^*_p(t)$, $N_{R,p}(t)$ and $N_{i}(t)$ to denote molecular counts. Here we will use the corresponding lowercase notation --- $x_p(t)$, $x^*_p(t)$, $n_{R,p}(t)$ and $n_{i}(t)$ --- to denote the concentration. Let $V = w^3$ be the volume of a voxel, then we have $x_p(t) = \frac{X_p(t)}{V}$ etc. Also, when concentration is used, the reaction rates are expressed in terms of reaction rate constant rather than propensity function constants. For the activation reaction \eqref{cr:on}, we denote the reaction rate constant by $\hat{k}_+$ and it can be shown that the reaction rate constant $\hat{k}_+$ and propensity function constant $k_+$ are related by $\hat{k}_+ =  V k_+$. For the inactivation reaction \eqref{cr:off} and diffusion between voxels, the rate constants are the same as the propensity function constants so we will not define new notation for them. We begin by explaining how the mean quantities can be computed. We will focus on the mixed configuration because it is the more complicated case and it provides us with insights on the property of the approximate demodulator. 

\subsection{Mean molecular counts and reference signal} 
\label{sec:LNA:mean}
We first explain how the mean concentration of all species in the mixed configuration can be computed. For notation, we will add an overhead bar, e.g. $\bar{x}^*_1(t), \bar{X}^*_1(t)$ etc, to denote the mean quantity. 

The derivation is based on the reaction rate law \cite{Erban:2009us}. We will use the concentration of \cee{X^*} in receiver voxel 1 as an example. The concentration of this species is affected by two reactions and two diffusion events: (i) activation reaction at a rate of $\hat{k}_+ \bar{n}_{R,1}(t) \bar{x}_1(t)$; (ii) deactivation reaction at a rate of $k_- \bar{x}^*_1(t)$; (iii) diffusion of \cee{X^*} from receiver voxel 1 to receiver voxel 2 at a rate of $d_r \bar{x}^*_1(t)$; (iv) diffusion of \cee{X^*} from receiver voxel 2 to receiver voxel 1 at a rate of $d_r \bar{x}^*_2(t)$. Therefore the rate of change of $\bar{x}^*_1(t)$ is:
\begin{align}
\frac{d \bar{x}^*_1(t)}{dt} =& \hat{k}_+ \bar{n}_{R,1}(t) \bar{x}_1(t) - k_- \bar{x}^*_1(t) - d_r \bar{x}^*_1(t) + d_r \bar{x}^*_2(t)
\end{align} 
We can use the same method to write down the ODEs that govern the evolution of all species in the system. 
\iftcom
	The full system of ODEs is in Appendix C of our technical report \cite{riaz2019using}. 
\else
	The full system of ODEs is in Appendix \ref{app:LNA}. 
\fi
The solution of this system of ODEs will give us the mean concentration. The mean species count can then be computed by multiplying the mean concentration by the voxel volume. For the mixed configuration, we require $\beta_{k,p}(t) = {\mathbf E}[X_p(t) N_{R,p}(t) | k]$ and we propose to approximate this by
\begin{align}
 \frac{d \beta_{k,p}(t)}{dt} =& V^2 \left( \bar{x}_p(t)  \frac{d \bar{n}_{R,p}(t)}{dt}   +  \frac{d \bar{x}_p(t) }{dt}   \bar{n}_{R,p}(t)   \right)
\end{align}


\subsection{Second order statistics}
\label{sec:LNA:cov}
This section explains how the second order statistics of the molecular counts can be computed. We begin with using a simple example to explain the intuition behind the computation. Consider a chemical reaction that takes place at a constant mean rate of $r$ reactions per unit volume per unit time, then the number of reactions $\rho$ that occurs in a time interval $\tau$ in a volume $V$ is Poisson distributed with mean $Vr\tau$. By using Gaussian approximation, we can write $\rho$ as $V r \tau + \sqrt{V r\tau} \xi$ where $\xi$ is a standard Gaussian random variable and we have used the fact that the variance of a Poisson random variable is equal to its mean. One key idea of the LNA is to approximate the reaction rate using this method. However, additional approximations are required because the reaction rate are nonlinear and one also needs to approximate the square root term. Let us use $x^*_1(t)$ as an example. For LNA, the second order statistics of $x^*_1(t)$ is computed from $\sqrt{V} \widetilde{x}^*_1(t)$ where $\widetilde{x}^*_1(t)$ is a Gaussian random variable whose evolution is governed by:
\begin{align}
\frac{d \widetilde{x}^*_1(t)}{dt} =& 
\hat{k}_+ \bar{n}_{R,1}(t) \widetilde{x}_1(t) + \hat{k}_+ \widetilde{n}_{R,1}(t) \bar{x}_1(t)  - k_- \widetilde{x}^*_1(t) - d_r \widetilde{x}^*_1(t) + d_r \widetilde{x}^*_2(t) + \nonumber \\
& \sqrt{\hat{k}_+ \bar{n}_{R,1}(t) \bar{x}_1(t)} \gamma_{\rm ar,rv1}(t) + \sqrt{k_- \bar{x}^*_1(t)} \gamma_{dr,RV1}(t) + \sqrt{d_r \bar{x}^*_1(t)} \gamma_{d,X*,1to2}(t) + \sqrt{d_r \bar{x}^*_2(t)} \gamma_{d,X*,2to1}(t)
\label{eq:LNA:2nd}
\end{align} 
where $\gamma_{\rm ar,rv1}(t)$, $\gamma_{dr,RV1}$, $\gamma_{d,X*,1to2}(t)$, $\gamma_{d,X*,1to2}(t)$ and $\gamma_{d,X*,2to1}(t)$ are continuous-time standard Gaussian white noise associated with the reactions. Note that linearisation has been applied to the reaction rate $\hat{k}_+ n_{R,1}(t) x^*_1(t)$ in the first line of \eqref{eq:LNA:2nd}. One can see that, if the mean concentrations are given, then \eqref{eq:LNA:2nd} is a linear stochastic differential equation (SDE) in $\widetilde{x}^*_1(t)$, $\widetilde{n}_{R,1}(t)$ etc. We can use similar method to write down the SDEs for other species in the system. With this set of SDEs, we can compute the covariance matrix for vectored Gaussian distribution of the species concentrations by solving a Lyapunov differential equation \cite{Gardiner,Cardelli:2016bp}.  

\subsection{Mean and covariance of $Z_k(t)$} 
Having obtained the mean and covariance of the Gaussian distribution associated with concentrations of the species in the system, we can now compute the joint probability distribution of $Z_0(t)$, $Z_1(t)$, ... , $Z_{K-1}(t)$. We can see from  \eqref{eqn:part:zkp} and \eqref{eqn:demod:mixed} that $Z_k(t)$ is linear in the stochastic quantities, i.e. $\left[ \frac{dX^{*}_{p}(t)}{dt} \right]_+$ and $X^{*}_{P}(t)$, therefore the $Z_k(t)$'s are jointly Gaussian distributed. The key problem here is to estimate the statistical property of $\left[ \frac{dX^{*}_{p}(t)}{dt} \right]_+$. Since the increment of $X^{*}_{p}(t)$ can be due to an activation reaction or the diffusion of an \cee{X_*} into the voxel, we model this term by using the SDE:
\begin{align}
\left[ \frac{dX^{*}_{p}(t)}{dt} \right]_+ =& \sqrt{V} \left(
\hat{k}_+ \bar{n}_{R,1}(t) \widetilde{x}_1(t) + \hat{k}_+ \widetilde{n}_{R,1}(t) \bar{x}_1(t) + d_r \widetilde{x}^*_2(t) + \right. \nonumber \\
& \left. \sqrt{\hat{k}_+ \bar{n}_{R,1}(t) \bar{x}_1(t)} \gamma_{\rm ar,rv1}(t) + \sqrt{d_r \bar{x}^*_2(t)} \gamma_{d,X*,2to1}(t). \right)
\label{eq:LNA:2nd_dZ*}
\end{align}  
By combining this with the Lyapunov differential equation for calculating the covariance of the concentration, we can compute the joint probability distribution of $Z_0(t)$, $Z_1(t)$, ... , $Z_{K-1}(t)$. 

When the number of symbols $K$ is 2, the BER can be computed as follows. If Symbol 0 is transmitted, bit error occurs if $Z_0(t) < Z_1(t)$. Since $Z_0(t)$ and $Z_1(t)$ are jointly Gaussian distributed, the random variable $Y(t) = Z_0(t) - Z_1(t)$ is Gaussian distributed, therefore the BER equals to the probability of $Y(t) < 0$. The computation of the BER for Symbol 1 is similar.

\section{Numerical Examples}
\label{sec:results}
\label{sec:SECTION_6_Numerical_Examples}
This section presents numerical results to understand the performance of the partitioned and mixed configurations. We first describe the methodology and then the results. 

\subsection{Methodology}
We assume the propagation medium has the shape of a rectangular prism divided into voxels with edge length $w$. Different numerical studies will use different medium size and we will also study the impact of the parameter $w$, so we will specify their numerical values when the present the studies.

The diffusion coefficient $D$ of the propagation medium is 1 $\mu$m$^2$s$^{-1}$. Unless otherwise stated, we assume an absorbing boundary condition where signalling molecules may leave the surface of boundary voxel at a rate $\frac{d}{50}$. 

The transmitter is assumed to use $K = 2$ symbols.  Unless otherwise stated, each symbol is represented by an emission pattern which is generated by a chemical reaction of the form:
\begin{align}
\cee{
\phi & ->[r_k] S  \label{cr:input}  
}
\end{align}
where $r_k$ is the rate of production of signalling molecules \cee{S} when symbol $k$ is transmitted. We assume that symbols 0 and 1 cause, respectively, 10 and 40 signalling molecules to be generated per second on average by the transmitter. The reaction rate constants for the reactions \eqref{cr:general_on} is 0.005 $\mu$m$^3$s$^{-1}$ and for reaction \eqref{cr:general_off} is 1s$^{-1}$. 
Note that the diffusion and reaction parameters are the same as those we used in our previous work \cite{chou2015markovian}, and are realistic for a biological environment. 

The number of receptors will vary from experiment to experiment and will be stated for each experiment. 

Unless otherwise stated, the molecular circuits at the receiver voxel is the circuit given in \eqref{cr:all}. 

In the following experiments, for the partitioned configuration, the total number of \cee{X} and \cee{X^*} in each receiver voxel is given by the parameter $M$. For the mixed configuration, the value of $M$ should be interpreted as the total number of \cee{X} and \cee{X^*} in a receiver voxel at the beginning of the simulation. Note that this parameter is common to all receiver voxels. 
\begin{figure}
\centering
\begin{minipage}{1\textwidth}
  \centering    
    \includegraphics[scale=0.45]{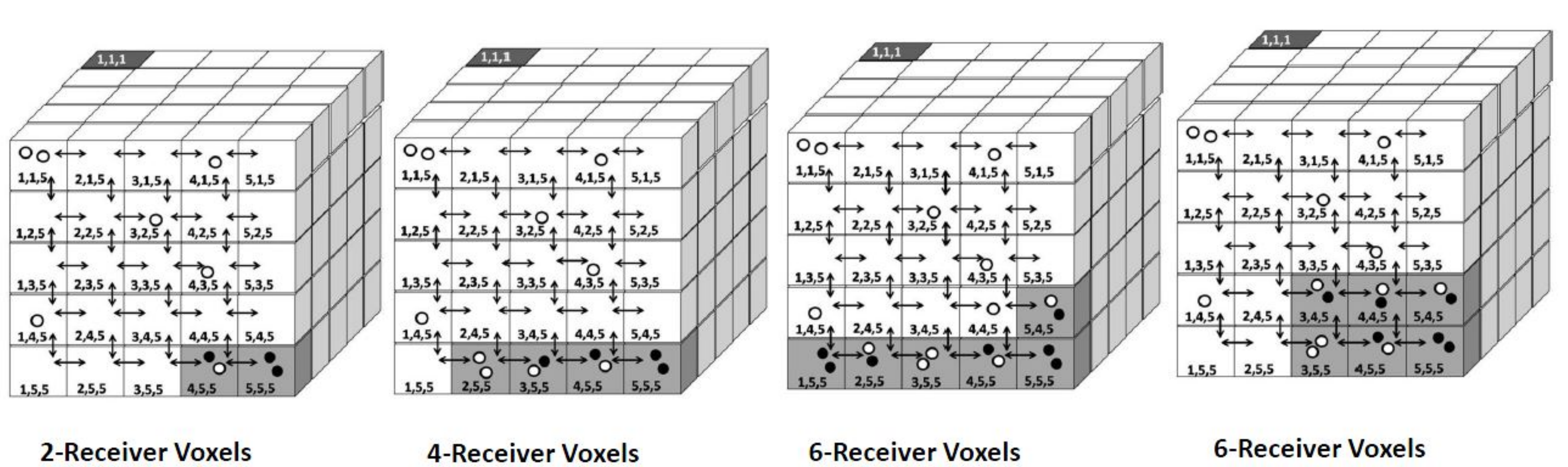}
        \caption{Illustrating the receiver voxel locations. Note: Gray voxels are receiver voxels; dark voxels are transmitter voxels; empty circle are signalling molecules; and, filled circles are receptors.}
    \label{diff_rx_loc}
\end{minipage} 
\end{figure}
The parameter $D_r$ is the diffusion coefficient for \cee{X} and \cee{X^*}. In the experiments, we will use the inter-voxel diffusion rate $d_r = \frac{D_r}{w^2}$ (in s) instead. Note that $d_r = 0$ is the same as the partitioned configuration.  

Since our interest is in the demodulation performance, we assume that there is no inter-symbol interference (ISI). Note that if ISI is present, we can deal with it using the decision feedback algorithm in our earlier work in \cite{chou2015markovian}. 

We use Stochastic Simulation Algorithm (SSA) \cite{Gillespie:1977ww} to simulate the CTMP that models both diffusion and reaction of molecules in the system. 

The approximate demodulation filter \eqref{eqn:generalized_partitioned} requires the mean $\alpha_{k,p}(t) =  E[N_{R,p}(t) | k]$ while the filter \eqref{eqn:demod:mixed} requires the mean $\beta_{k,p} = E[x_1(t)N_{R,p}(t) | k]$. We will use SSA simulation to estimate these means by running SSA simulation 500 times and compute the average. 

We numerically integrate the approximate demodulation filters to obtain $Z_0(t)$ and $Z_1(t)$. We use the initial condition $Z_k(0) = 0$ for all $k$ which means that all symbols are equally probable in the system. We will use BER as the performance metric. Each BER value is estimated using 300 independent SSA runs. 

The BER at time $t$ is determined as follows: If the transmission symbol is 0 (resp. 1), then a bit-error occurs at time $t$ if $Z_1(t) > Z_0(t)$ (resp. $Z_0(t) > Z_1(t)$).




\subsection{Partitioned case: Comparing the approximate filter \eqref{eqn:part:zkp} against the optimal filter \eqref{MAP_generalized_partitioned}}
\label{sec:num:sub:part}

 The aim of this section is to verify that the approximate filter \eqref{eqn:part:zkp} for computing the log-posteriori probability for the partitioned case is a good approximation of the optimal filter \eqref{MAP_generalized_partitioned}. 
 
 The optimal demodulation filter \eqref{MAP_generalized_partitioned} requires the computation of ${\mathbf E}[N_{R,p}(t) | k, {\cal X}_R^*(t)]$ which can be obtained by solving an optimal Bayesian filtering problem. Since filtering problems are computationally expensive to solve, we propose the approximate demodulation filter \eqref{eqn:part:zkp} which approximates ${\mathbf E}[N_{R,p}(t) | k, {\cal X}_R^*(t)]$ by ${\mathbf E}[N_{R,p}(t) | k]$. The aim of this subsection is to compare the accuracy of the optimal and approximate filters. 

In this comparison, we consider a transmission medium of 1$\mu$m $\times$ $\frac{1}{3}$$\mu$m $\times$ $\frac{1}{3}$$\mu$m. With  $w = \frac{1}{3}$ $\mu$m, the medium consists of an array of $3 \times 1 \times 1$ voxels. We will index these three voxels sequentially by indices 1, 2 and 3. We assume Voxel 1 is the transmitter; and Voxels 2 and 3 are the receiver voxels. A reflecting boundary condition is assumed. 

The reason why we have chosen to use such a small number of voxels is because of the dimensionality of the Bayesian filtering problem. The filtering problem requires us to compute the probability of the state vector $N(t) = (N_1(t),N_2(t),N_3(t))$ where $N_i(t)$ is the number of signalling molecules in the voxel with index $i$. If each voxel can have a maximum of 100 signalling molecules at a time, then there are approximately 10$^6$ possible $N(t)$ vectors and the filtering problem has to estimate the probability ${\rm P}[N(t) | k, {\cal X}_R^*(t)]$ for each possible $N(t)$ vector. Although there are approximation techniques to solve the Bayesian filtering problem, that would introduce inaccuracies. The use of small number of voxels will allow us to compute ${\rm P}[N(t) | k, {\cal X}_R^*(t)]$ precisely. 

For this experiment, the $K = 2$ symbols are deterministic. Symbol $k$ emits $s_k$ signalling molecules at times 0, 0.2 and 0.4, where $s_0 = 8$ and $s_1 = 20$. The number of $M$ per voxel is 10. The simulation time is about 2 seconds. 

\begin{figure}
    \centering
    \begin{subfigure}[t]{0.45\textwidth}
        \centering
        \includegraphics[scale=0.35]{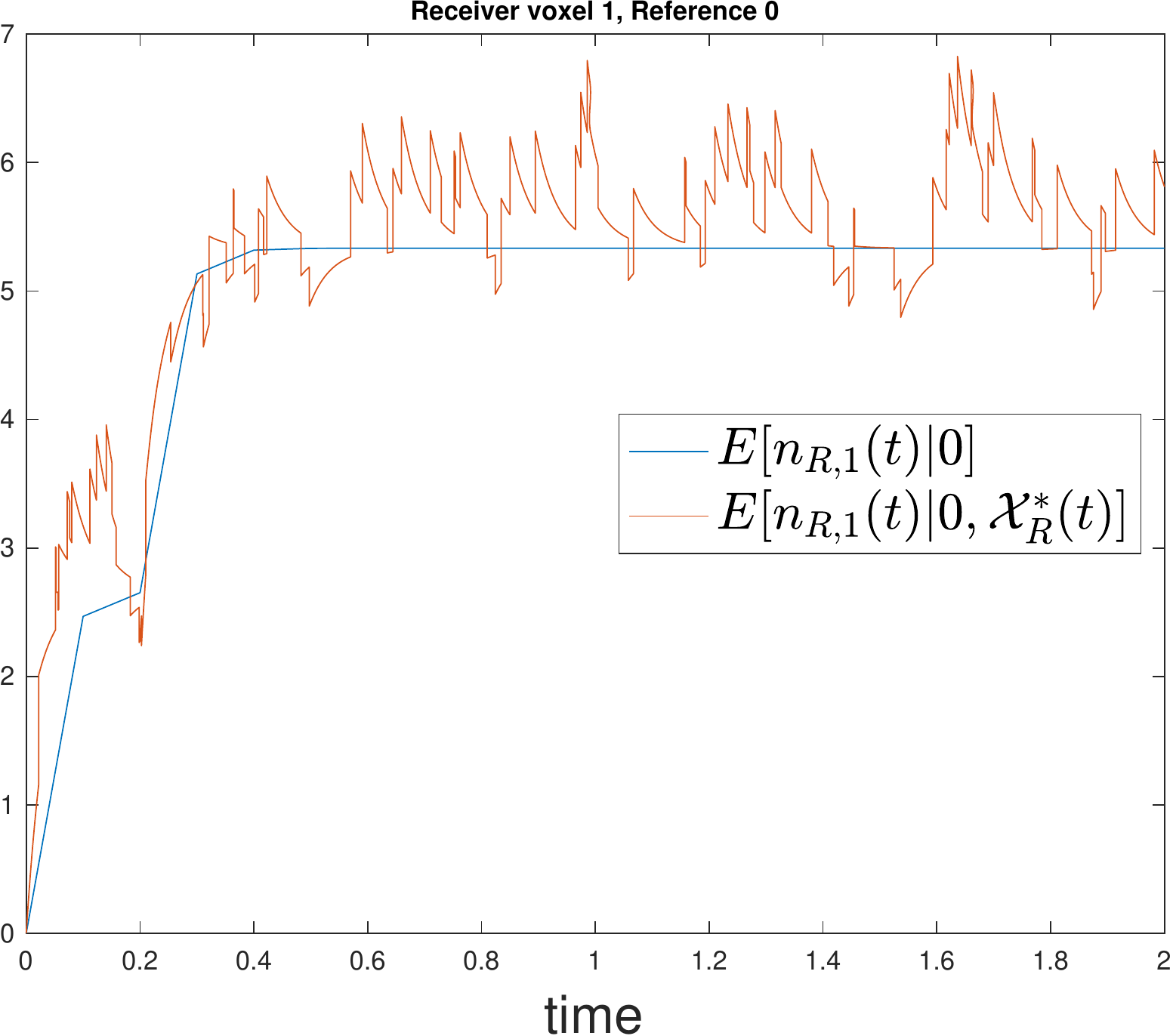}
        \caption{}
        \label{fig:approx_bayes_part_mean_ref0_rv1}
    \end{subfigure}
     \begin{subfigure}[t]{0.45\textwidth}
        \centering        \includegraphics[scale=0.35]{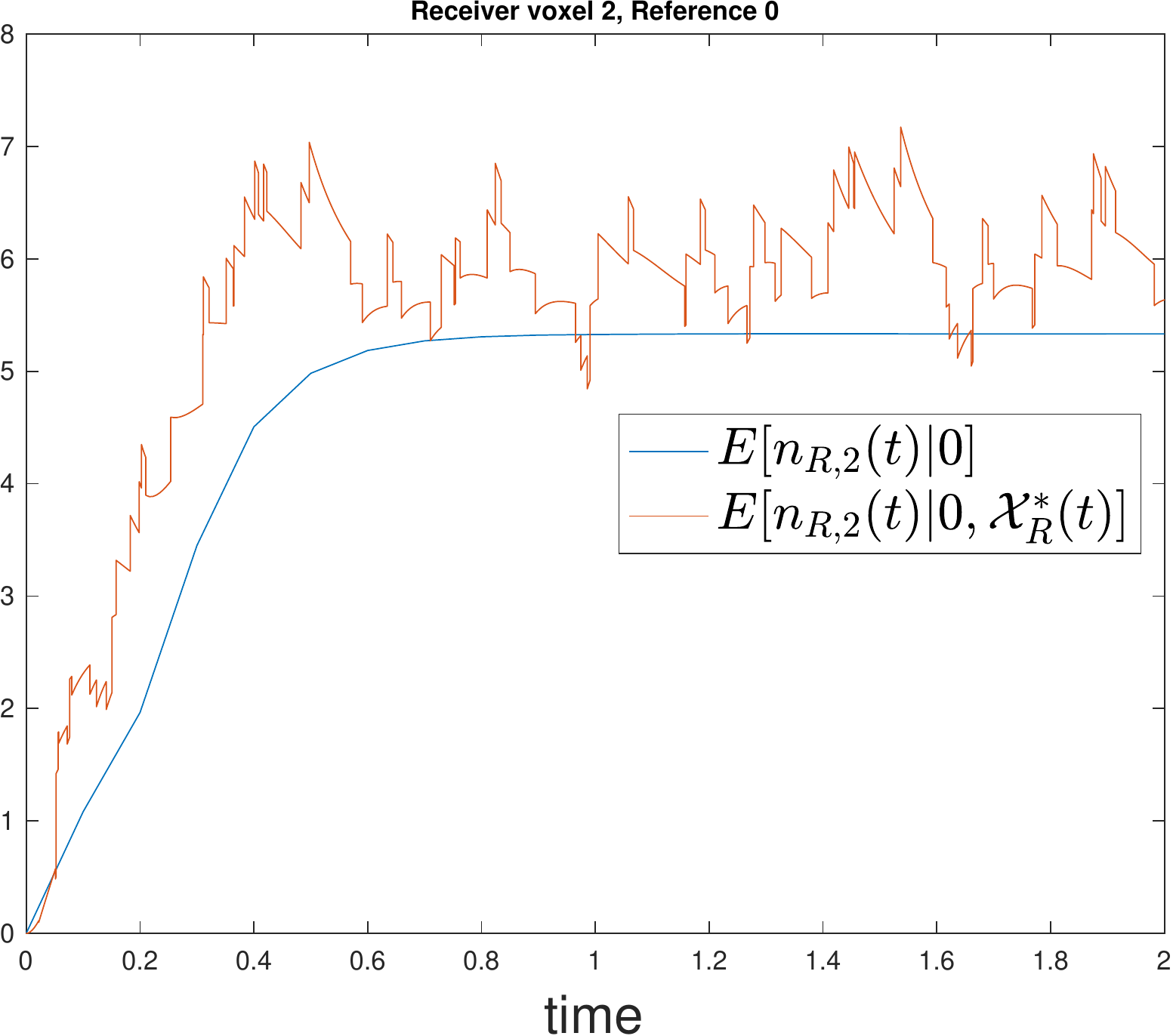}
        \caption{}
        \label{fig:approx_bayes_part_mean_ref0_rv2}
    \end{subfigure}

    \begin{subfigure}[t]{0.45\textwidth}
        \centering        \includegraphics[scale=0.35]{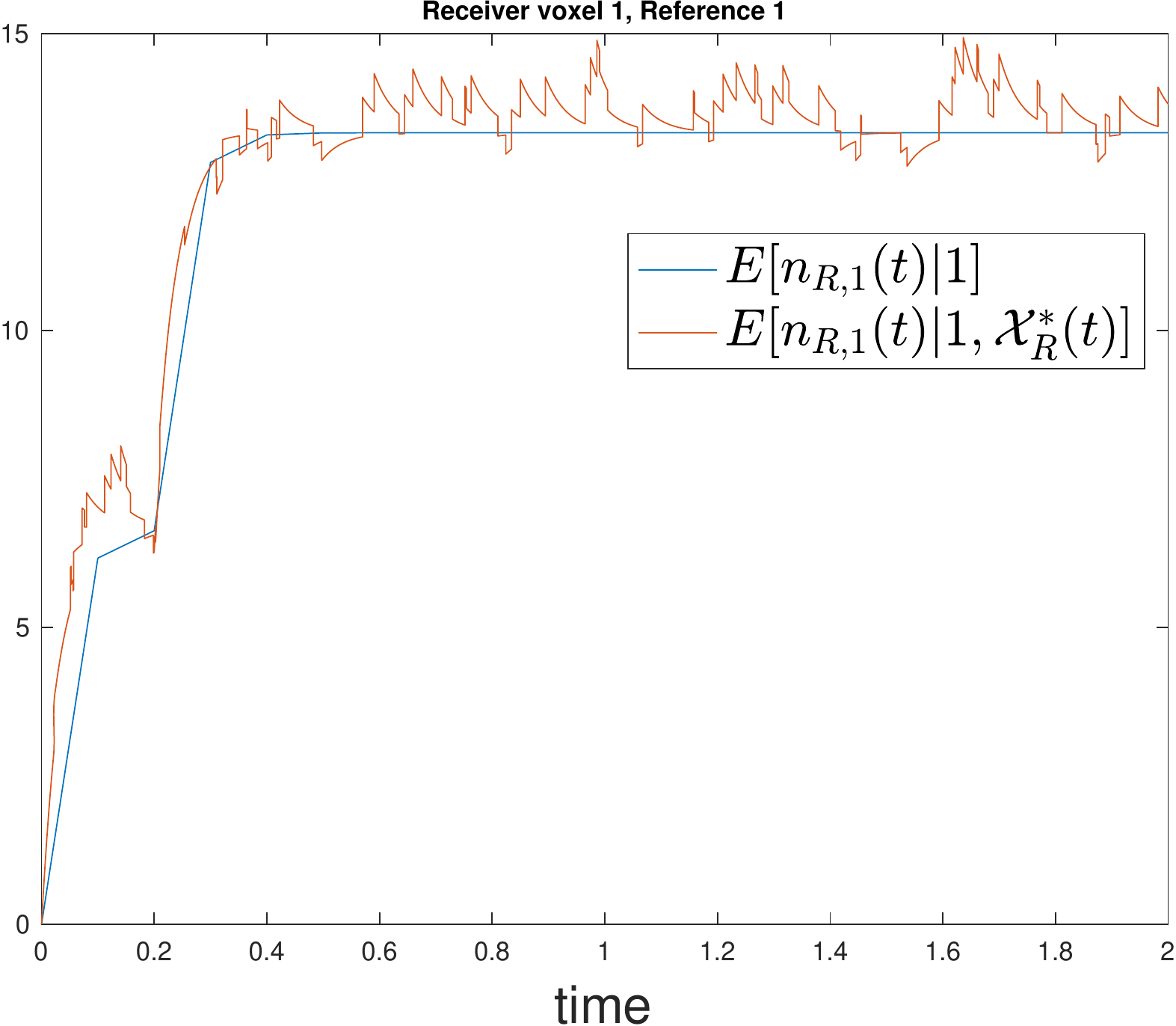}
        \caption{}
        \label{fig:approx_bayes_part_mean_ref1_rv1}
    \end{subfigure}
     \begin{subfigure}[t]{0.45\textwidth}
        \centering        \includegraphics[scale=0.35]{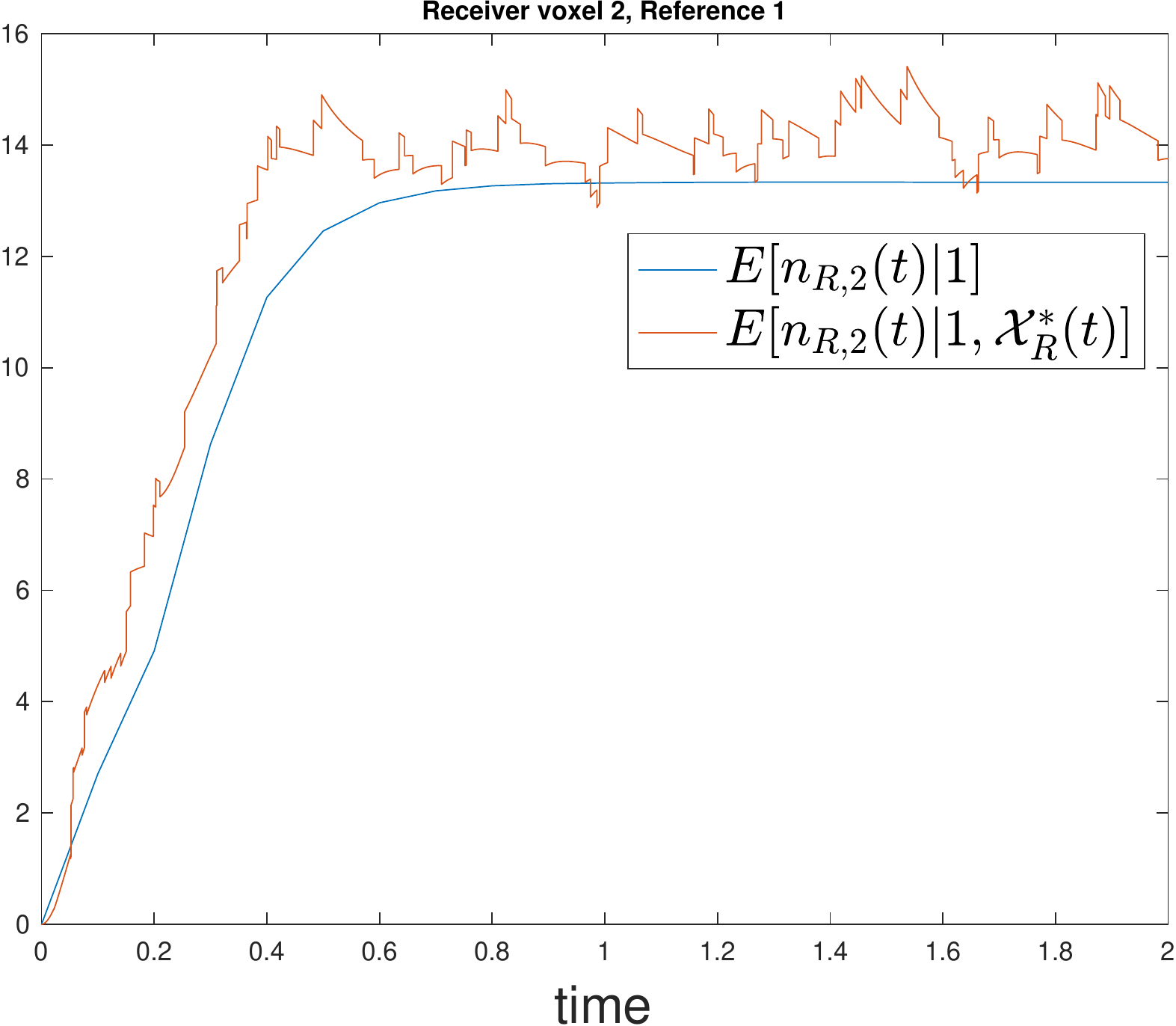}
        \caption{}
        \label{fig:approx_bayes_part_mean_ref1_rv2}
    \end{subfigure}  
    
    \begin{subfigure}[t]{0.45\textwidth}
         \centering       \includegraphics[scale=0.35]{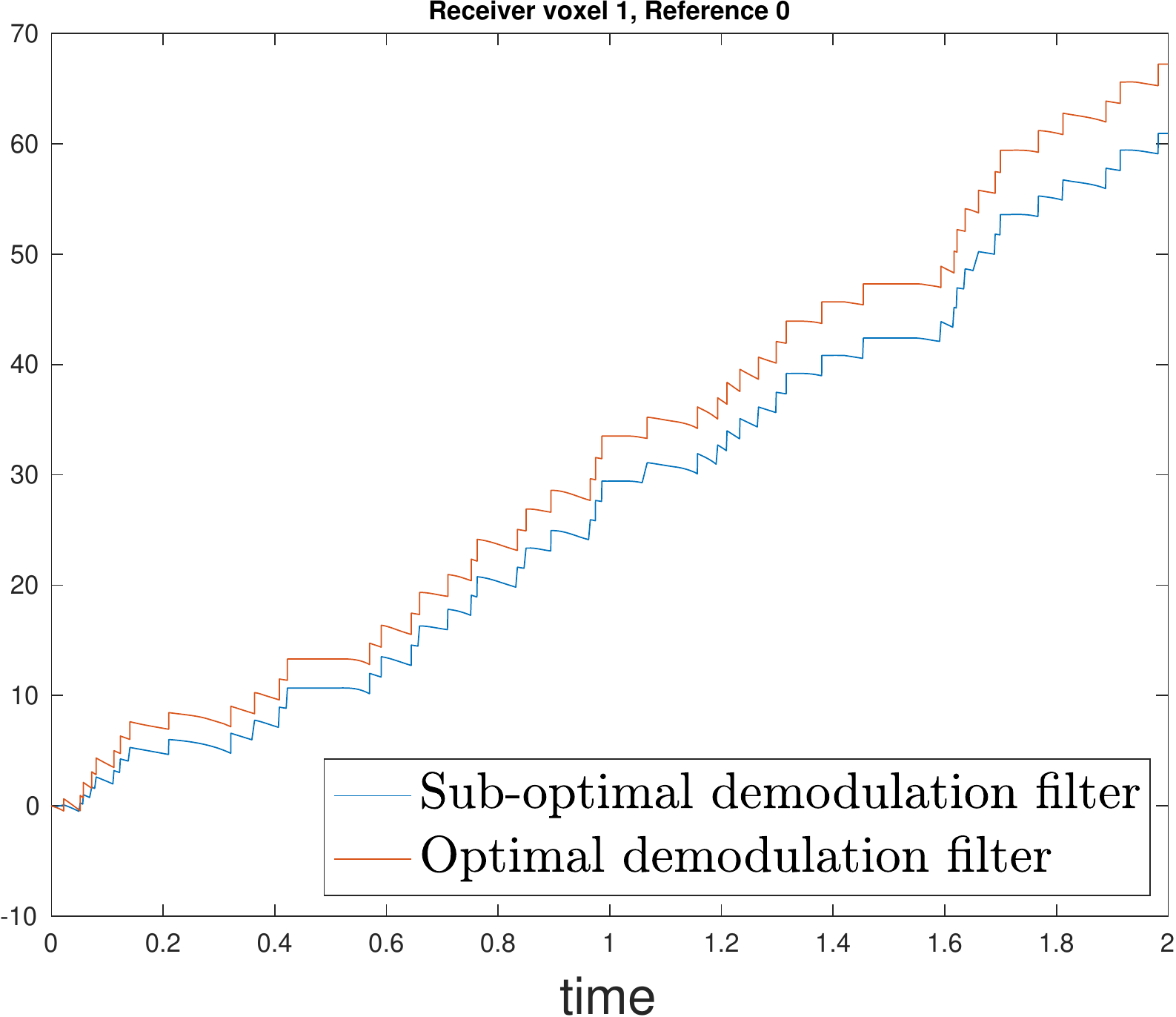}
        \caption{}
        \label{fig:approx_bayes_part_L_ref0_rv1}
    \end{subfigure}
     \begin{subfigure}[t]{0.45\textwidth}
        \centering        \includegraphics[scale=0.35]{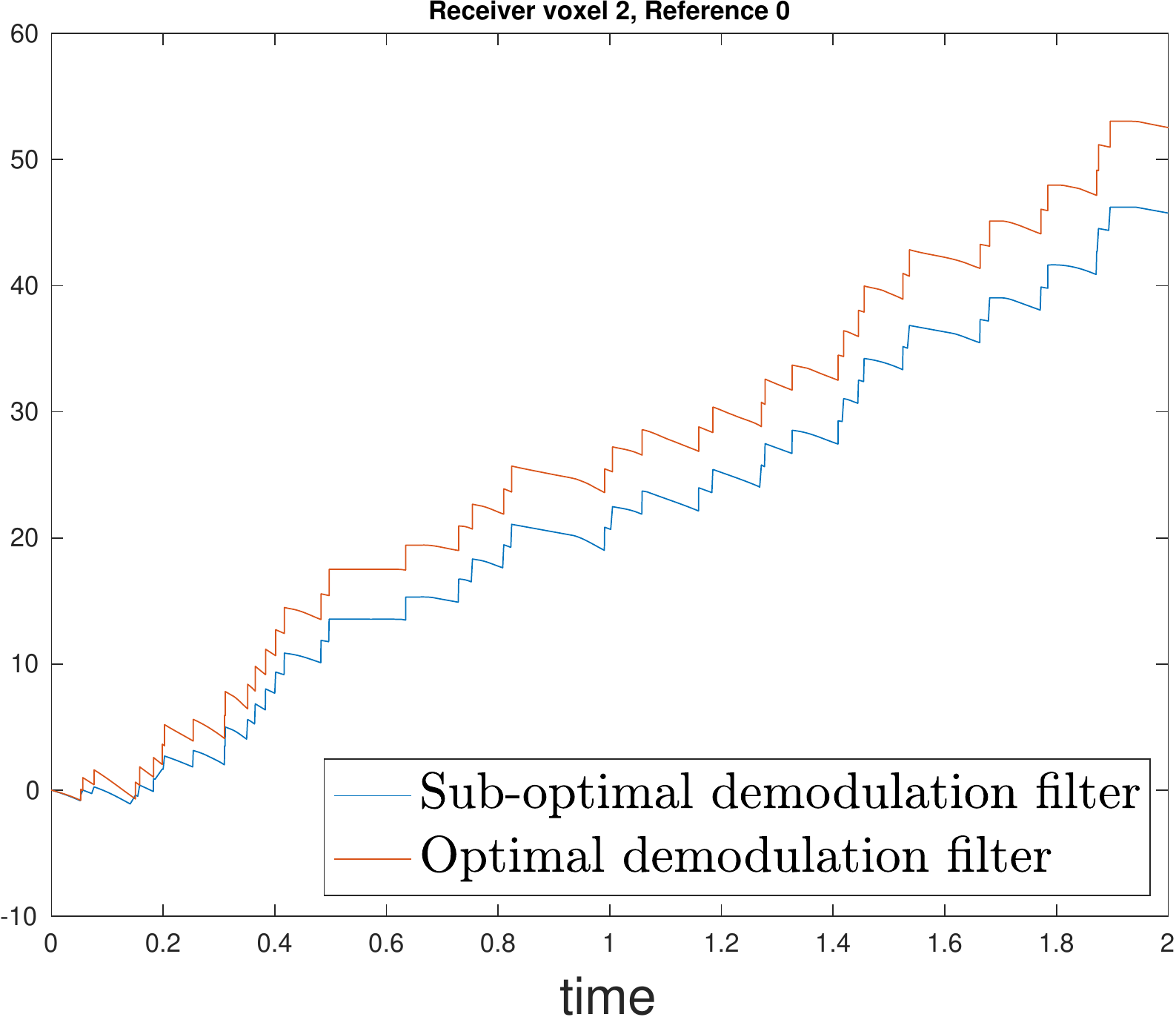}
        \caption{}
        \label{fig:approx_bayes_part_L_ref0_rv2}
    \end{subfigure}

    \begin{subfigure}[t]{0.45\textwidth}
        \centering        \includegraphics[scale=0.35]{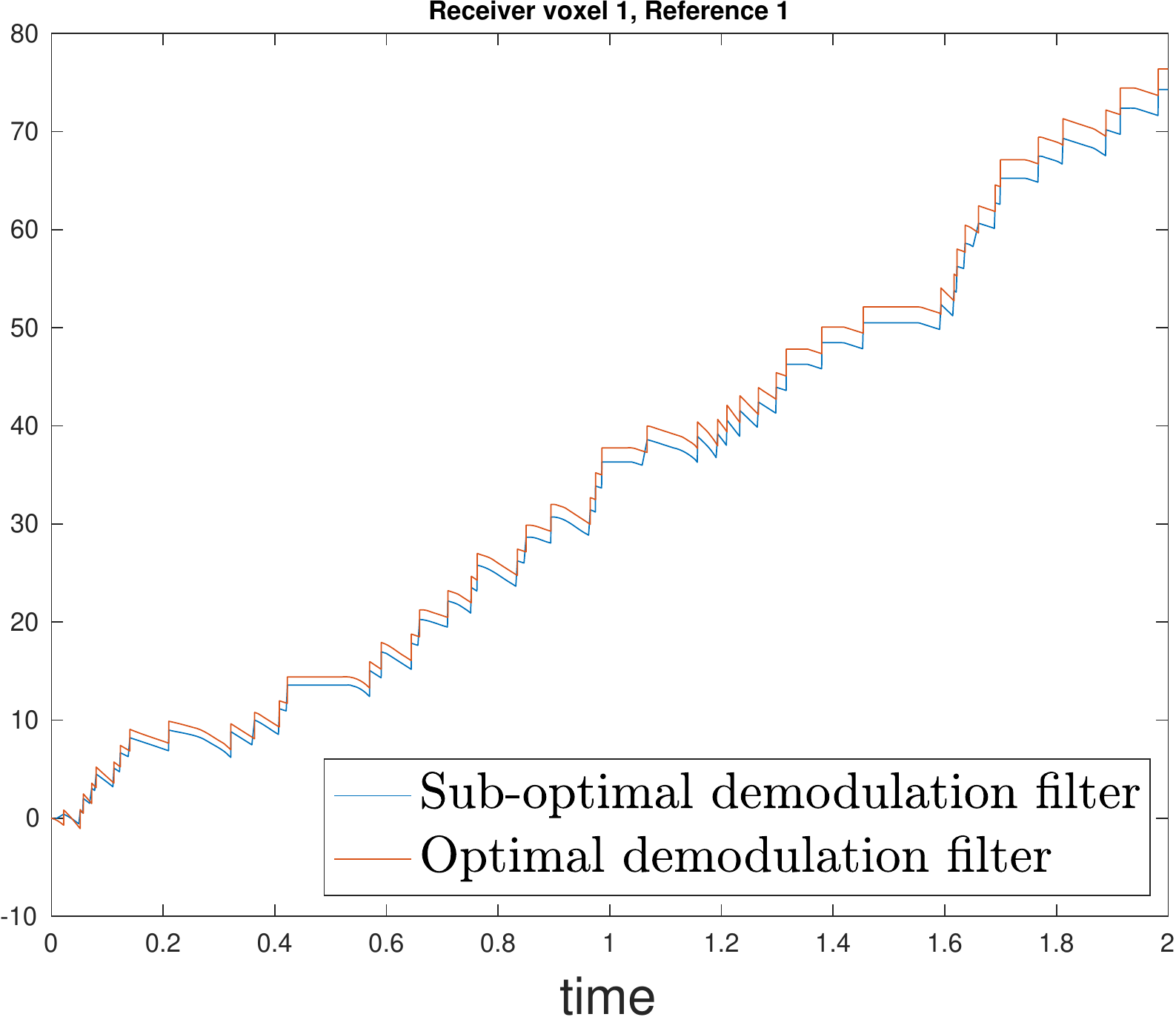}
        \caption{}
        \label{fig:approx_bayes_part_L_ref1_rv1}
    \end{subfigure}
     \begin{subfigure}[t]{0.45\textwidth}
         \centering       \includegraphics[scale=0.35]{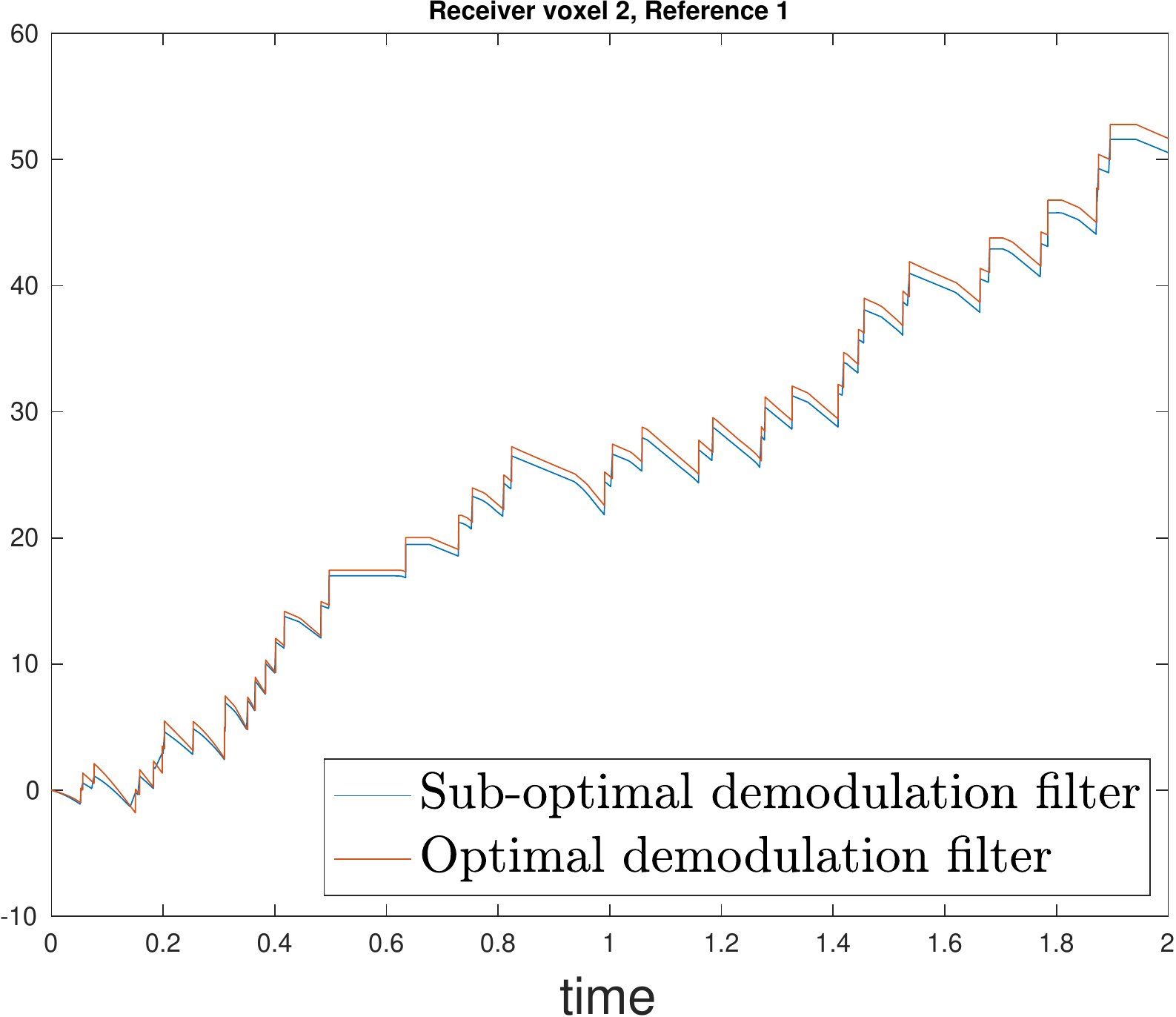}
        \caption{}
        \label{fig:approx_bayes_part_L_ref1_rv2}
    \end{subfigure}      
    
    \caption{Comparing approximate filter \eqref{eqn:part:zkp} and the optimal filter \eqref{MAP_generalized_partitioned}}
    \label{fig:approx_bayes_part}
\end{figure}

Figs.~\ref{fig:approx_bayes_part_mean_ref0_rv1} to \ref{fig:approx_bayes_part_mean_ref1_rv2} compare ${\mathbf E}[N_{R,p}(t) | k]$ and ${\mathbf E}[N_{R,p}(t) | k, {\cal X}_R^*(t)]$ for $p = 1,2$ and $k = 0,1$ when Symbol 1 was transmitted. It can be seen that the approximation is quite accurate. The results for Symbol 0 are similar. 

Eqs.~\eqref{eqn:part:zkp} and \eqref{MAP_generalized_partitioned} show that the log-posteriori probability of the receiver can be computed by summing up the log-posteriori probabilities computed from each voxel. Figs.~\ref{fig:approx_bayes_part_L_ref0_rv1} to \ref{fig:approx_bayes_part_L_ref1_rv2} compare $L_0(t)$ and $L_1(t)$ computed from the two voxels when Symbol 1 was transmitted. It can again be seen that the approximation is quite accurate. The results for Symbol 0 are similar.



In the rest of this section, we will use the approximate demodulation filter \eqref{eqn:part:zkp} because of its lower computational complexity.

\subsection{Mixed case: Comparing the approximate filter \eqref{eqn:demod:mixed} against the optimal filter \eqref{eqn:demod:mixed:opt}}

The aim of this subsection is to demonstrate that the approximate filter \eqref{eqn:demod:mixed} for the mixed configuration is an accurate approximation of the optimal version \eqref{eqn:demod:mixed:opt}. 

We use the same voxel configuration as subsection \ref{sec:num:sub:part}. We allow \cee{X} and \cee{X^*} to diffuse between the receiver voxels with a $D_r = 0.2 D$. For the filtering problem, we need to compute the probability of the state $N(t) = (N_1(t),N_2(t),N_3(t),X_{R,1}(t),X_{R,2}(t))$ where $N_i(t)$ is the number of signalling molecules in the voxel with index $i$ and $X_{R,j}(t)$ is the number of \cee{X} molecules in receiver voxel $j$.

Initially, each receiver voxel has $4$ \cee{X} molecules. The transmission symbols are similar to those in subsection \ref{sec:num:sub:part} except that $s_0 = 10$ and $s_1 = 15$. 

\begin{figure}
    \centering
    \begin{subfigure}[t]{0.45\textwidth}
        \centering
        \includegraphics[scale=0.35]{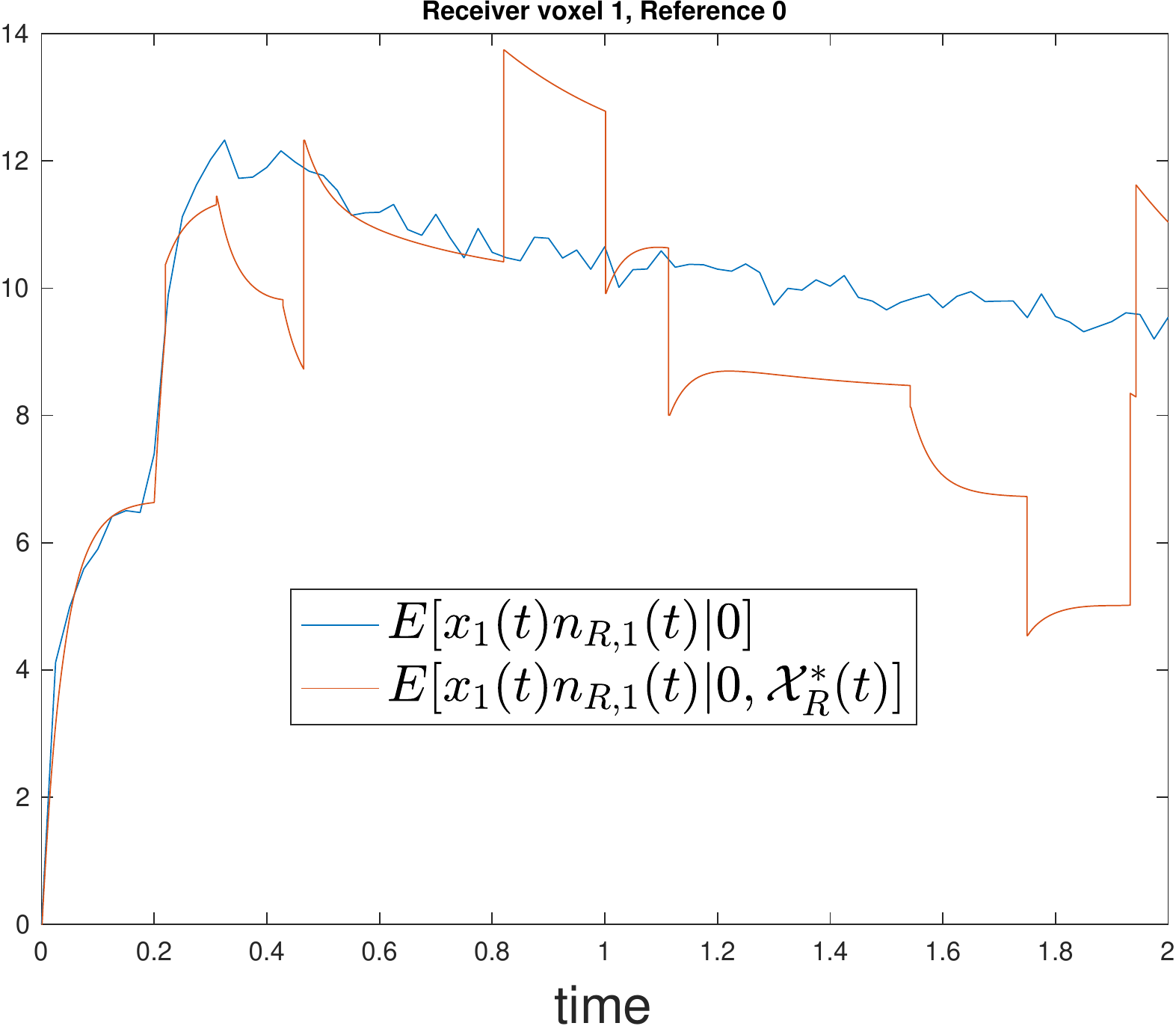}
        \caption{}
        \label{fig:approx_bayes_mixed_mean_ref0_rv1}
    \end{subfigure}
     \begin{subfigure}[t]{0.45\textwidth}
        \centering        \includegraphics[scale=0.35]{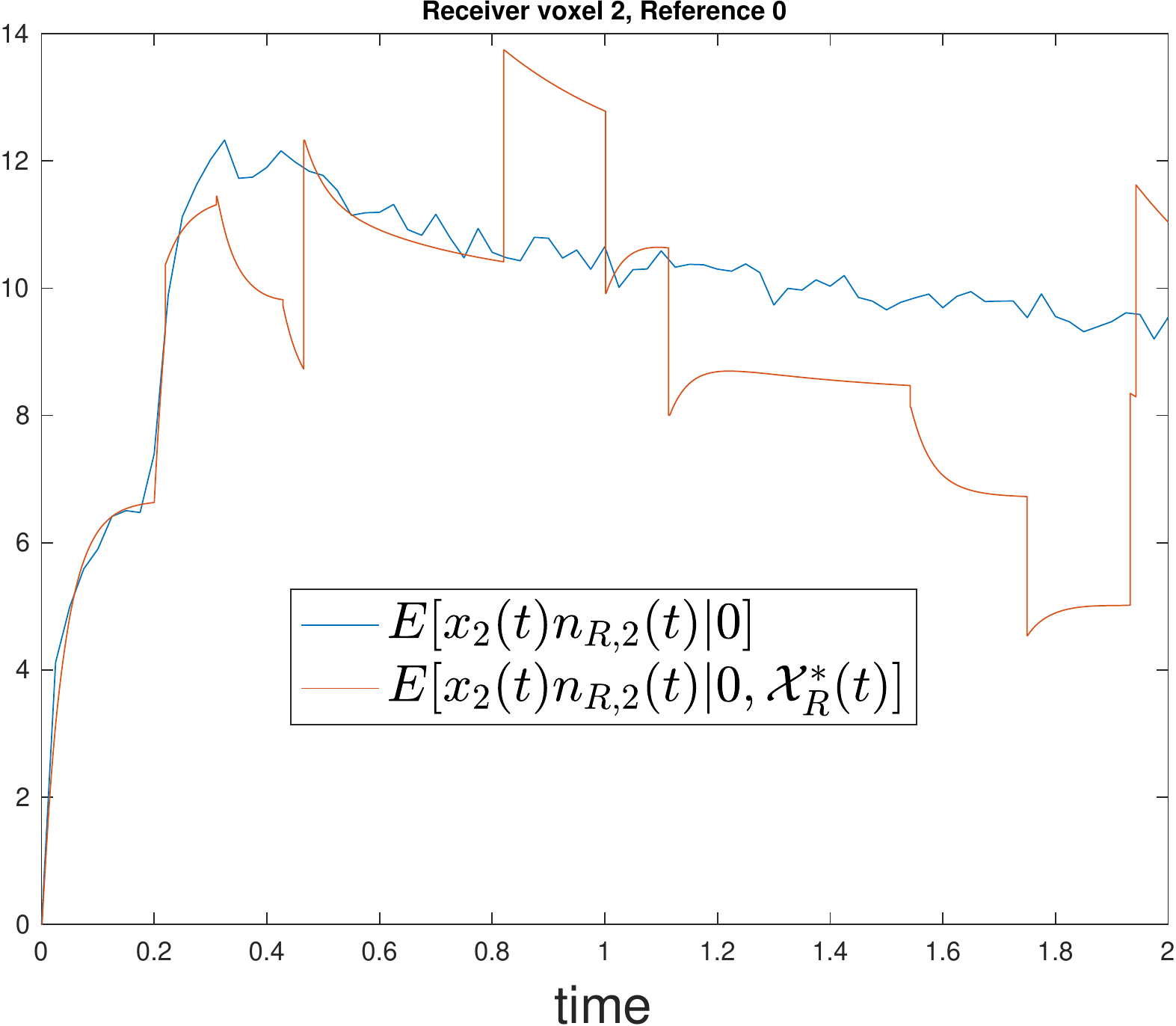}
        \caption{}
        \label{fig:approx_bayes_mixed_mean_ref0_rv2}
    \end{subfigure}

    \begin{subfigure}[t]{0.45\textwidth}
        \centering        \includegraphics[scale=0.35]{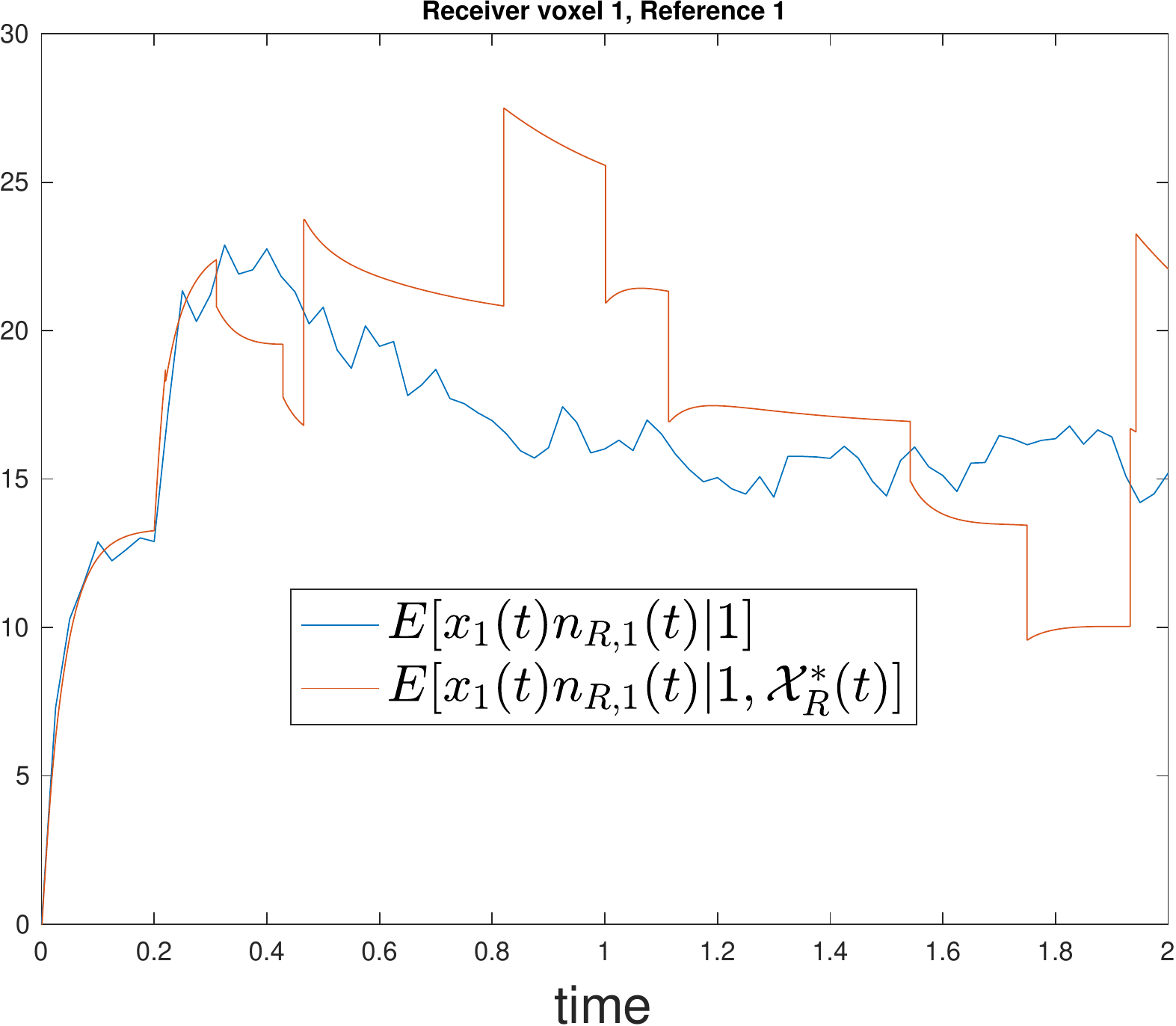}
        \caption{}
        \label{fig:approx_bayes_mixed_mean_ref1_rv1}
    \end{subfigure}
     \begin{subfigure}[t]{0.45\textwidth}
        \centering        \includegraphics[scale=0.35]{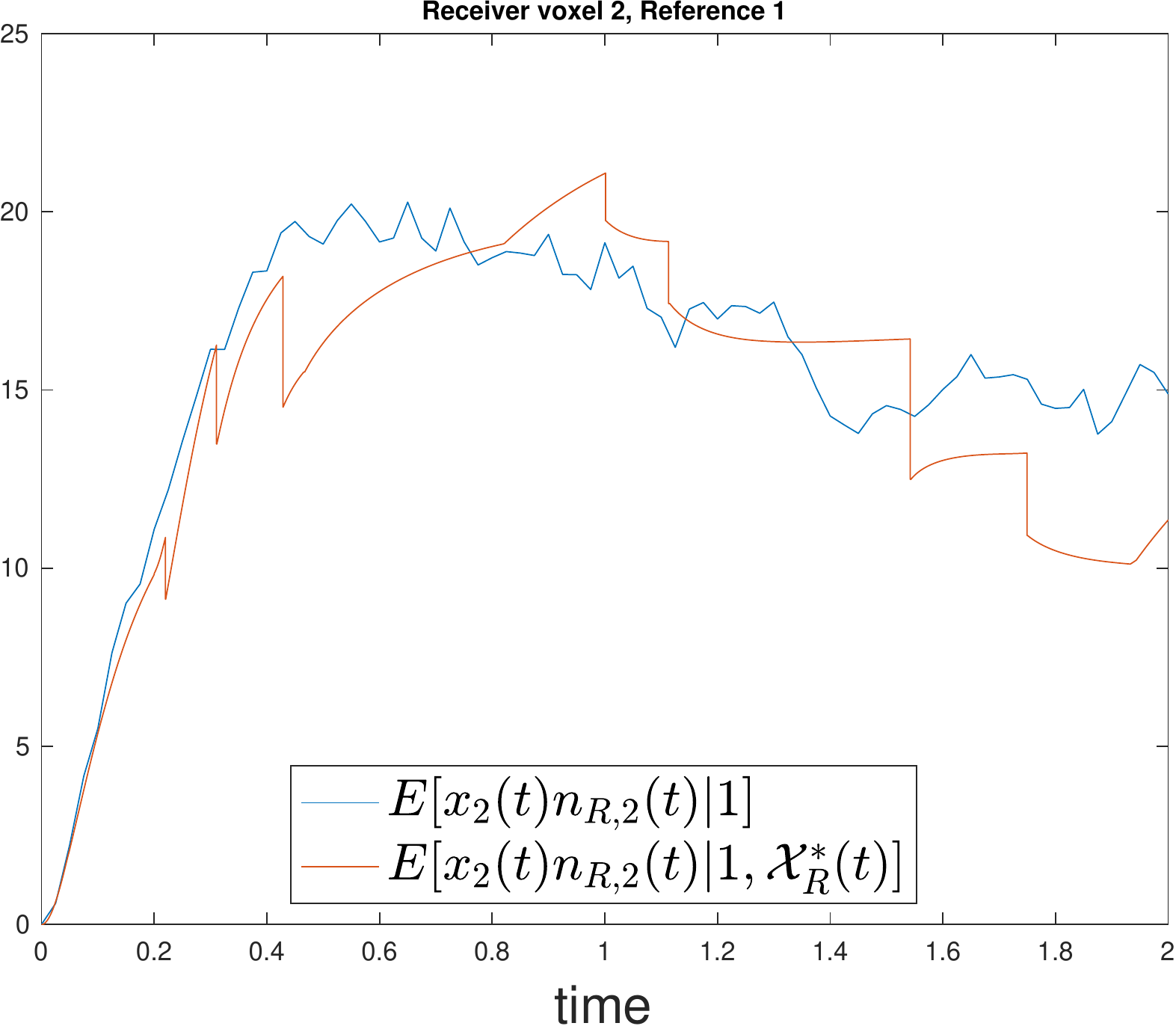}
        \caption{}
        \label{fig:approx_bayes_mixed_mean_ref1_rv2}
    \end{subfigure}  
    
    \begin{subfigure}[t]{0.45\textwidth}
         \centering       \includegraphics[scale=0.35]{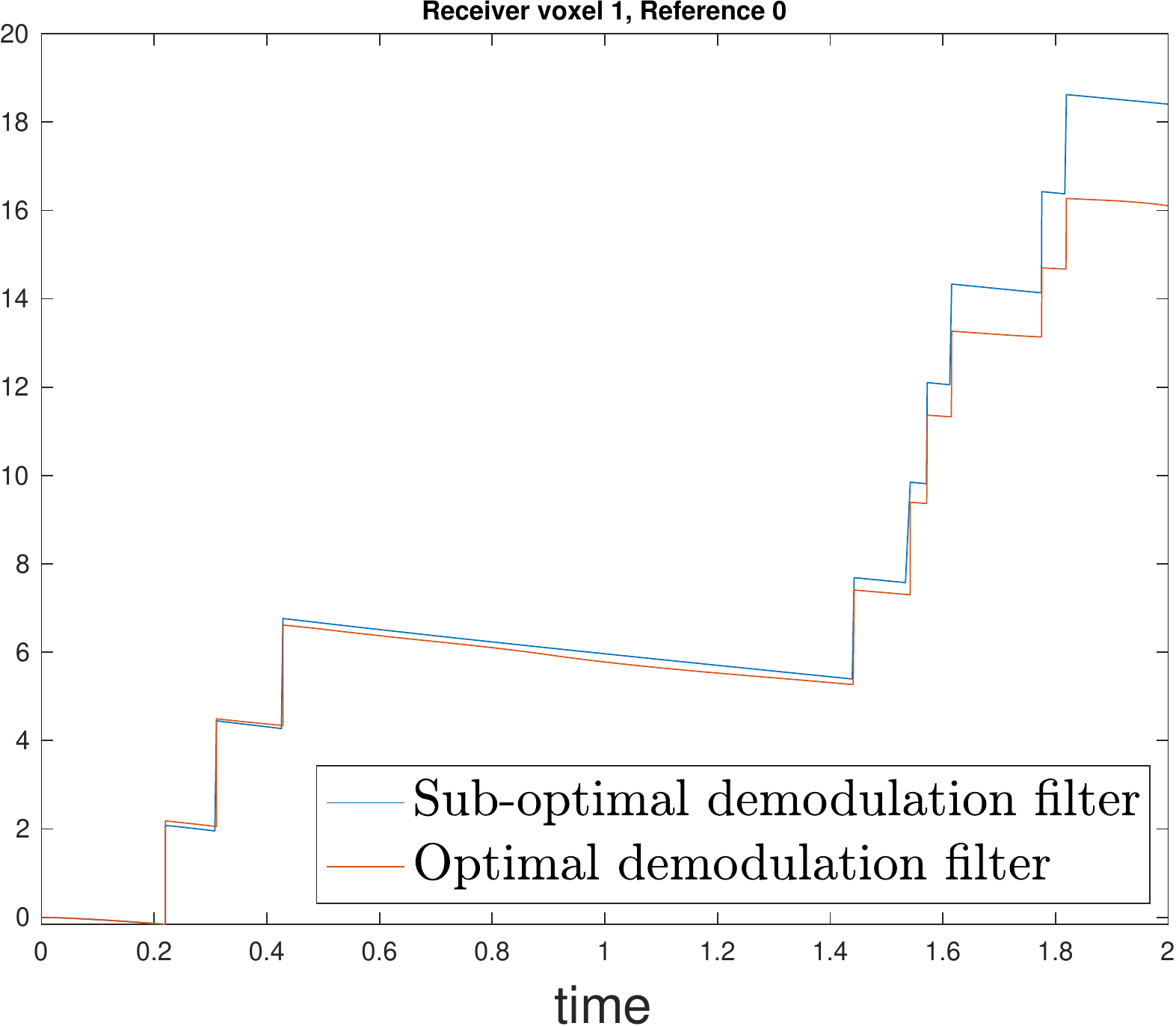}
        \caption{}
        \label{fig:approx_bayes_mixed_L_ref0_rv1}
    \end{subfigure}
     \begin{subfigure}[t]{0.45\textwidth}
        \centering        \includegraphics[scale=0.35]{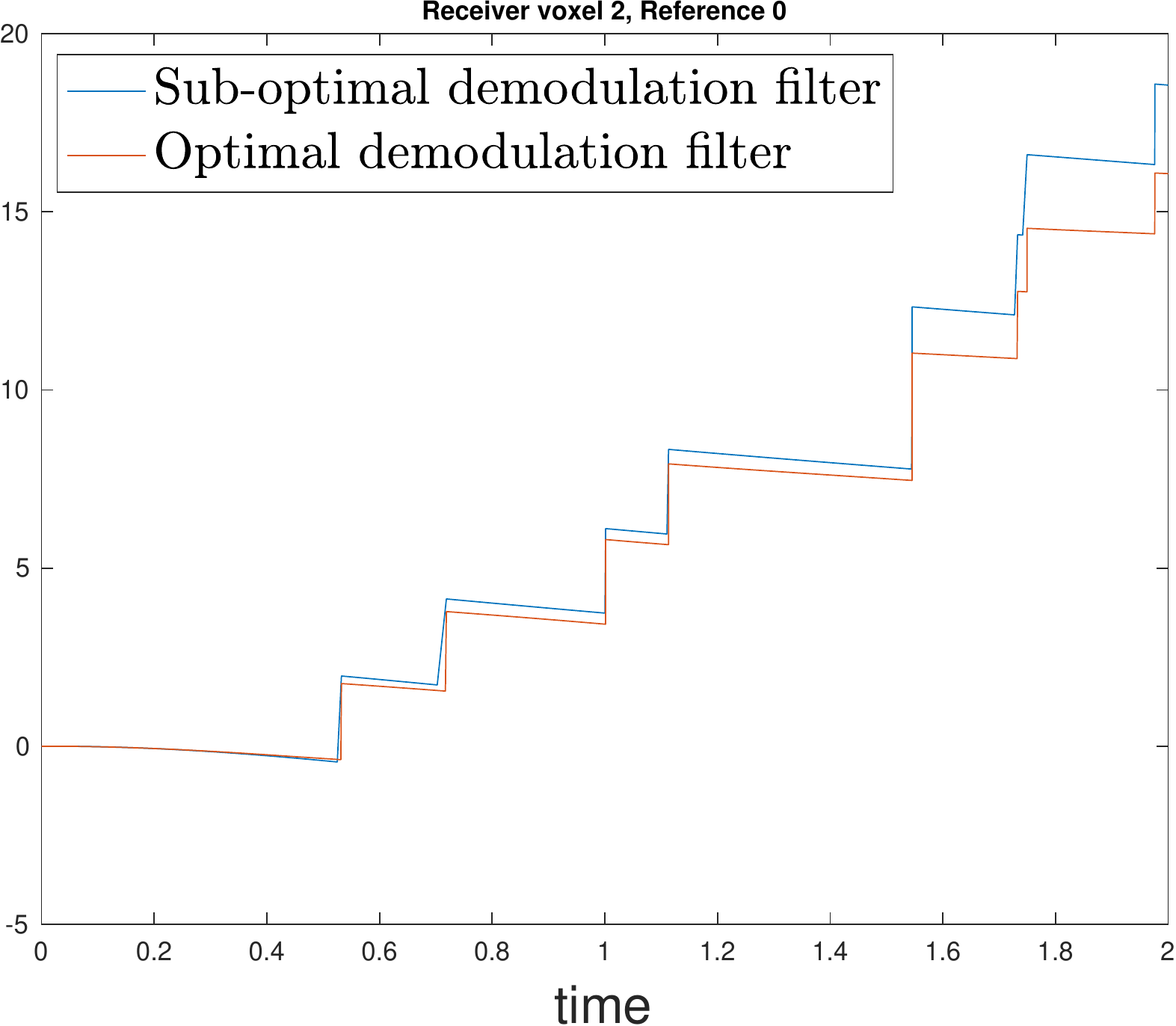}
        \caption{}
        \label{fig:approx_bayes_mixed_L_ref0_rv2}
    \end{subfigure}

    \begin{subfigure}[t]{0.45\textwidth}
        \centering        \includegraphics[scale=0.35]{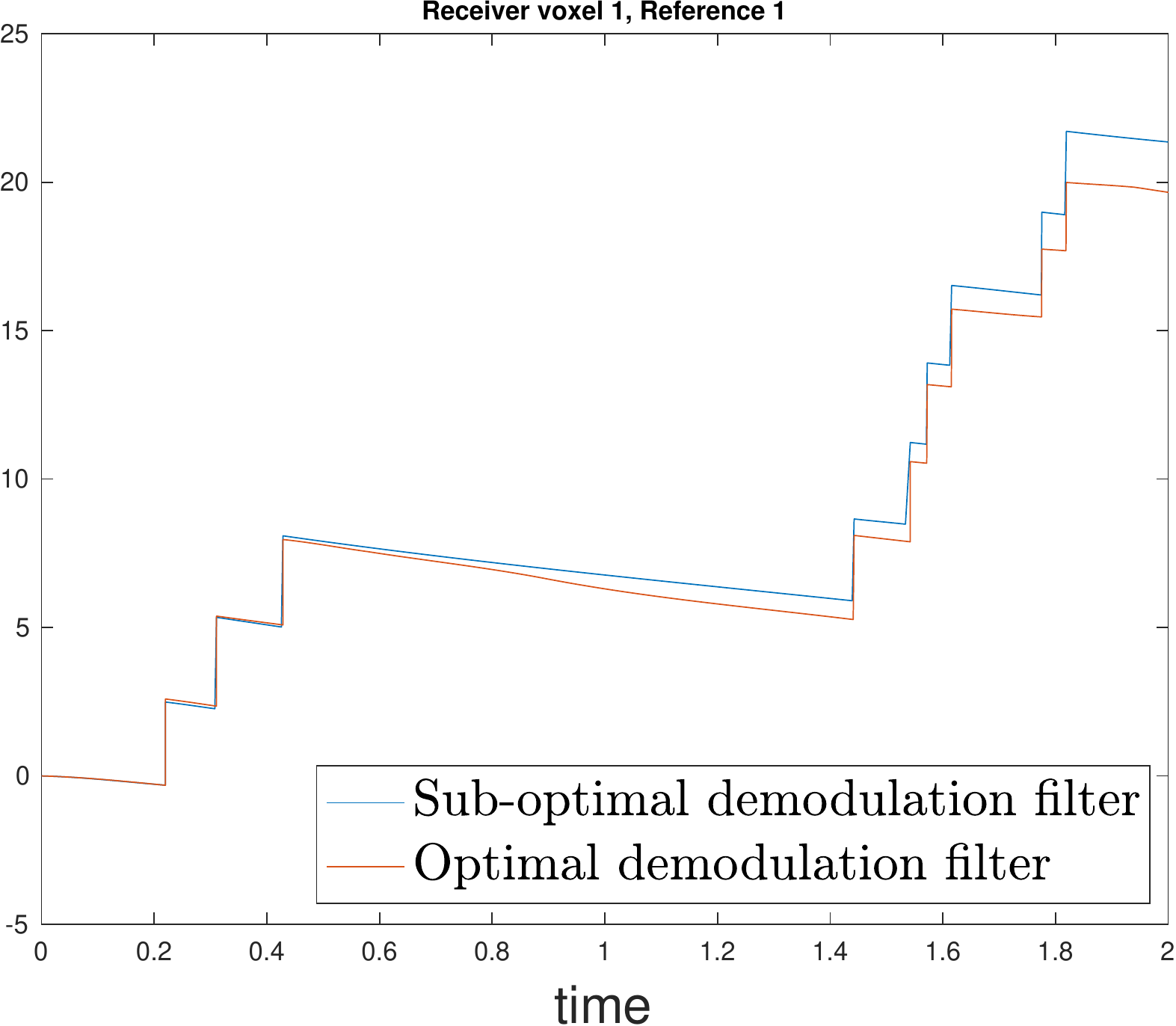}
        \caption{}
        \label{fig:approx_bayes_mixed_L_ref1_rv1}
    \end{subfigure}
     \begin{subfigure}[t]{0.45\textwidth}
         \centering       \includegraphics[scale=0.35]{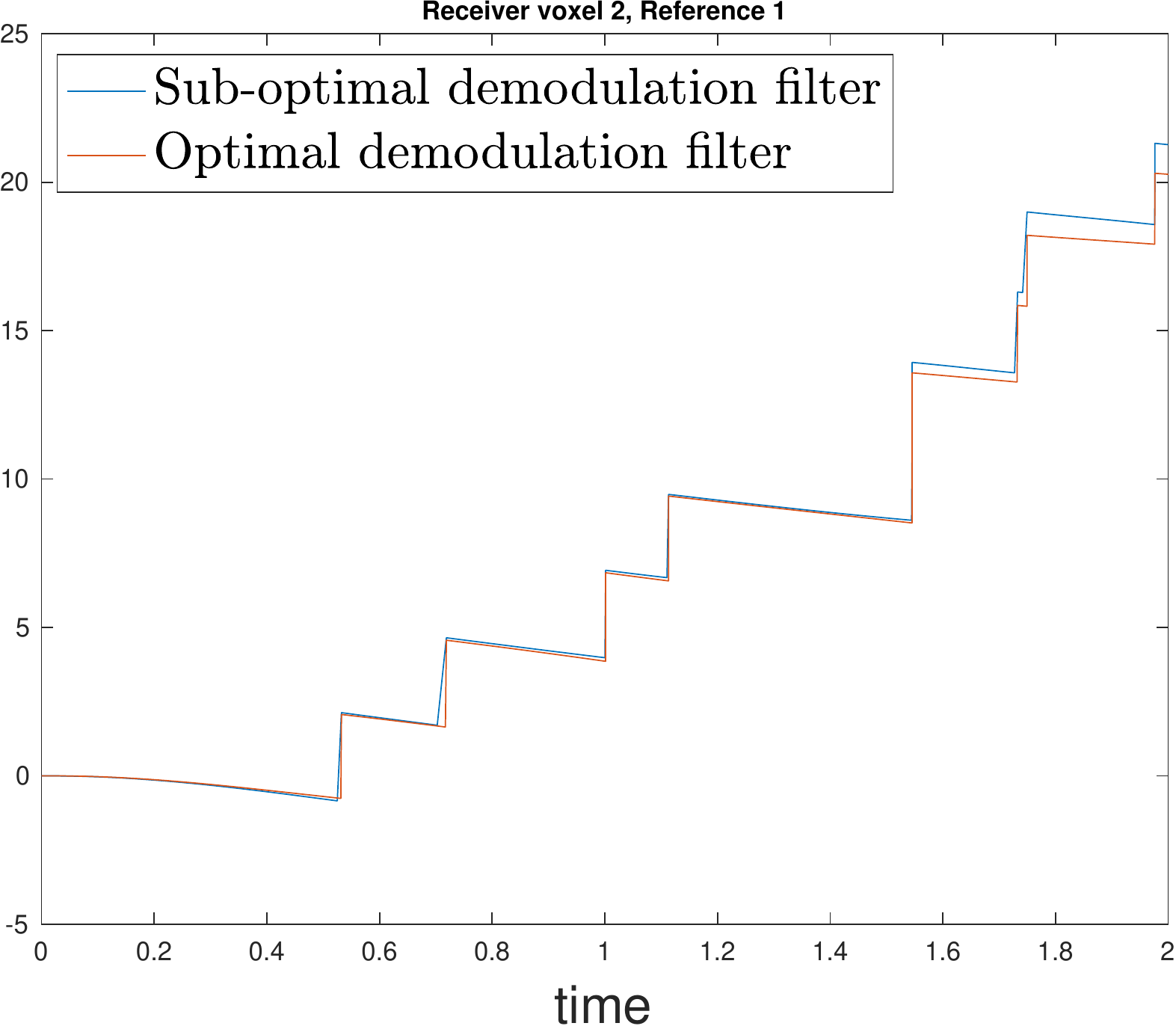}
        \caption{}
        \label{fig:approx_bayes_mixed_L_ref1_rv2}
    \end{subfigure}      
    
    \caption{Comparing approximate filter \eqref{eqn:demod:mixed} and the optimal filter \eqref{eqn:demod:mixed:opt}.}
    \label{fig:approx_bayes_mixed}
\end{figure}

Figs.~\ref{fig:approx_bayes_mixed_mean_ref0_rv1} to \ref{fig:approx_bayes_mixed_mean_ref1_rv2} compare ${\mathbf E}[X_p(t) N_{R,p}(t) | k]$ and ${\mathbf E}[X_p(t) N_{R,p}(t) | k, {\cal X}_R^*(t)]$ for $p = 1,2$ and $k = 0,1$ when Symbol 1 was transmitted. Note that we computed ${\mathbf E}[X_p(t) N_{R,p}(t) | k]$ by averaging the results from 500 SSA simulations. It can be seen that the approximation is reasonably accurate. The results for Symbol 0 are similar. 

Eqs.~\eqref{eqn:demod:mixed} and \eqref{eqn:demod:mixed:opt} show that the log-posteriori probability of the receiver can be computed by summing up the log-posteriori probabilities computed from each voxel. Figs.~\ref{fig:approx_bayes_mixed_L_ref0_rv1} to \ref{fig:approx_bayes_mixed_L_ref1_rv2} compare $L_0(t)$ and $L_1(t)$ computed from the two voxels when Symbol 1 was transmitted. It can again be seen that the approximation is quite accurate. The results for Symbol 0 are similar.

\begin{figure}
    \centering
       \begin{subfigure}[t]{0.45\textwidth}
        \centering       
        \includegraphics[width=9cm,height=5cm]{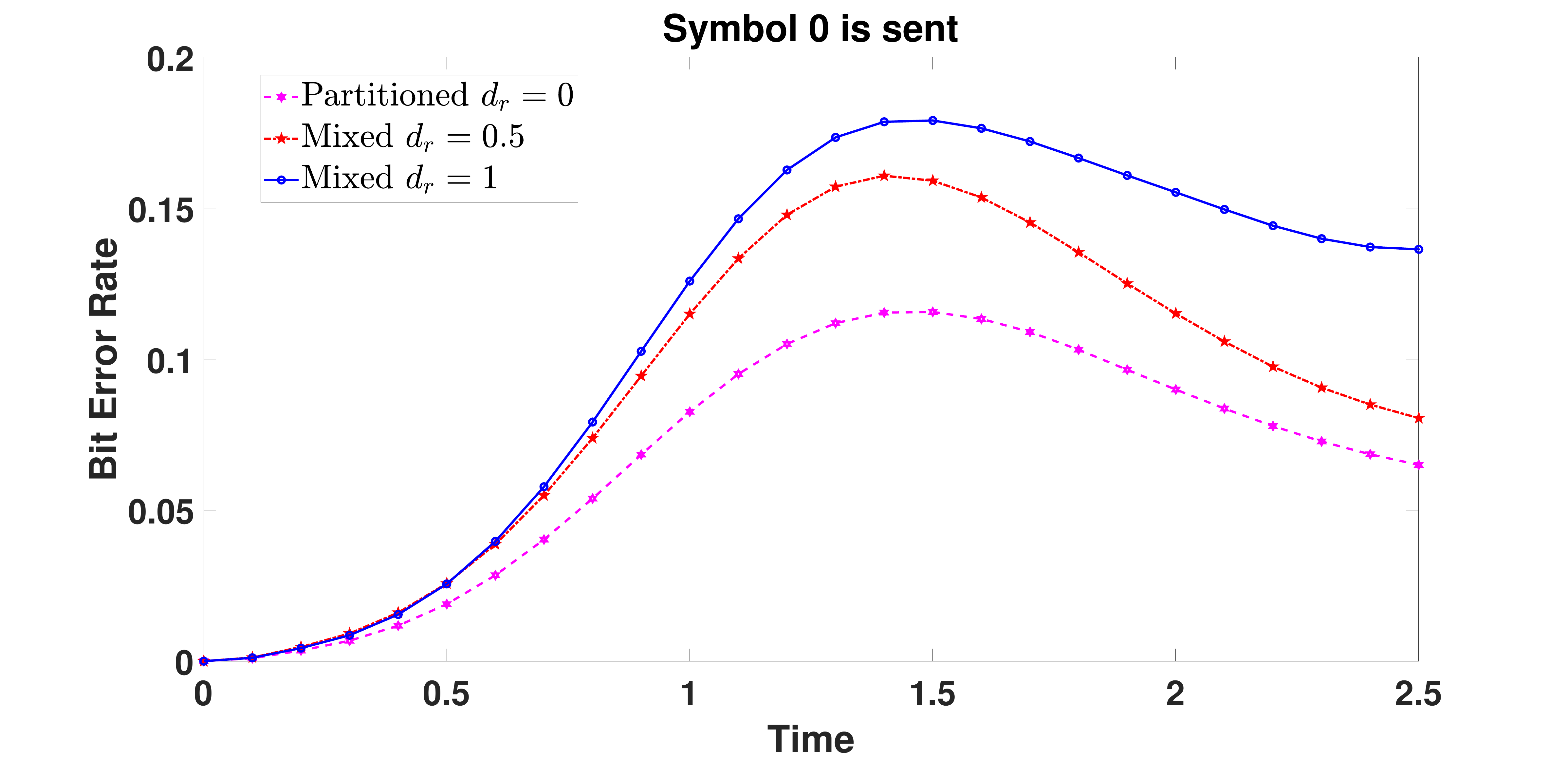}
        \caption{}
        \label{result_1}
    \end{subfigure}
     \begin{subfigure}[t]{0.45\textwidth}
        \centering        \includegraphics[width=9cm,height=5cm]{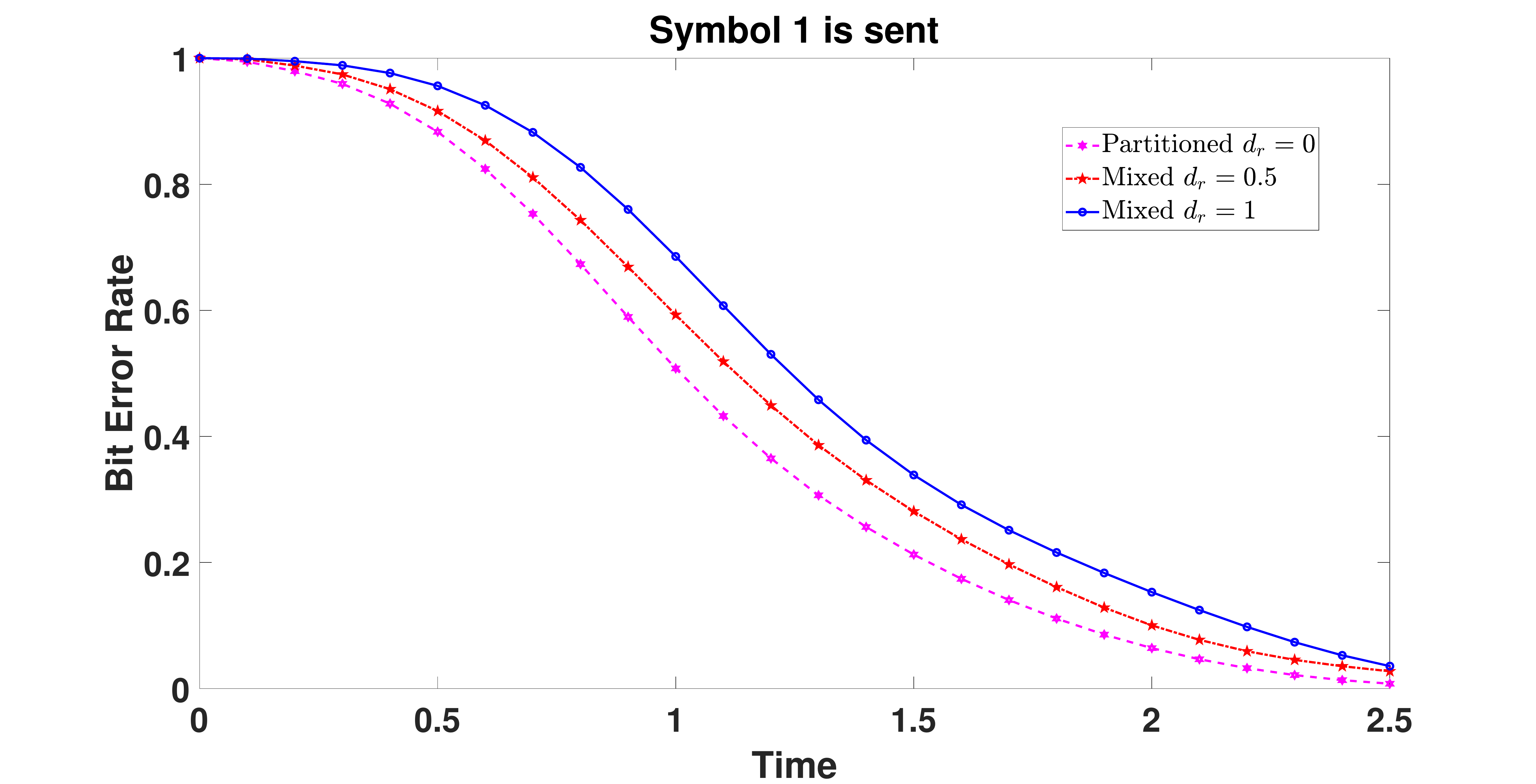}
        \caption{}
        \label{result_2}
    \end{subfigure}
     \caption{Comparing BER for Mixed and Partitioned configurations}
    \label{result_1_result_2}
\end{figure}

 \begin{figure}
    \centering  
        \begin{subfigure}[t]{0.45\textwidth}
        \centering       \includegraphics[width=9cm,height=5cm]{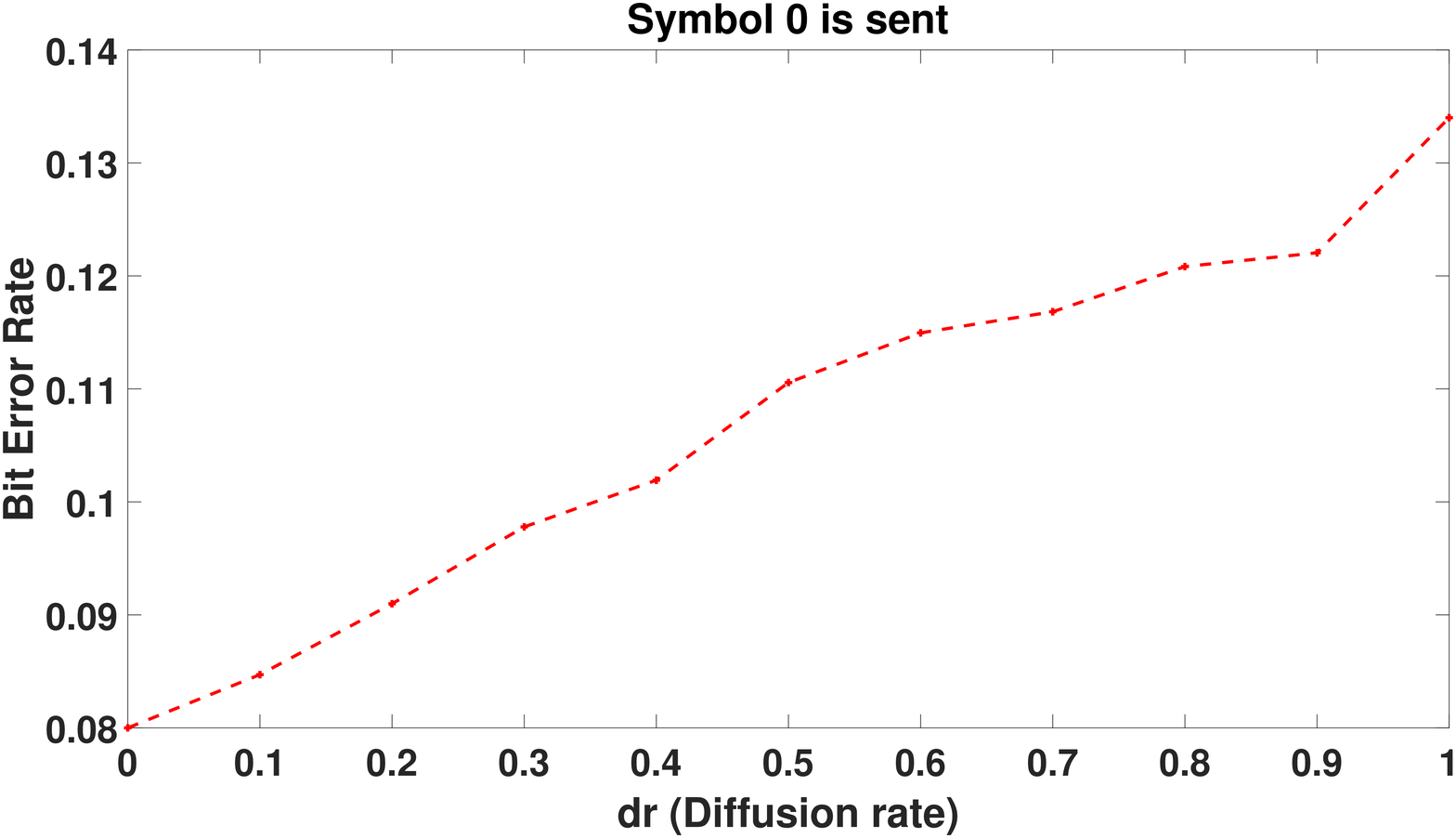}
        \caption{}
        \label{result_3}
    \end{subfigure}
         \begin{subfigure}[t]{0.45\textwidth}
        \centering     \includegraphics[width=9cm,height=5cm]{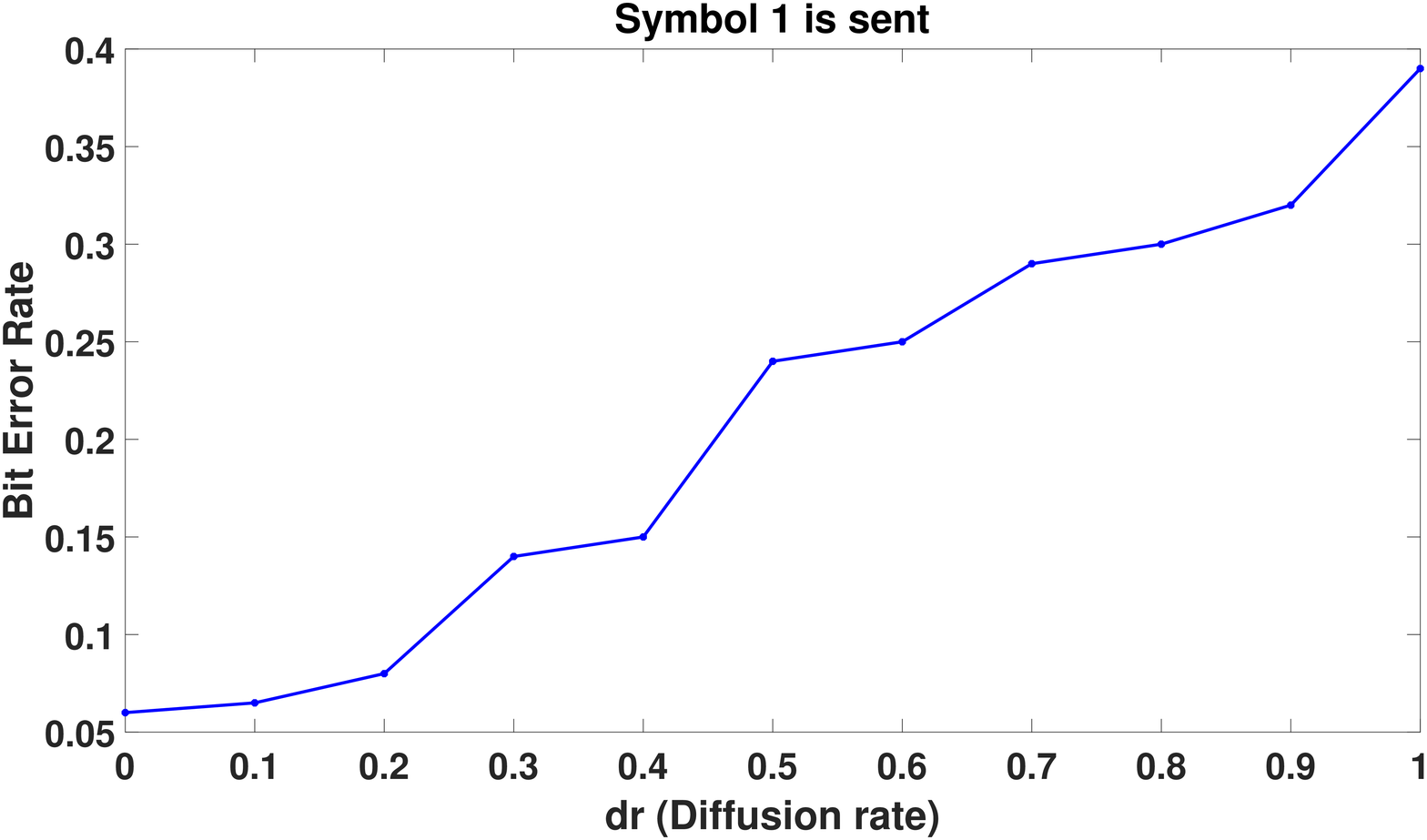}
        \caption{}
 \label{result_4}
    \end{subfigure}  
    \caption{Comparing BER for different values of diffusion coefficient $d_r$}
    \label{result_3_result_4}
\end{figure}

    \begin{figure}
    \centering  
        \begin{subfigure}[t]{0.45\textwidth}
         \centering      \includegraphics[width=9cm,height=5cm]{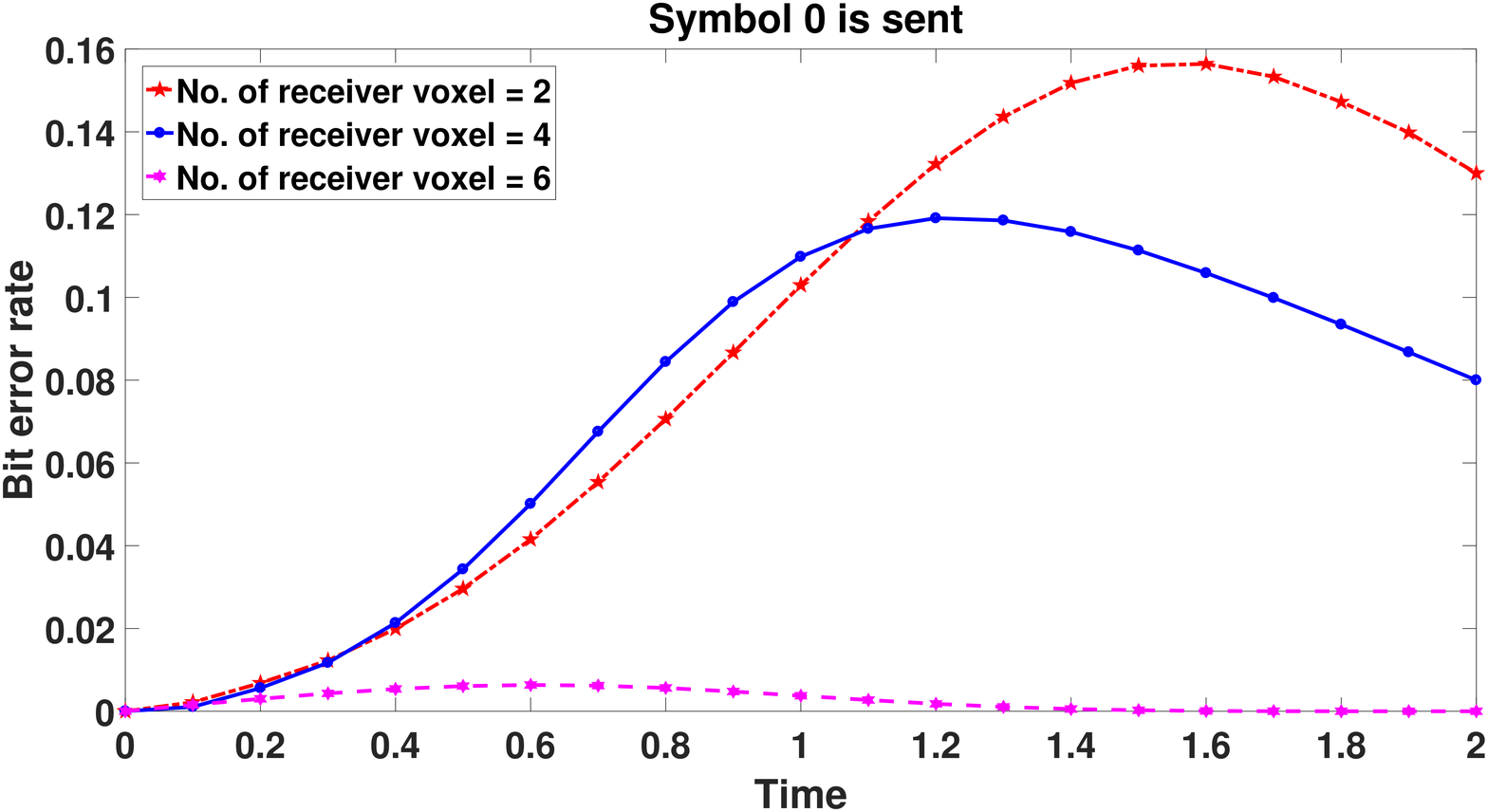}
        \caption{}
 \label{result_5}
         \end{subfigure}
     \begin{subfigure}[t]{0.45\textwidth}
        \centering       \includegraphics[width=9cm,height=5cm]{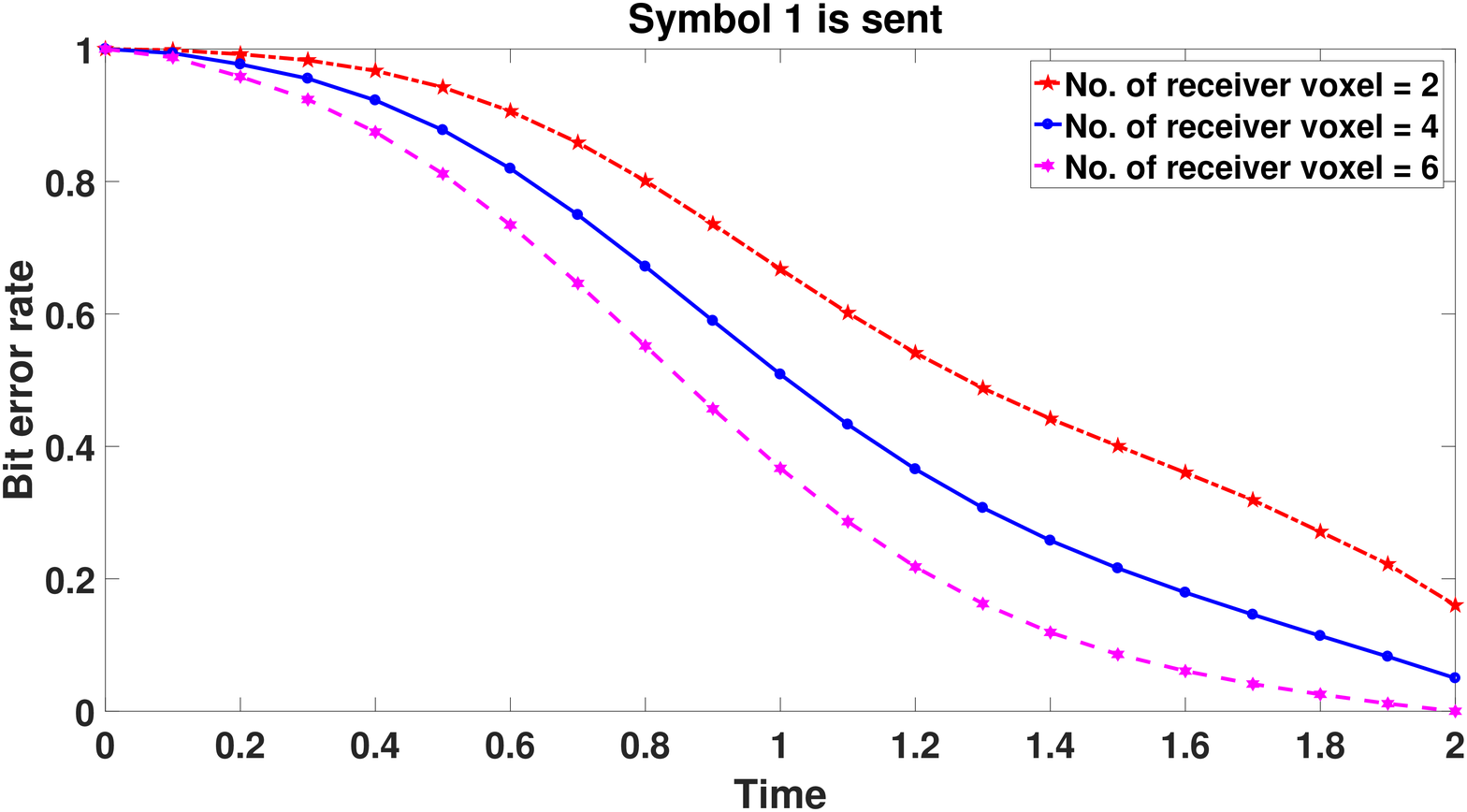}
       \caption{}
 \label{result_6}
    \end{subfigure}
       \caption{Comparing BER for different number of receiver voxels}
    \label{result_5_result_6}
\end{figure}
\begin{figure}
    \centering  
    \begin{subfigure}[t]{0.45\textwidth}
        \centering        \includegraphics[width=9cm,height=5cm]{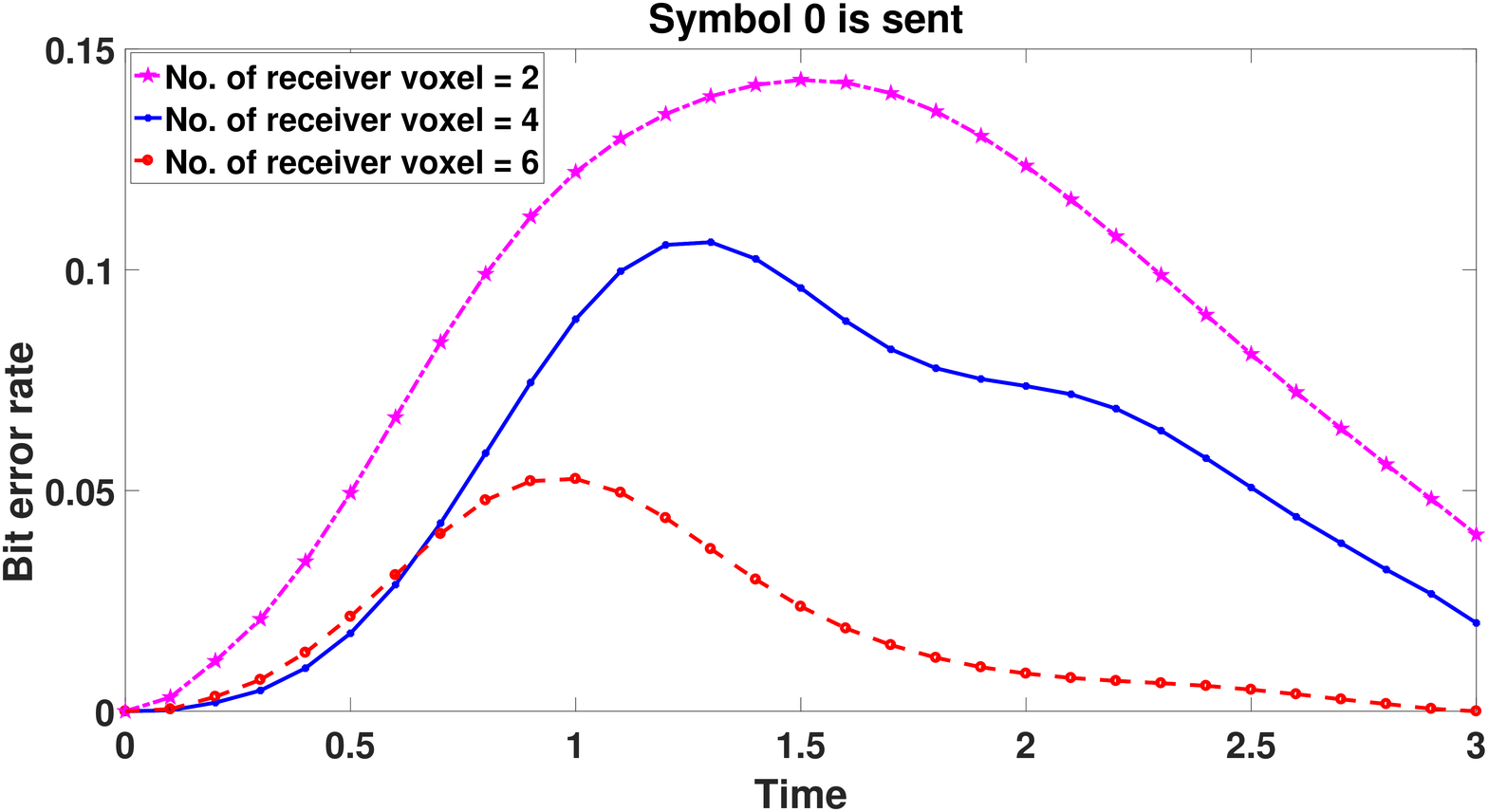}
       \caption{}
 \label{result_7}
    \end{subfigure}
     \begin{subfigure}[t]{0.45\textwidth}
         \centering        \includegraphics[width=9cm,height=5cm]{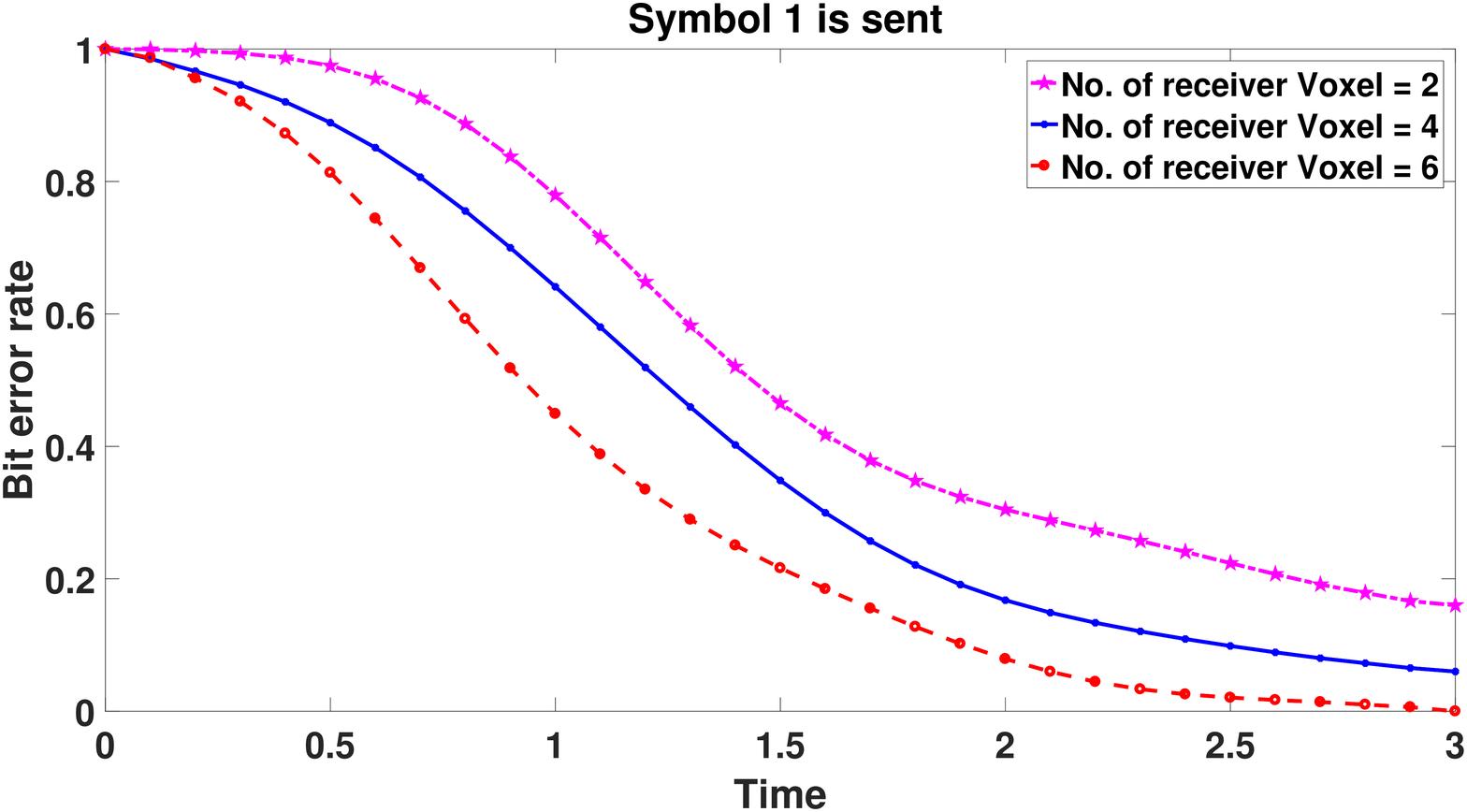}
       \caption{}
       \label{result_8}
    \end{subfigure}      
    \caption{Comparing BER for different number of voxels when total number of receptor for all receiver voxels i.e. M is fix.}
    \label{result_7_result_8}
\end{figure}
\subsection{BER for mixed and partitioned configurations}
\label{sec:expt:mixed:part}
The aim of this experiment is to compare the BER for mixed and partitioned configurations for different receiver locations. We assume that the size of propagation medium is 1$\frac{2}{3}\mu$m$\times$1$\frac{2}{3}\mu$m$\times$1$\frac{2}{3}\mu$m and the size of voxel is $\frac{1}{3}$$\mu$m$^{3}$ (i.e. $w= \frac{1}{3}$ $\mu$ m). This forms a grid of $5 \times 5 \times 5$ voxels. 

We use a 3-tuple $(x,y,z)$ where $1 \leq x, y, z \leq 5$ to identify the locations of the voxels. The transmitter voxel is located at (1,1,1). We use two receiver voxels and place them at (4,5,5) and (5,5,5). The transmitter and receiver voxel locations are depicted in the left-most picture in Fig.~\ref{diff_rx_loc}.We use $M = 10$ and $P = 2$. 

We study the impact of $d_r$ on the BER. Three different values for $d_r$ are used: 0, 0.5 and 1. The SSA simulations is performed up to time 2.5. Fig.~\ref{result_1_result_2} shows the BER for Symbols 0 and 1. It can be seen that a lower $d_r$ leads to a lower BER. In particular, the partitioned configuration leads to the lowest BER. 

\iftcom 

    The above simulations assume that the receiver voxels are placed at a diagonally opposite corner to the transmitter voxel. We have also simulated other configurations where the receiver voxels are placed at other locations in the medium and the results are similar, see our technical report \cite{riaz2019using}
\else 
        The above simulations assume that the receiver voxels are placed at a diagonally opposite corner to the transmitter voxel. For two receiver voxel configuration, We have also performed simulations with the receiver voxels at different locations within the medium. The different receiver voxel placements that we have used are: (3,4,4), (4,4,4); (2,3,3), (3,3,3); and (4,1,1),(5,1,1). Fig.~\ref{different_receiver_locations} shows BER is lower for smaller $d_r$. 
\fi 
\subsection{BER for different values of diffusion coefficient}
\label{sec:num:sub:mixed}
The aim of this experiment is to further study the impact of diffusion of the receiver species on BER. We choose $M = 10$ and $P = 2$. We vary $d_r$ $\mu$m$^2$s$^{-1}$ from 0 to 1 with an increment of 0.1. We used the BER at time 2.5 sec for comparison. Figs. \ref{result_3} and \ref{result_4} show how BER varies with $d_r$ for, respectively, Symbols 0 and 1. It shows that BER increases monotonically with $d_r$. 

We can consider the partitioned configuration as perfect isolation of receptors into clusters where there is a cluster per voxel and the mixed configuration as imperfect isolation where larger values of $d_r$ means farther away from perfect isolation. The results in this section show that our approximate demodulators offer a gradual degradation in performance with $d_r$.

\subsection{Impact of the number of receiver voxels with fixed $M$}
This section studies the impact of the number of receiver voxels on BER for partitioned configuration. We maintain $M = 10$ and we use three different receivers, with 2, 4 and 6 voxels. The total number of receptors for these receivers are therefore 20, 40 and 60 respectively. The voxel locations are: (4,5,5), (5,5,5) for 2 receiver voxels; (2,5,5), (3,5,5), (4,5,5) and (5,5,5) for 4 receiver voxels; (5,4,5), (1,5,5), (2,5,5), (3,5,5), (4,5,5) and (5,5,5) for 6 receiver voxels.  The transmitter and receiver voxel locations are depicted in Fig.~\ref{diff_rx_loc}a-Fig.~\ref{diff_rx_loc}c.
Fig.~\ref{result_5_result_6} shows that a higher number of voxels leads to a lower BER. 

\iftcom 
  \textcolor{blue}{
            We have also simulated the two different configurations for 6 receiver voxels case by changing the location of receiver voxels in the medium and the results follow the same trend, see our techincal report  \cite{riaz2019using}}
   
\else 

We have also simulated two other configurations for the 6 receiver voxel case by changing the location of receiver voxels in the medium. The different receiver voxel positions that we have used are: 
\begin{itemize}
    \item (3,4,5), (4,4,5), (5,4,5), (3,5,5), (4,5,5), (5,5,5) which is depicted in Fig.~\ref{diff_rx_loc}d; and, 
    \item (3,4,1), (4,4,1), (5,4,1), (3,5,1), (4,5,1), (5,5,1)
\end{itemize} 
Fig.~\ref{different_receiver_locations_with_six_receiver_voxel} shows that the results follow the same trend.
\fi 

\subsection{Impact of the number of receiver voxels with a fixed total number of receptors}

This section studies the impact of the number of receiver voxels on BER. We use three different number of voxels per receiver, namely 2, 4 and 6 voxels. We maintain the total number of receptors in each receiver at 60. Therefore, the number of receptors per voxel for the three receivers are 30, 15 and 10. Figs.~\ref{result_7} and \ref{result_8} show that BER for Symbols 0 and 1 over time. It shows that in general, a higher number of voxels will lead to a lower BER. 


\begin{figure}
    \centering
       \begin{subfigure}[t]{0.45\textwidth}
        \centering       
        \includegraphics[width=9cm,height=5cm]{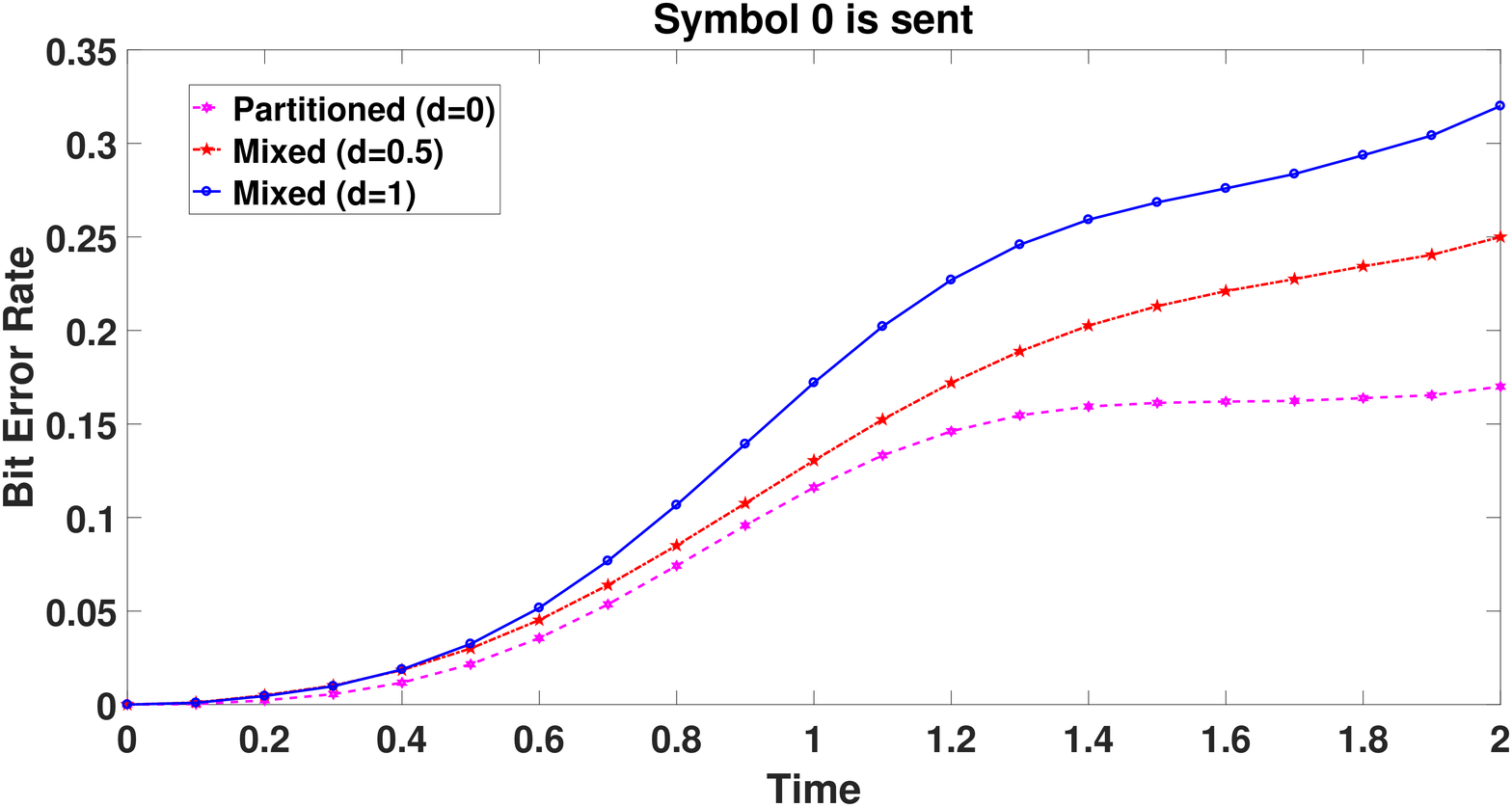}
        \caption{BER for Mixed and Partitioned for symbol 0}
        \label{result_9}
    \end{subfigure}
     \begin{subfigure}[t]{0.45\textwidth}
        \centering        \includegraphics[width=9cm,height=5cm]{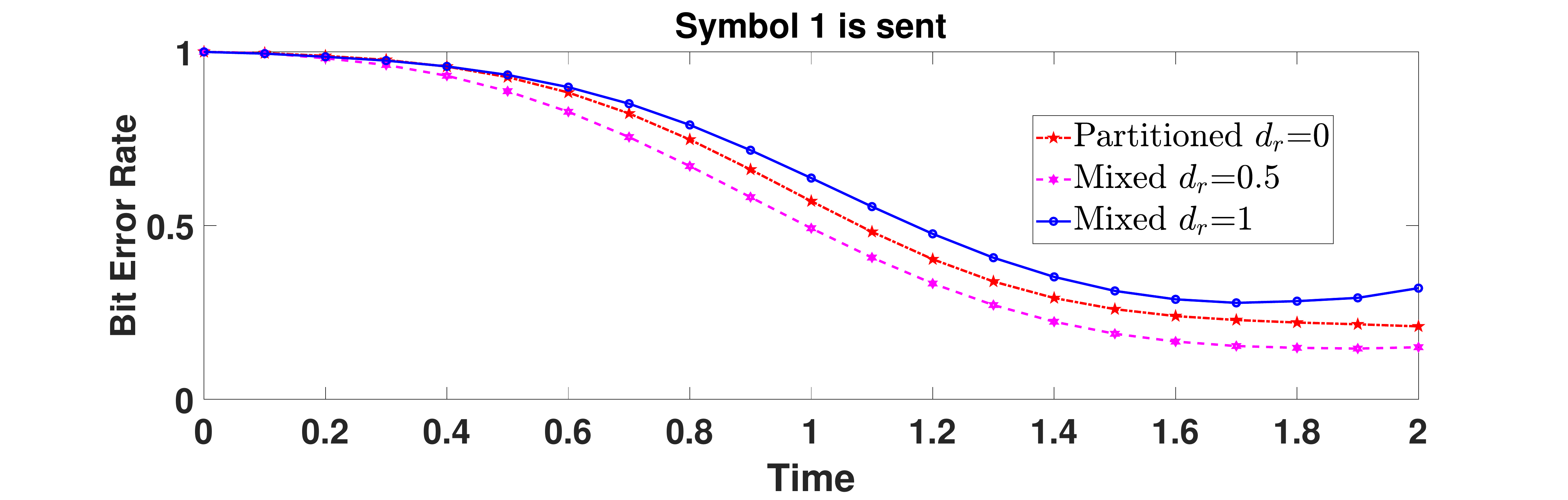}
        \caption{BER for Mixed and Partitioned for symbol 1}
        \label{result_10}
    \end{subfigure}
    \caption{Comparing BER for Mixed and Partitioned configuration for different molecular circuit}
    \label{fig:result_9_10}
\end{figure}
\iftcom
\else 
\begin{figure}
    \centering
       \begin{subfigure}[t]{0.45\textwidth}
        \centering       
        \includegraphics[width=9cm,height=5cm]{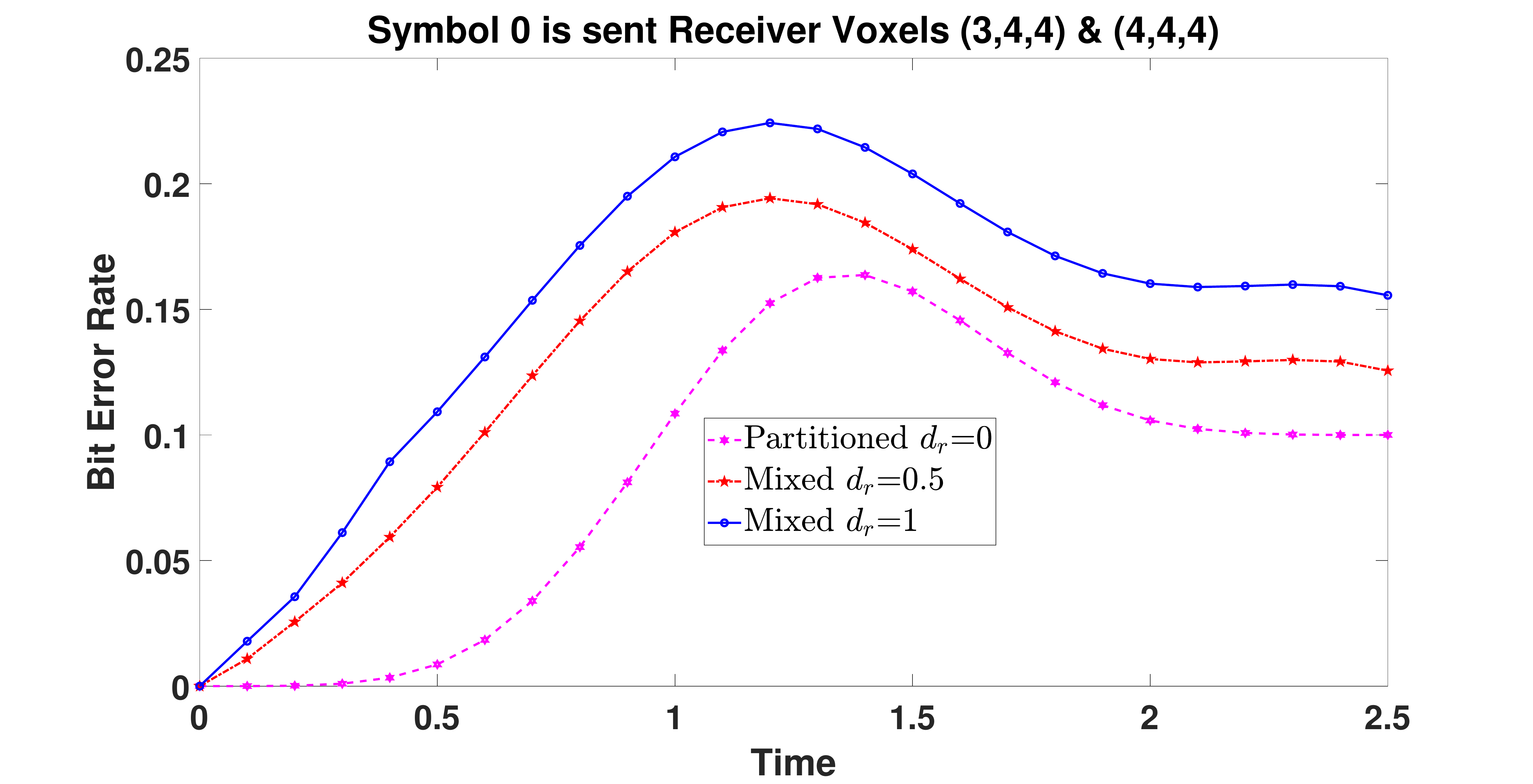}
        \caption{}
        \label{R2S0}
    \end{subfigure}
     \begin{subfigure}[t]{0.45\textwidth}
        \centering       
        \includegraphics[width=9cm,height=5cm]{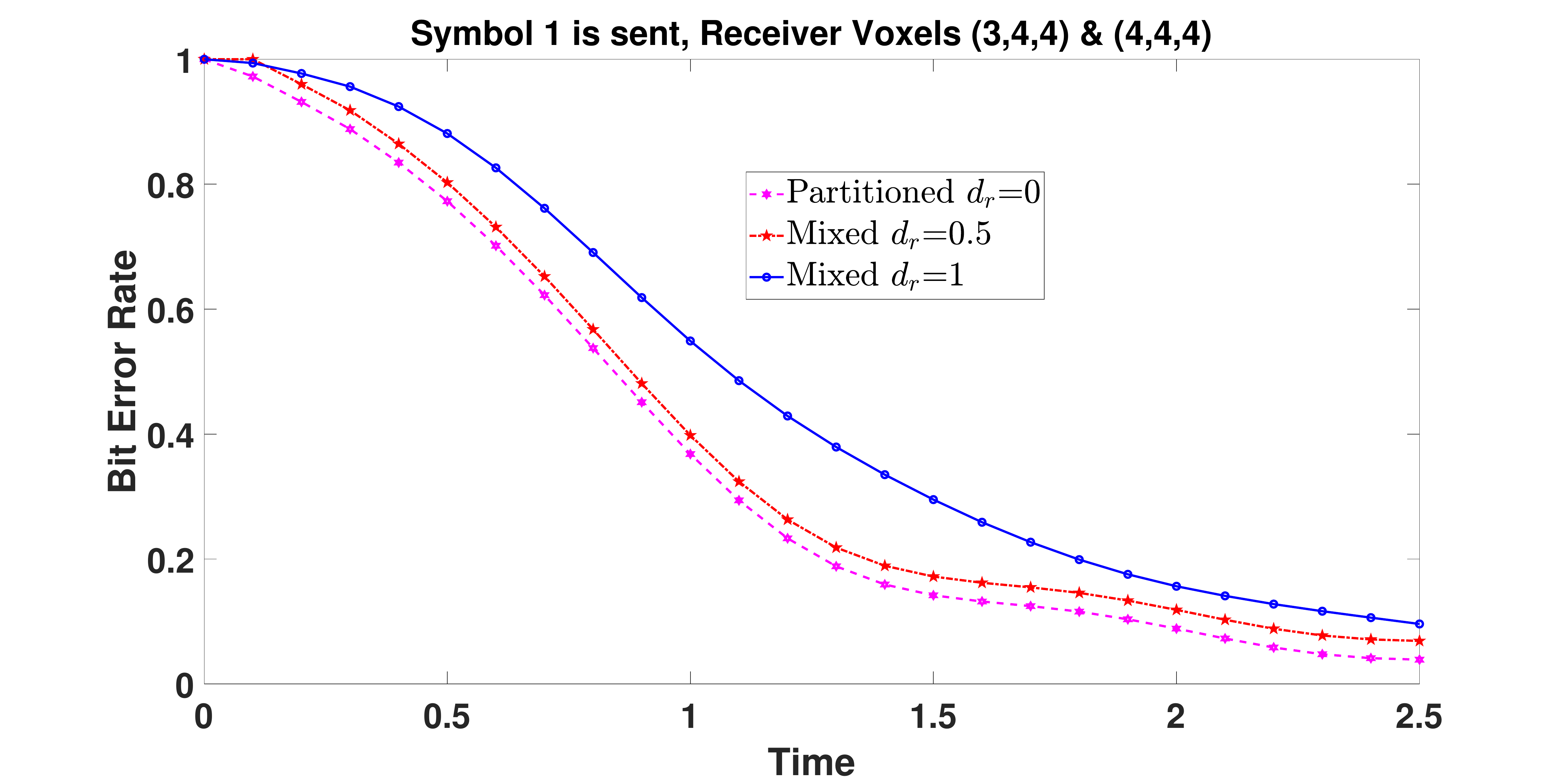}
        \caption{}
        \label{R2S1}
    \end{subfigure}
        \begin{subfigure}[t]{0.45\textwidth}
        \centering       
        \includegraphics[width=9cm,height=5cm]{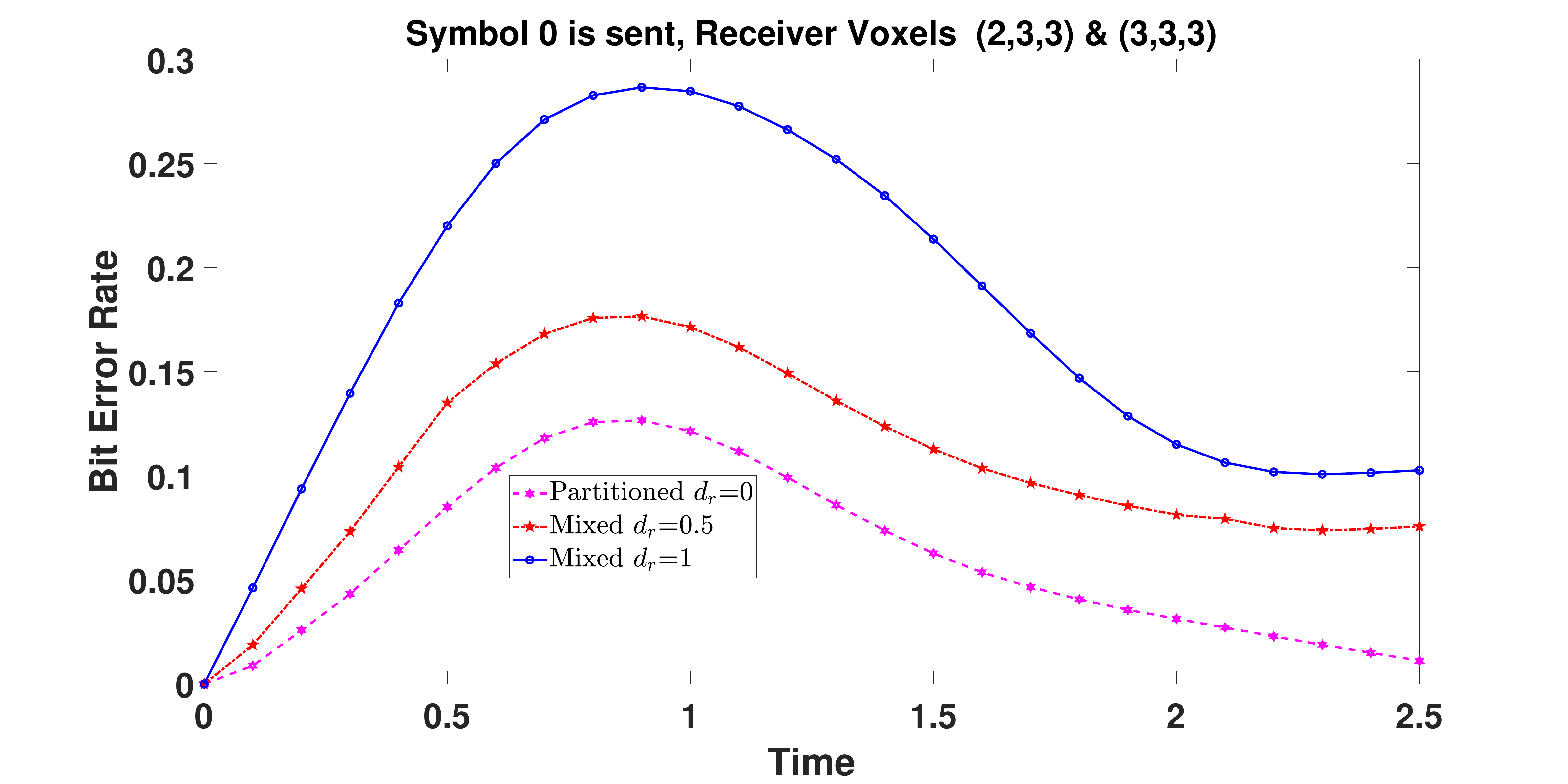}
        \caption{}
        \label{R3S0}
    \end{subfigure}
         \begin{subfigure}[t]{0.45\textwidth}
        \centering     
        \includegraphics[width=9cm,height=5cm]{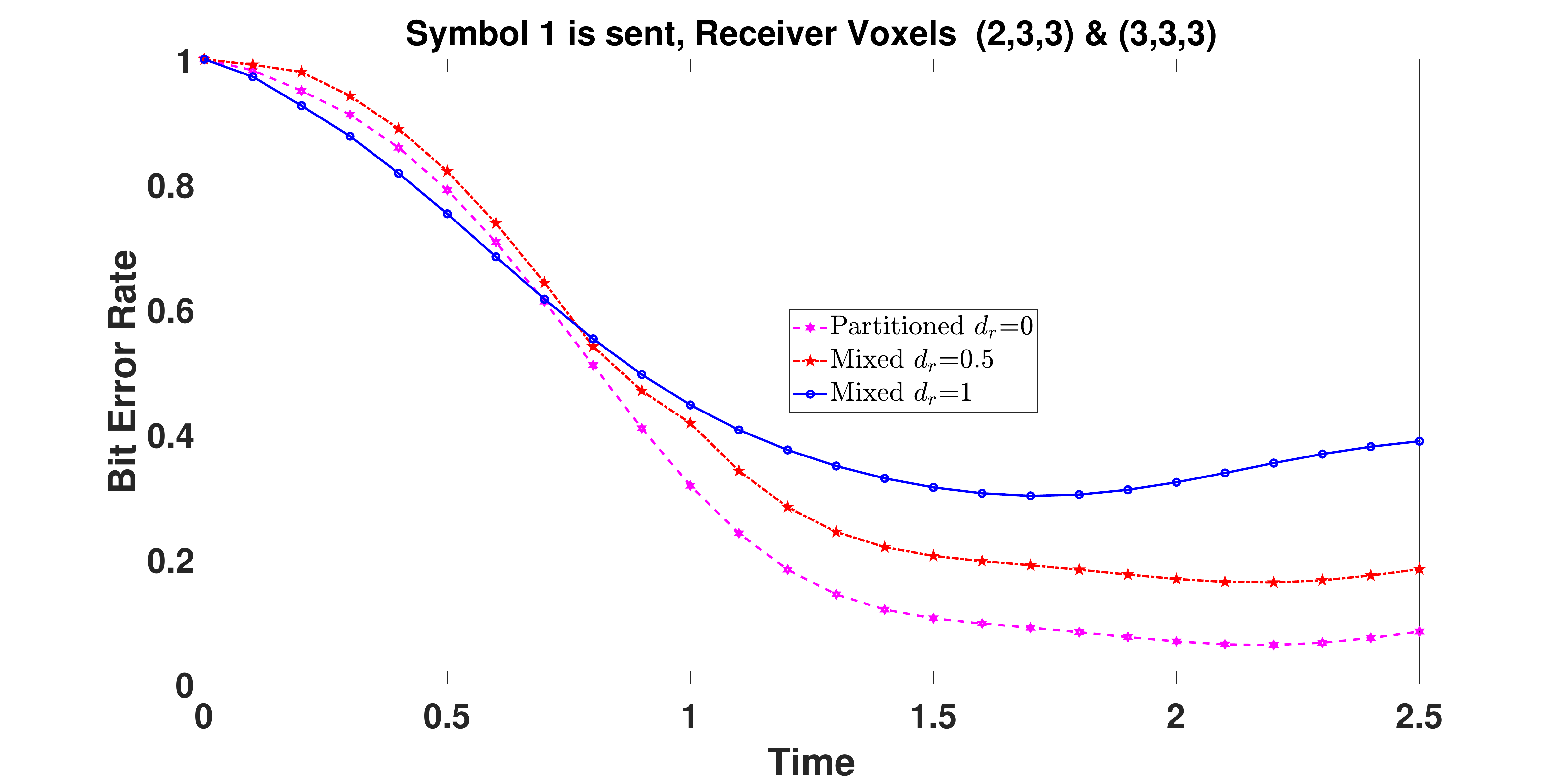}
        \caption{}
        \label{R3S1}
    \end{subfigure}  
        \begin{subfigure}[t]{0.45\textwidth}
         \centering      
         \includegraphics[width=9cm,height=5cm]{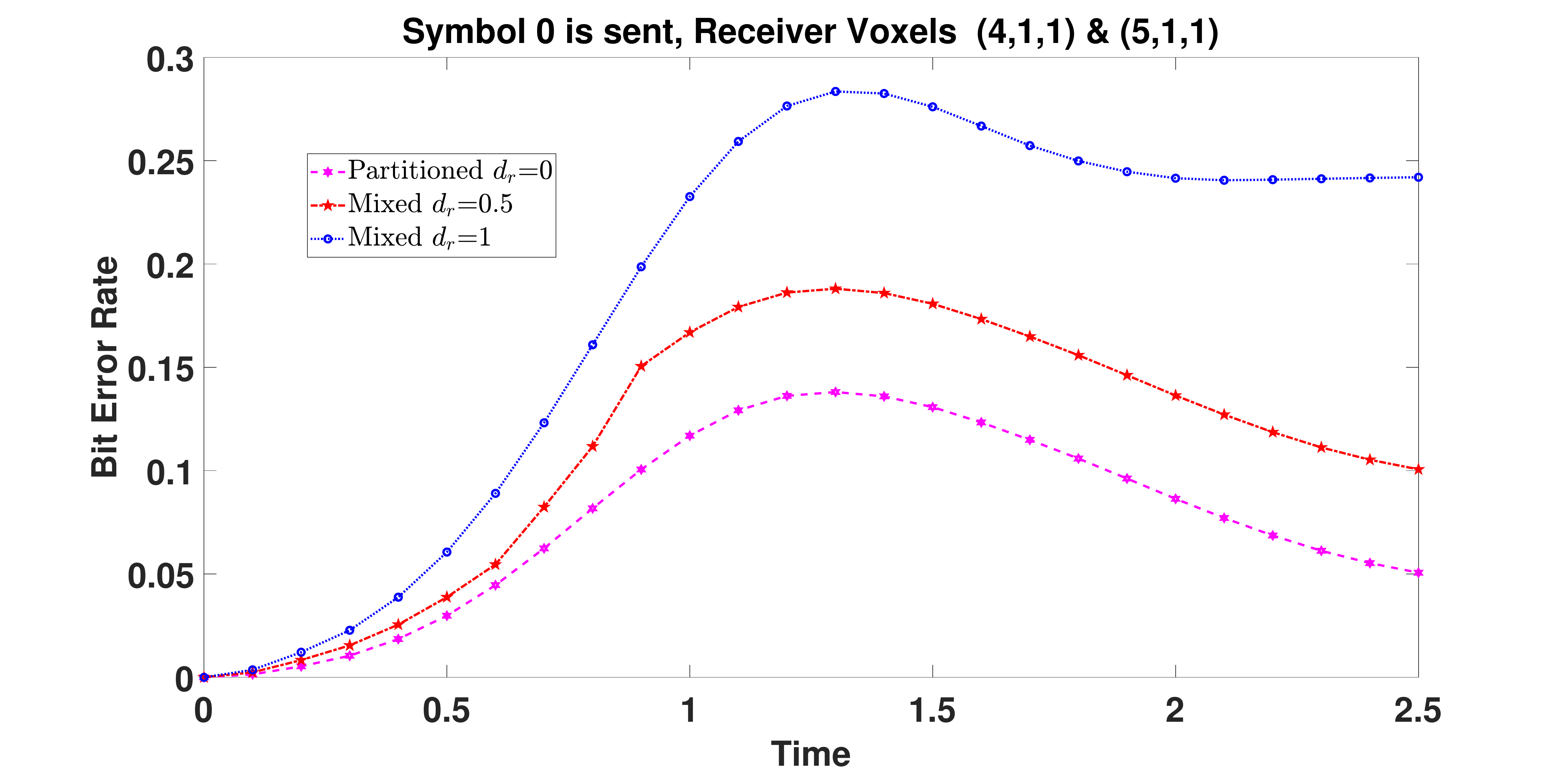}
        \caption{}
        \label{R5S0}
         \end{subfigure}
     \begin{subfigure}[t]{0.45\textwidth}
        \centering       
        \includegraphics[width=9cm,height=5cm]{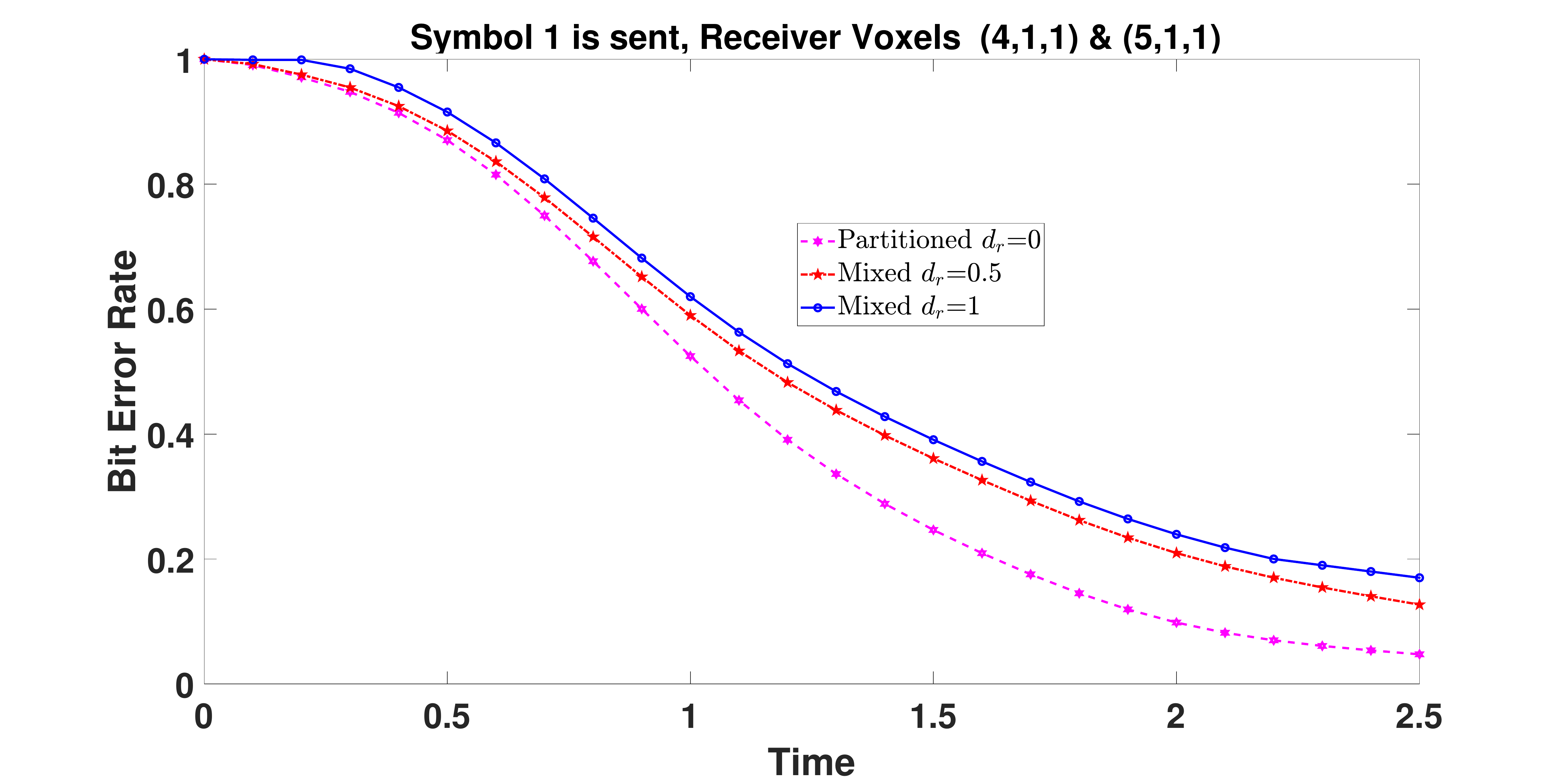}
       \caption{}
       \label{R5S1}
    \end{subfigure}
    \caption{Comparing BER for mixed and partitioned configurations for different receiver locations with two receiver voxels.}
    \label{different_receiver_locations}
\end{figure}
\fi

\iftcom
\else 
\begin{figure}
    \centering
       \begin{subfigure}[t]{0.45\textwidth}
        \centering       
        \includegraphics[width=9cm,height=5cm]{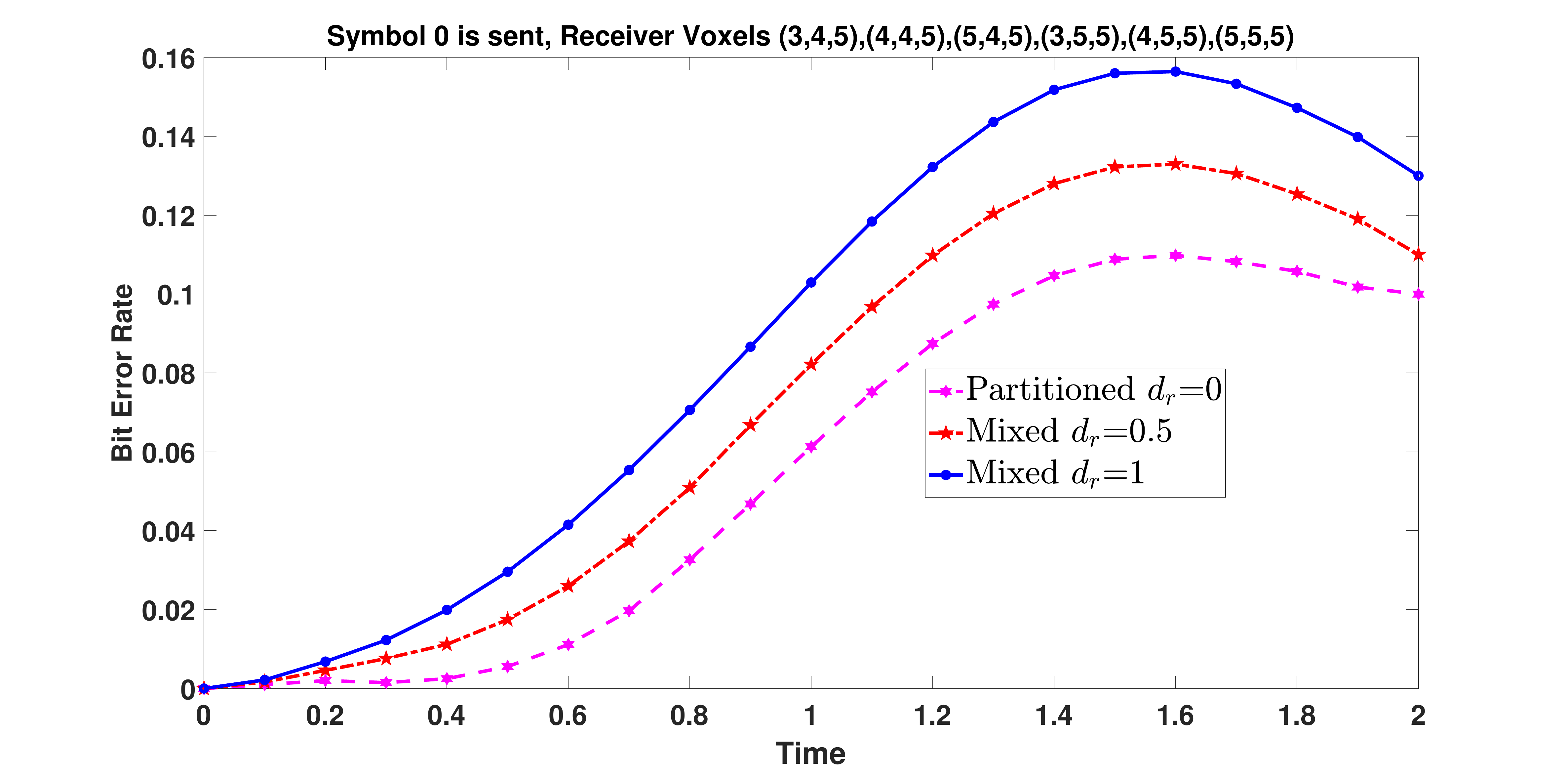}
        \caption{}
        \label{RS0_6}
    \end{subfigure}
     \begin{subfigure}[t]{0.45\textwidth}
        \centering       
        \includegraphics[width=9cm,height=5cm]{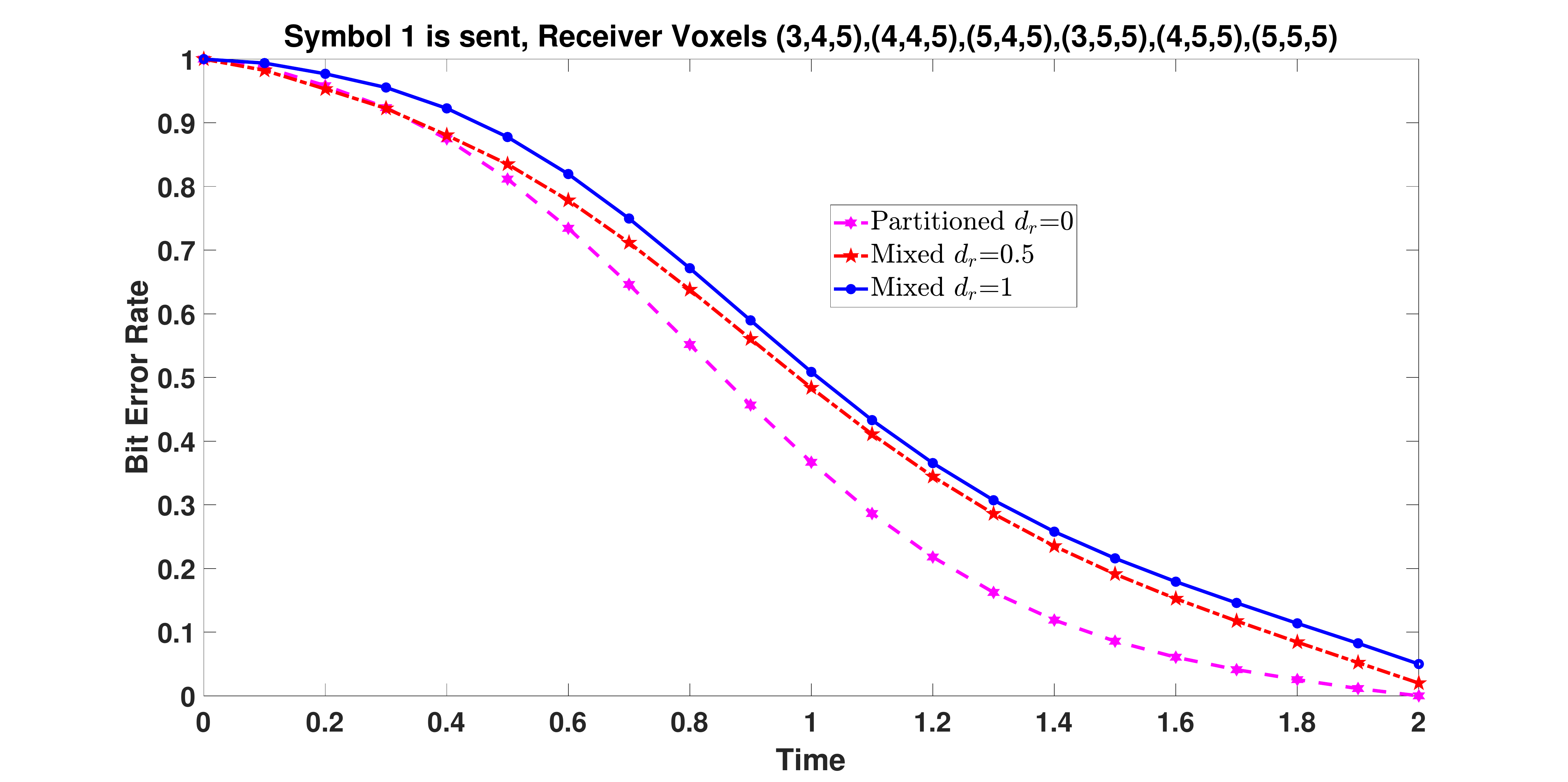}
        \caption{}
        \label{RS1_6}
    \end{subfigure}
        \begin{subfigure}[t]{0.45\textwidth}
        \centering       
        \includegraphics[width=9cm,height=5cm]{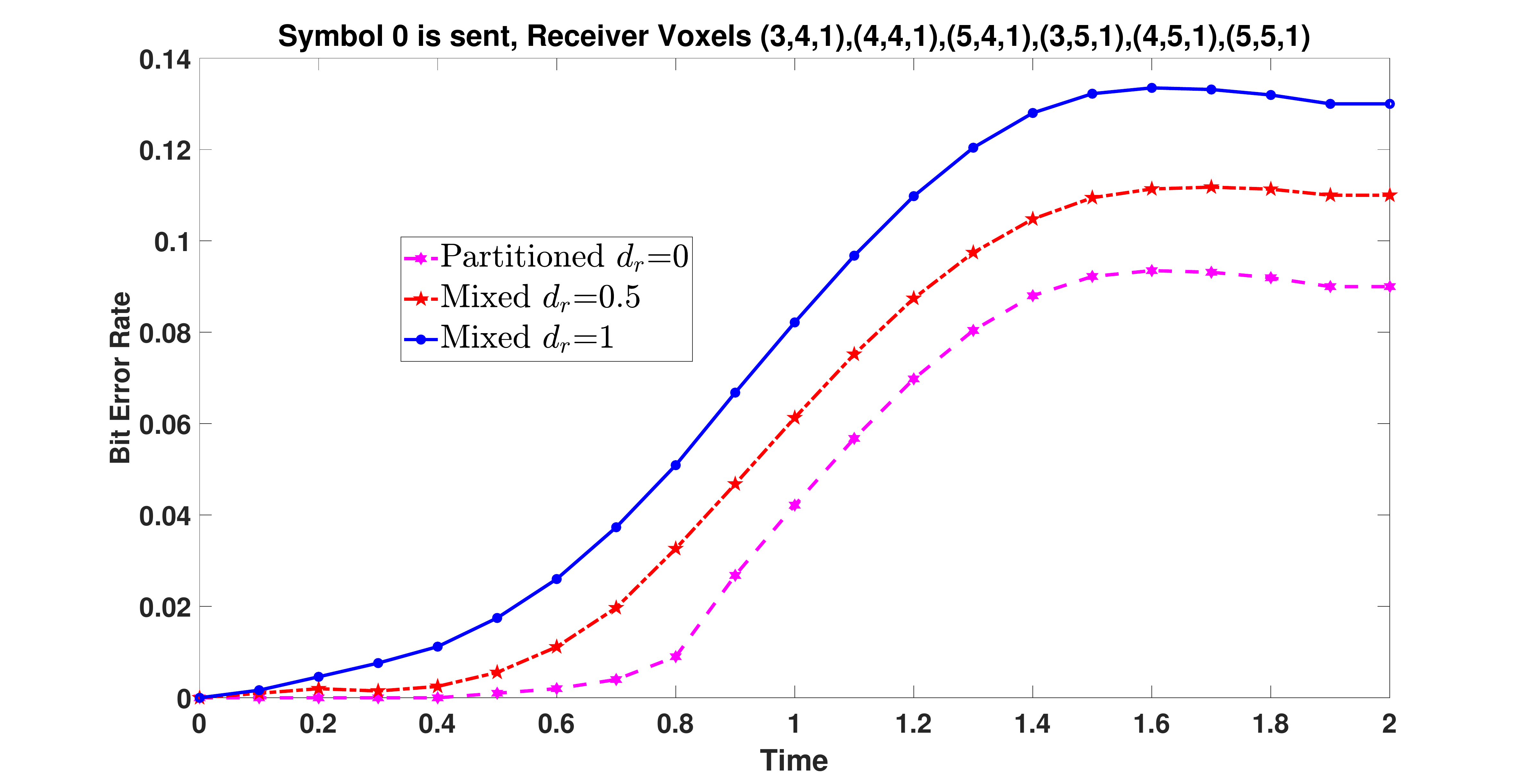}
        \caption{}
        \label{RS0a_6}
    \end{subfigure}
         \begin{subfigure}[t]{0.45\textwidth}
        \centering     
        \includegraphics[width=9cm,height=5cm]{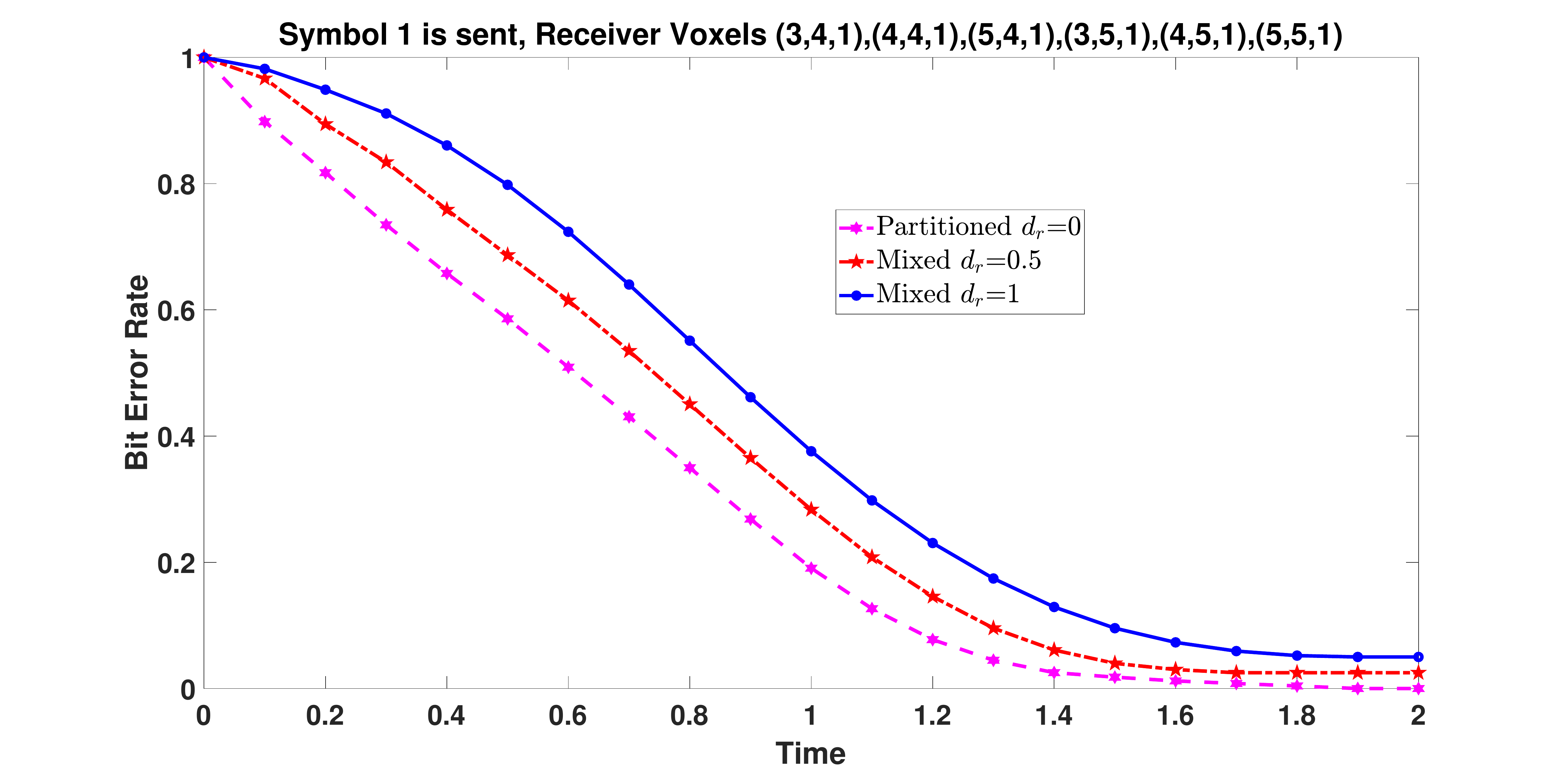}
        \caption{}
        \label{RS1a_6}
    \end{subfigure}  
    \caption{Comparing BER for mixed and partitioned configurations for different receiver locations with six receiver voxels.}
    \label{different_receiver_locations_with_six_receiver_voxel}
\end{figure}
\fi

\begin{figure}
    \centering
       \begin{subfigure}[t]{0.45\textwidth}
        \centering       
        \includegraphics[width=9cm,height=5cm]{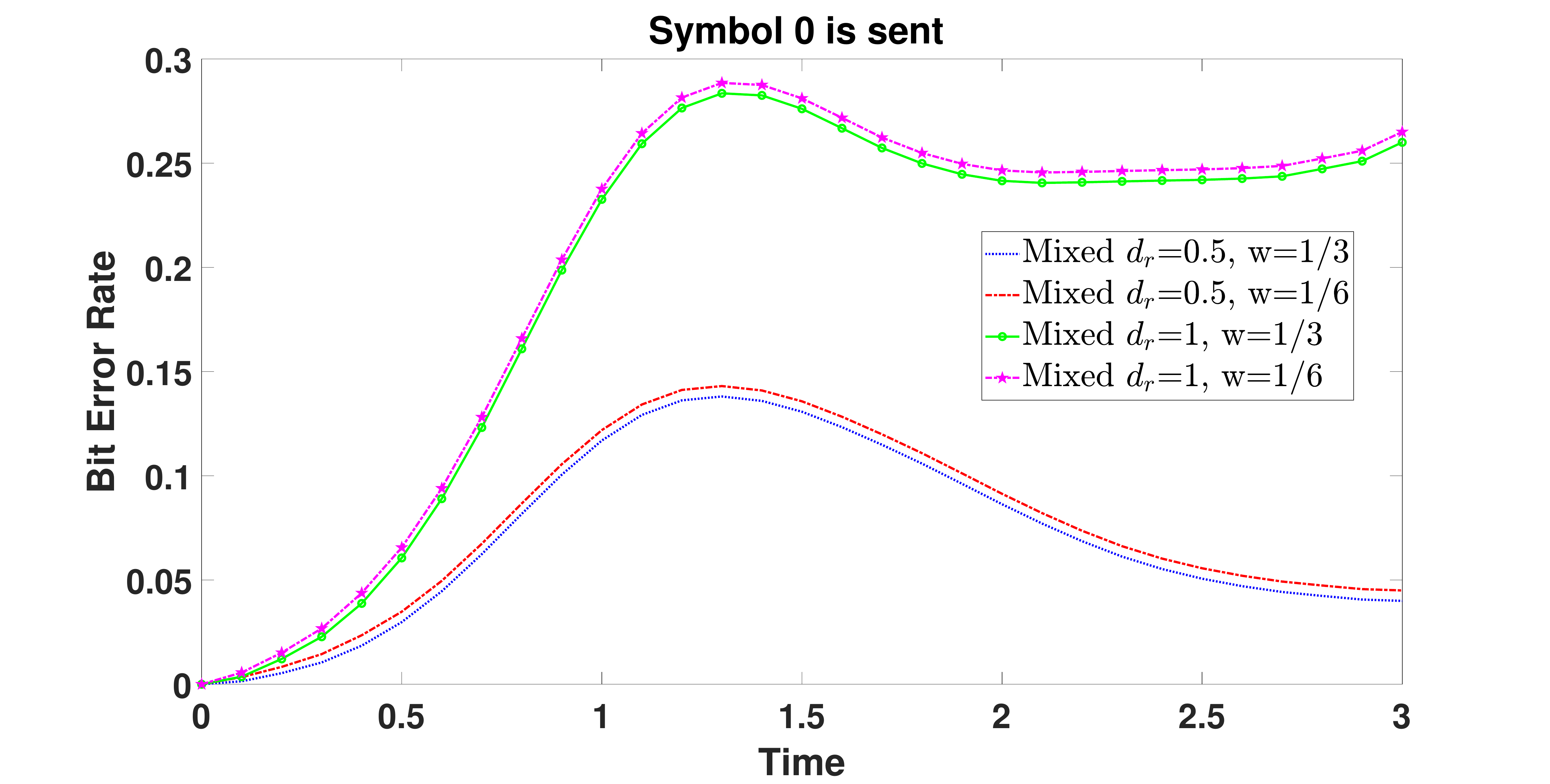}
        \caption{}
        \label{WMS00}
    \end{subfigure}
     \begin{subfigure}[t]{0.45\textwidth}
        \centering        \includegraphics[width=9cm,height=5cm]{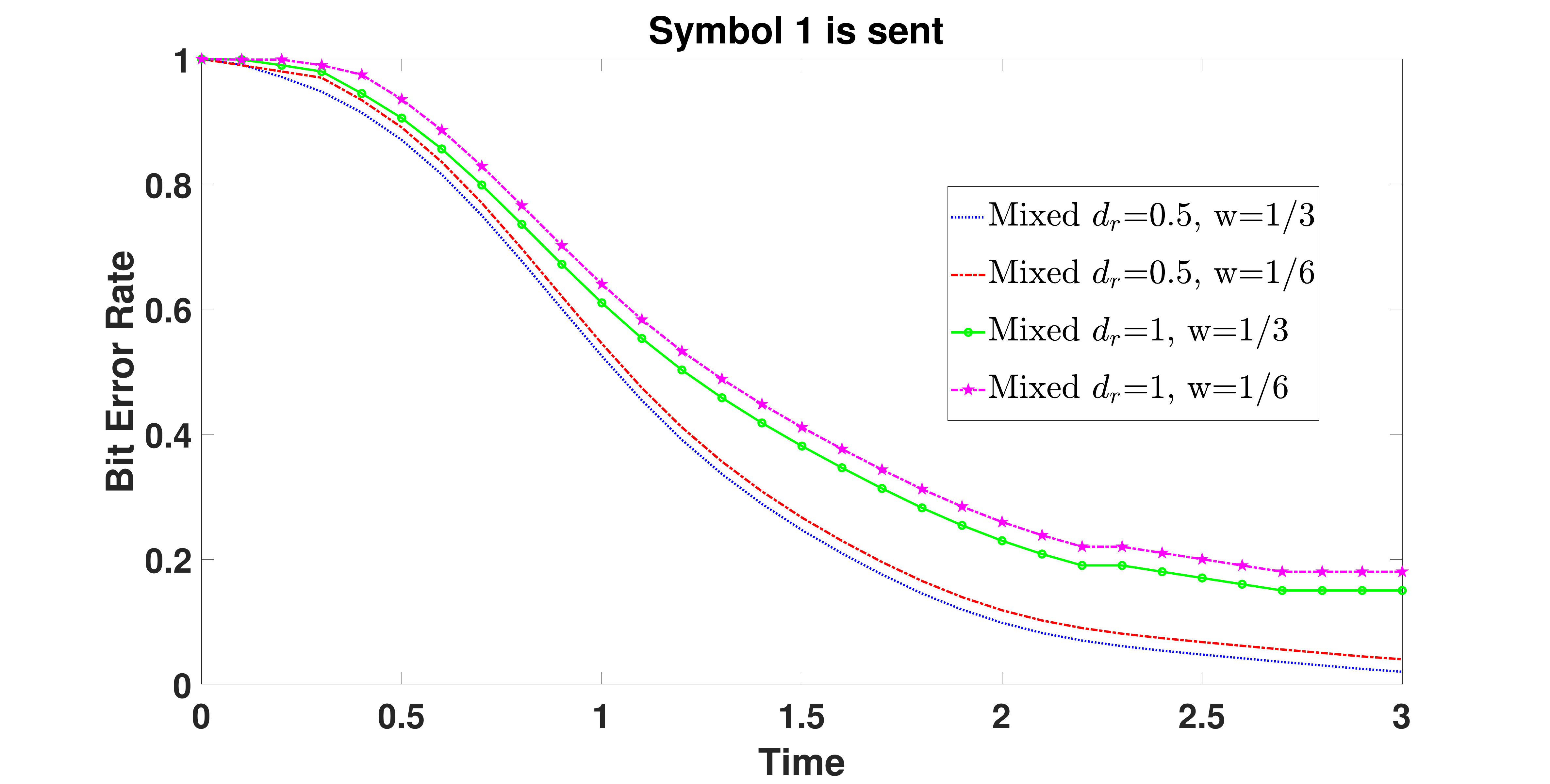}
        \caption{}
        \label{WMS11}
    \end{subfigure}
        \begin{subfigure}[t]{0.45\textwidth}
        \centering       \includegraphics[width=9cm,height=5cm]{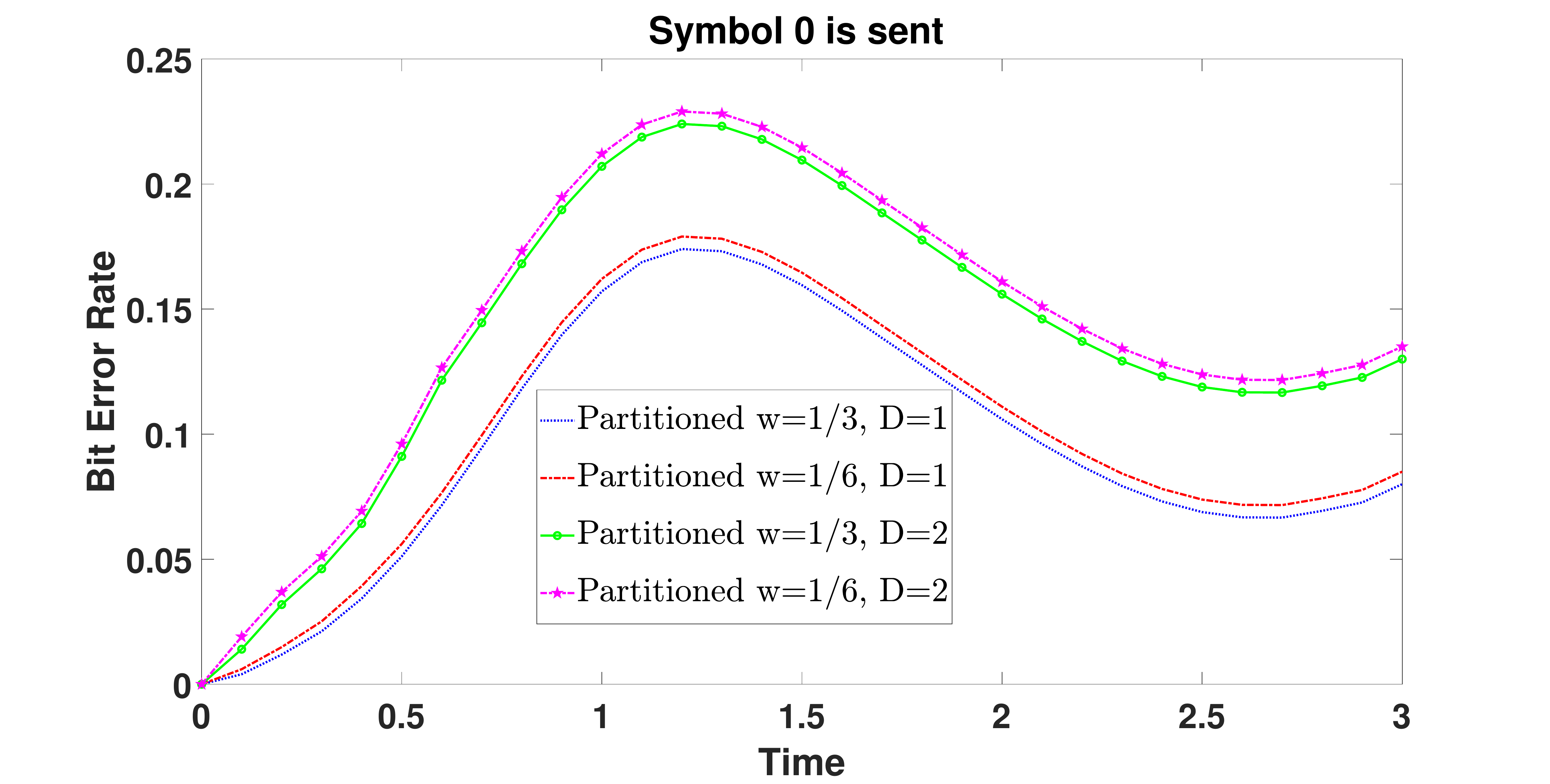}
        \caption{}
        \label{WPS00}
    \end{subfigure}
         \begin{subfigure}[t]{0.45\textwidth}
        \centering     \includegraphics[width=9cm,height=5cm]{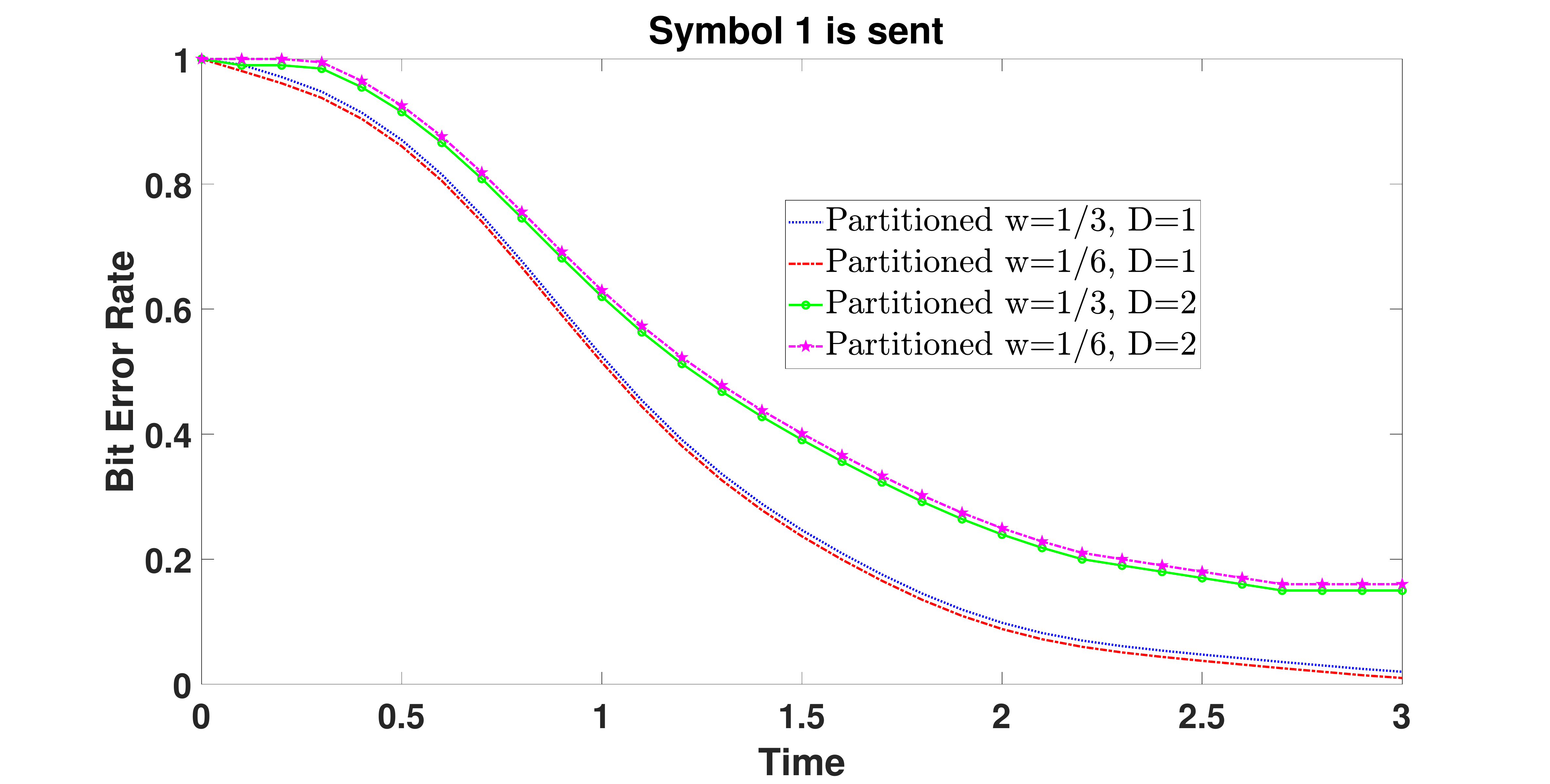}
        \caption{}
        \label{WPS11}
    \end{subfigure}  
   
    \caption{Impact of voxel size on BER}
    \label{WPS00_WPS11}
\end{figure}
\subsection{BER using a different molecular receiver circuit}
All the above experiments have been carried out with the receiver molecular circuit in \eqref{cr:all}. In this experiemnt, we use the following molecular circuit which was used in our earlier work \cite{awan2017generalized}:

\begin{align}
\cee{
S + E &<=>C[\tilde{\lambda}_1][\mu_1] C_{[1]} \label{eq:mc1:r1}  \\
S + C_{[1]}  &<=>C[\tilde{\lambda}_2][\mu_2] C_{[2]} \label{eq:mc1:r2}  }
\end{align}

where $E$ represent an unbound receptor with two binding sites.  In forward reaction \eqref{eq:mc1:r1}, $E$ can bind with $S$ molecule to form the complex $C_{[1]}$ whereas in forward reaction \eqref{eq:mc1:r2}, $C_{[1]}$ can bind with $S$ molecule to form the complex $C_{[2]}$. Furthermore, $\tilde{\lambda}_1$, $\mu_1$, $\tilde{\lambda}_2$ and $\mu_2$ are reaction rate constants. The complex $C_{[2]}$ is chosen as the output species. 

We assume $M = 2$ and $P = 2$. All other parameters remain the same. We use three values of $d_r$: 0, 0.5 and 1. 
Figs.~\ref{result_9} and \ref{result_10} show the BER for, respectively, Symbols 0 and 1. We witness the same trend as before where the BER increases with $d_r$.

\subsection{Impact of the voxel size}

We know from the literature on RDME that the voxel size has to be chosen correctly in order to produce correct simulation results \cite{Hellander:2017ea}, e.g. the paper  \cite{hellander2016reaction} mentions a lower bound on the voxel size. Here, we present an example to show that, if the voxel size is chosen correctly, then we are able to obtain consistent BER for different voxel sizes. 

Unless otherwise stated, we use the same parameter values as Section \ref{sec:expt:mixed:part}. We consider two voxel edge lengths: $w=\frac{1}{3}$ and $w=\frac{1}{6}$ while maintaining the same size for the medium, transmitter and receiver. We make the following adjustments: (i) A receiver consisting of 2 voxels for $w=\frac{1}{3}$ will have 16 voxels for $w = \frac{1}{6}$; (ii) Let $u$ be the rate at which signalling molecules are produced in the transmitter voxel when $w=\frac{1}{3}$, then they are produced at a rate of $\frac{u}{8}$ in each of the 8 transmitter voxels for $w=\frac{1}{6}$. In addition, we also use two diffusion coefficients $D$: 1 $\mu$m$^2$s$^{-1}$ and 2 $\mu$m$^2$s$^{-1}$. We assume $M$ = 40. Fig.~\ref{WPS00_WPS11} shows that the voxel dimension $w=\frac{1}{3}$ and $w=\frac{1}{6}$ give consistent BER. We remark that there is some recent work on automating the choice of voxel size for RDME, see \cite{Hellander:2017ea}.


\subsection{Analytical BER formula} 
\label{sec:num:ber}
In Section \ref{sec:LNA} we present a method to analytically compute the BER using LNA. The aim of this section is to study the accuracy of the method. We assume that the the medium is a cube of dimension ($\frac{2}{3}\mu$m)$^3$. The voxel edge length $w$ is $\frac{1}{3}\mu$m giving a medium shape of 2-by-2-by-2 voxels. The reason why we have chosen to use a small medium size is that, for the verification of the analytical formula, we will need to simulate many times to obtain an accurate estimation of the BER. The transmitter is located at voxel (1,1,1) and the receiver voxels are at (1,2,2) and (2,2,2). We use the same signalling molecule diffusion coefficient and reaction parameters as before. We assume $K = 2$. Since LNA works best when the system is in steady state, we create a steady state in the receiver by using short pulses (0.2s) as the transmission symbol and a much longer simulation time of 20s to allow the system to get to steady state. Also, a reflective boundary condition is used. 

\begin{figure}
    \centering
    \begin{subfigure}[t]{0.45\textwidth}
        \centering
        \includegraphics[scale=0.35]{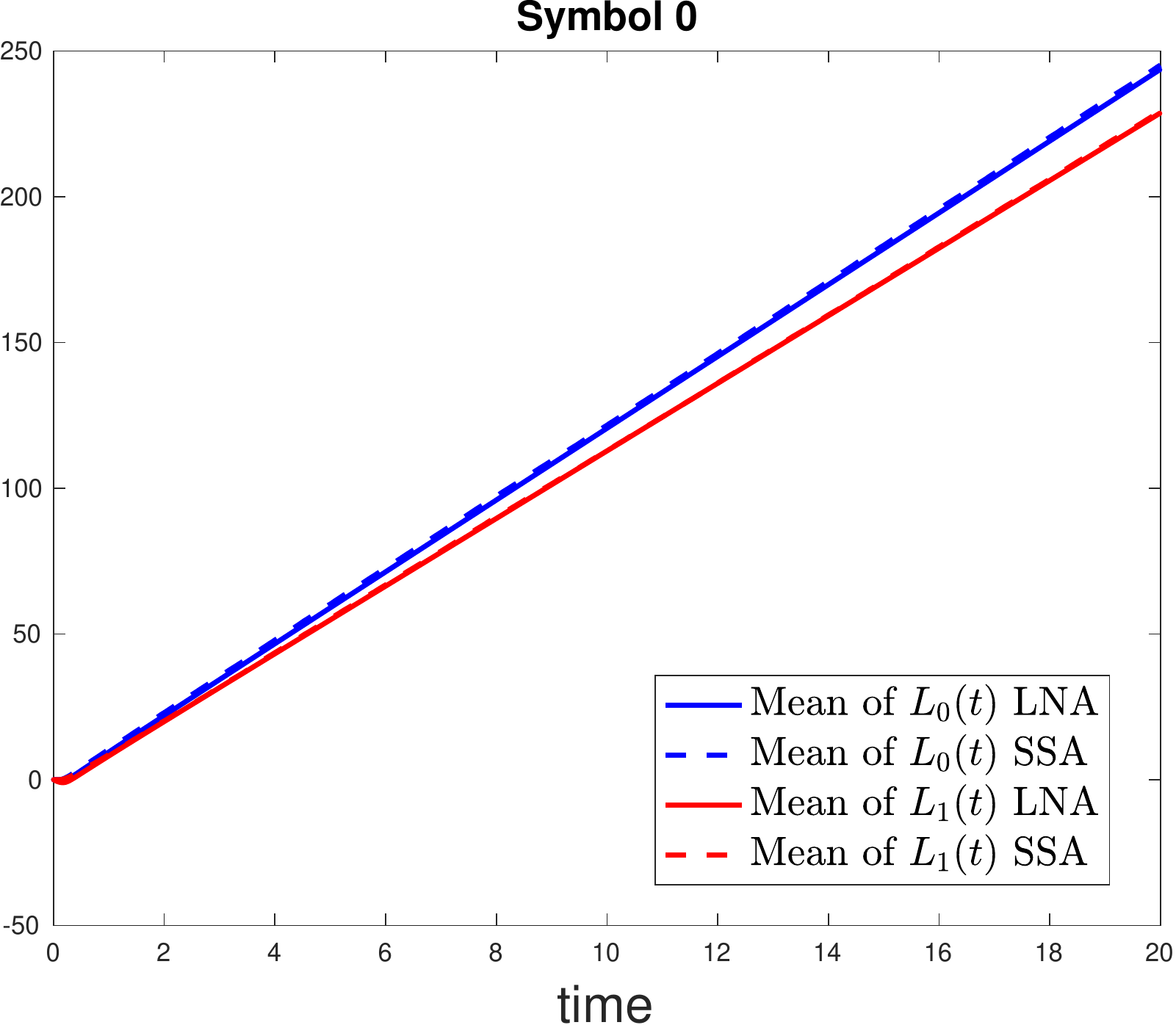}
        \caption{}
        \label{fig:lna_u0_mean}
    \end{subfigure}
     \begin{subfigure}[t]{0.45\textwidth}
        \centering        \includegraphics[scale=0.35]{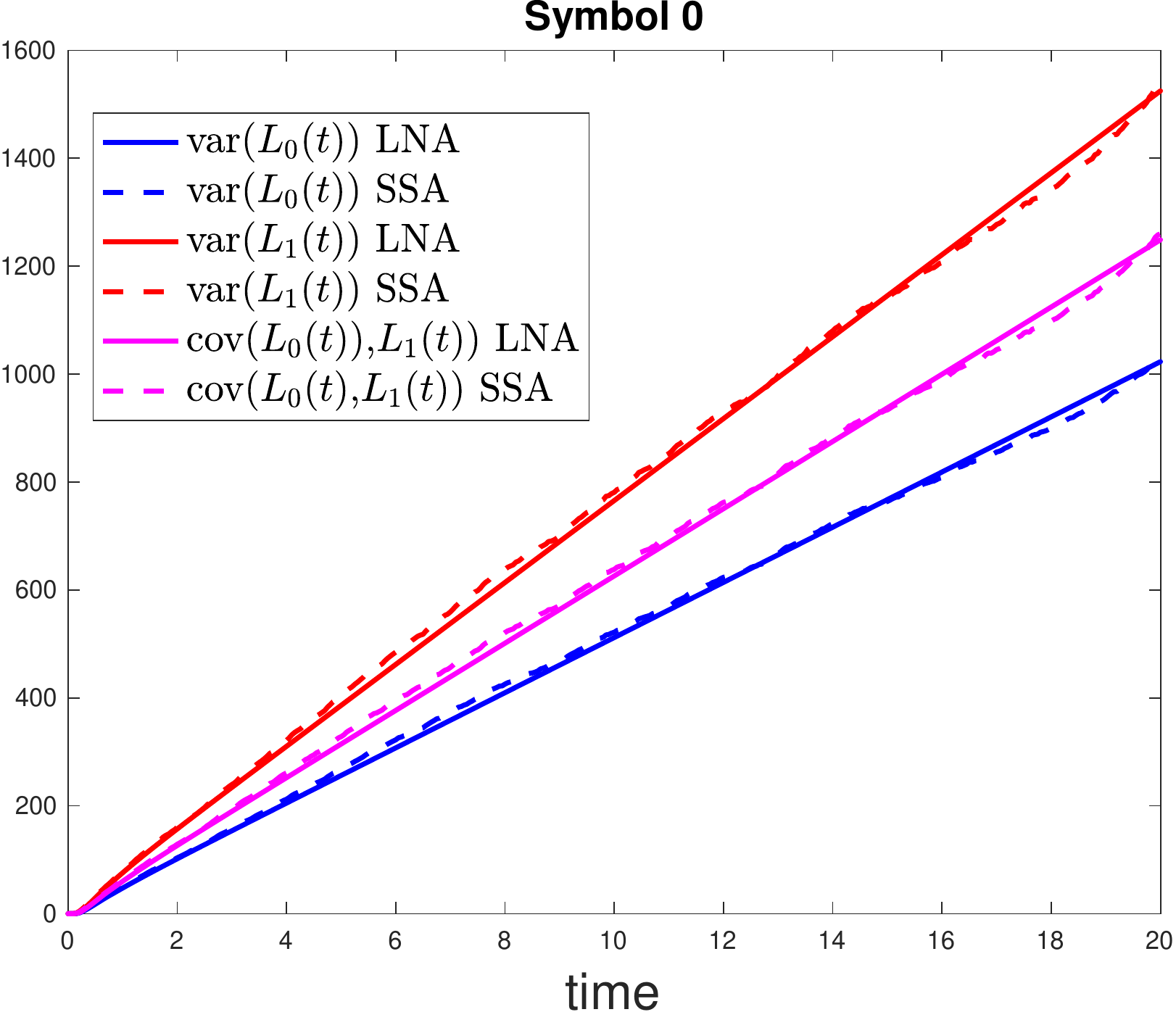}
        \caption{}
        \label{fig:lna_u0_var}
    \end{subfigure}
    \caption{Mean and covariance of $Z_0(t)$ and $Z_1(t)$ by LNA. Symbol 0.}  
    \label{fig:lna_u0}
\end{figure}

We first verify that LNA can be used to accurately compute the mean and the covariance matrix of $Z_0(t)$ and $Z_1(t)$ for the case where $d_r = 0.2$. The verification is done by performing SSA simulations 5000 times per input symbol. Fig.~\ref{fig:lna_u0} shows LNA can accurately estimate the mean and the covariance matrix of $Z_0(t)$ and $Z_1(t)$ for Symbol 0. Note that there are 3 distinct elements in the covariance matrix, which are the variance of $Z_0(t)$ and $Z_1(t)$ as well as the covariance of $Z_0(t)$ and $Z_1(t)$. 
\iftcom
 The results for Symbol 1 is similar, see our technical report \cite{riaz2019using} . 
\else
 The results for Symbol 1 is similar, see Fig.~\ref{fig:lna_u1}. 
\fi

\begin{figure}
\centering
\begin{minipage}{.5\textwidth}
  \centering    
    \includegraphics[scale=0.35]{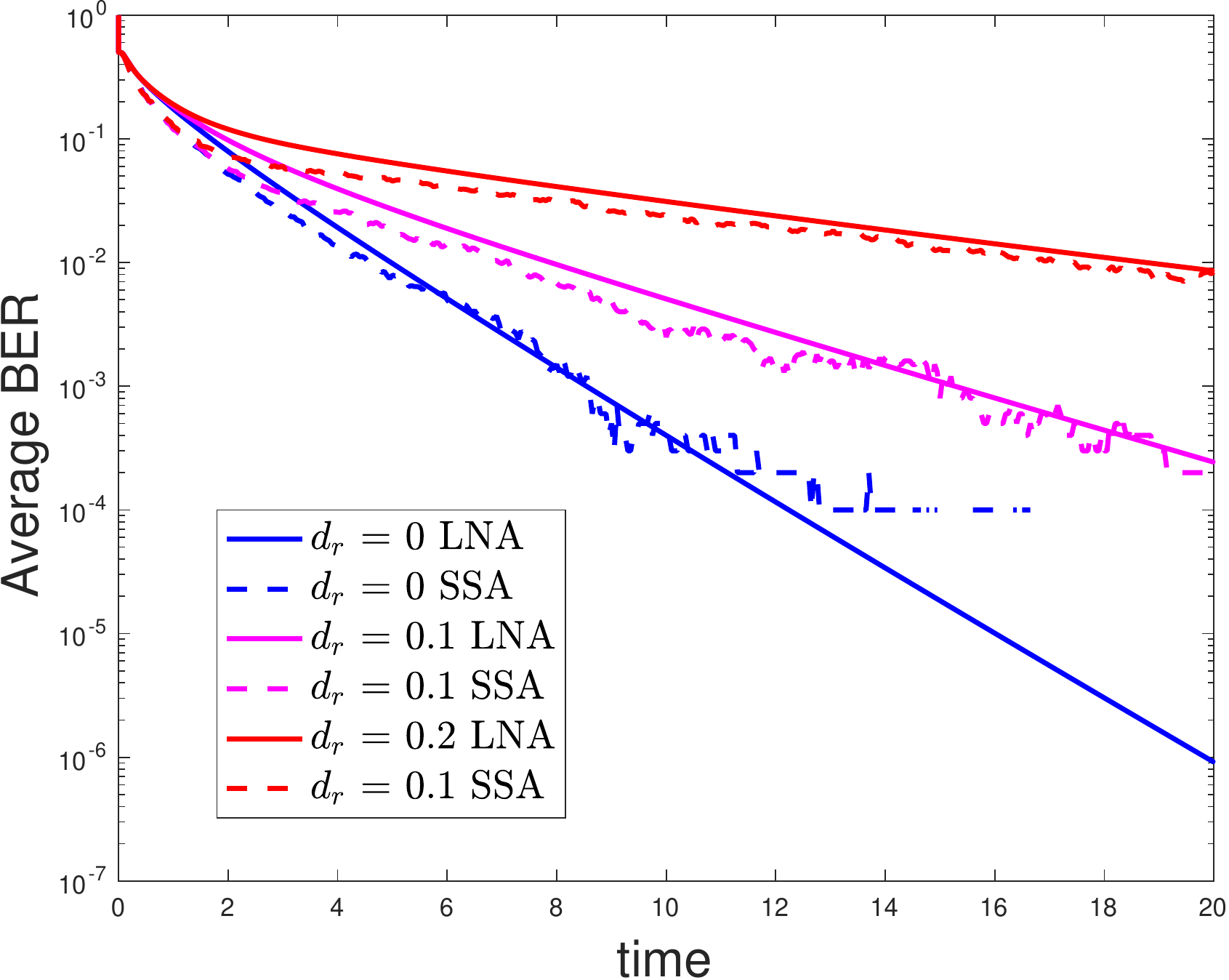}
    \caption{Average BER obtained from LNA. }
    \label{fig:lna_ber}
\end{minipage}%
\begin{minipage}{.5\textwidth}
  \centering
    \centering
    \includegraphics[scale=0.35]{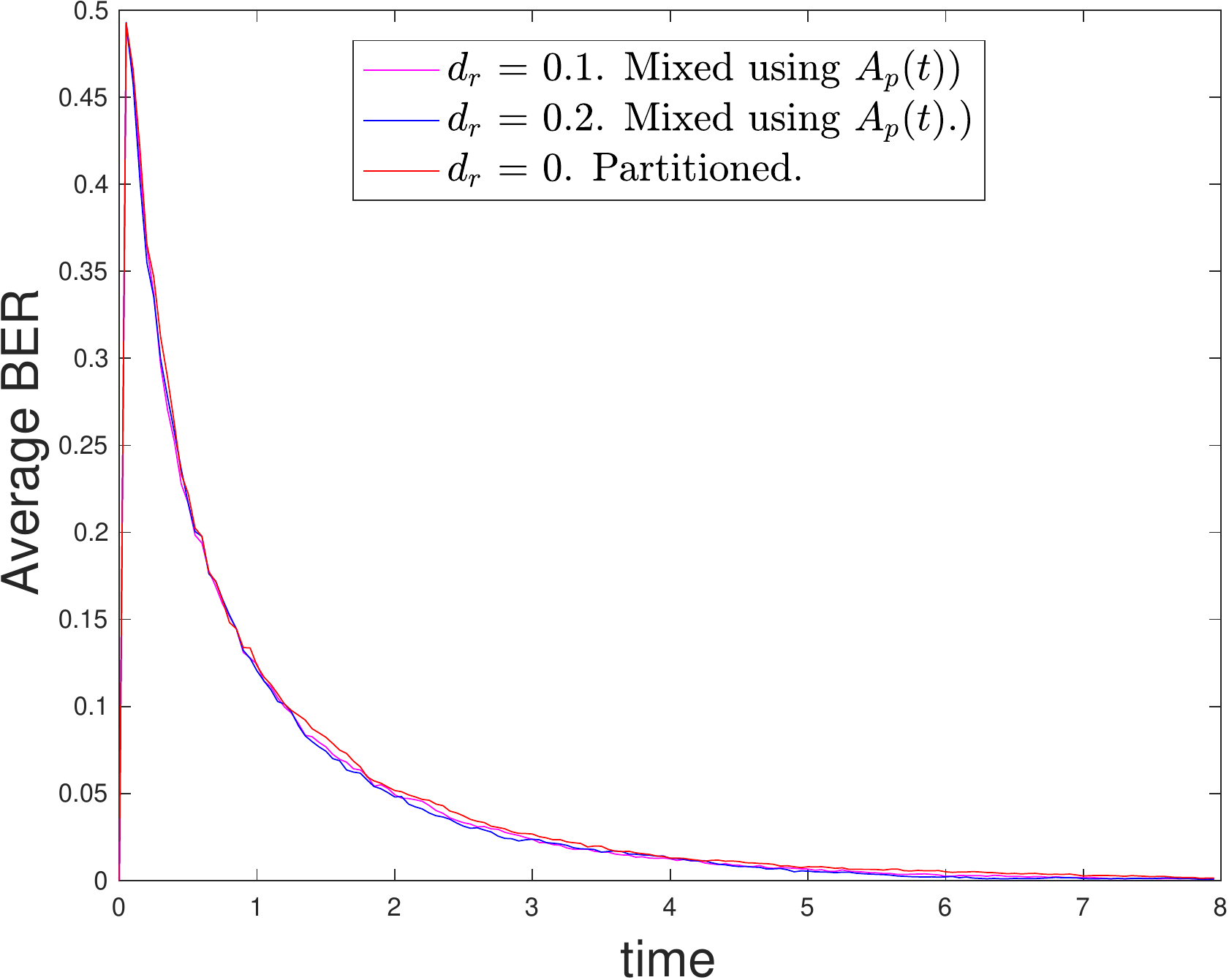}
    \caption{Comparing mixed case using $A_p(t)$ against the partitioned case.}
    \label{fig:possible}
\end{minipage}    
\end{figure}

Next we use mean and covariance matrix of $Z_0(t)$ and $Z_1(t)$ to estimate the BER. Fig.~\ref{fig:lna_ber} plots the average BER estimated by LNA for the cases $d_r = 0, 0.1, 0.2$, against those obtained from SSA. It can be seen the prediction obtained from LNA is fairly accurate. Note also that the BER estimation is more accurate in the later part of the simulation when the system is in steady state. Intuitively, this is because the approximation in LNA is based on approximating the number of reactions in a time interval, see the discussion in the beginning of Section \ref{sec:LNA:cov}. We remark that the BER estimation using SSA becomes inaccurate for $d_r = 0$ after time 10 because of the small BER and limited number of SSA simulations. 

\begin{remark}
A practical way of making use of the results in this section for molecular communications is as follows. The transmitter sends a symbol and the receiver runs until steady state. The receiver then makes a decision on the transmission symbol that is sent. Once the decision is made, then the receiver should reset itself to get ready for the next transmission symbol to be received. The resetting of the receiver can be done using chemical reactions, see our recent work \cite{riaz2019nanocom}. 
\end{remark}

\iftcom
\else
\begin{figure}
    \centering
    \begin{subfigure}[t]{0.45\textwidth}
        \centering
        \includegraphics[scale=0.35]{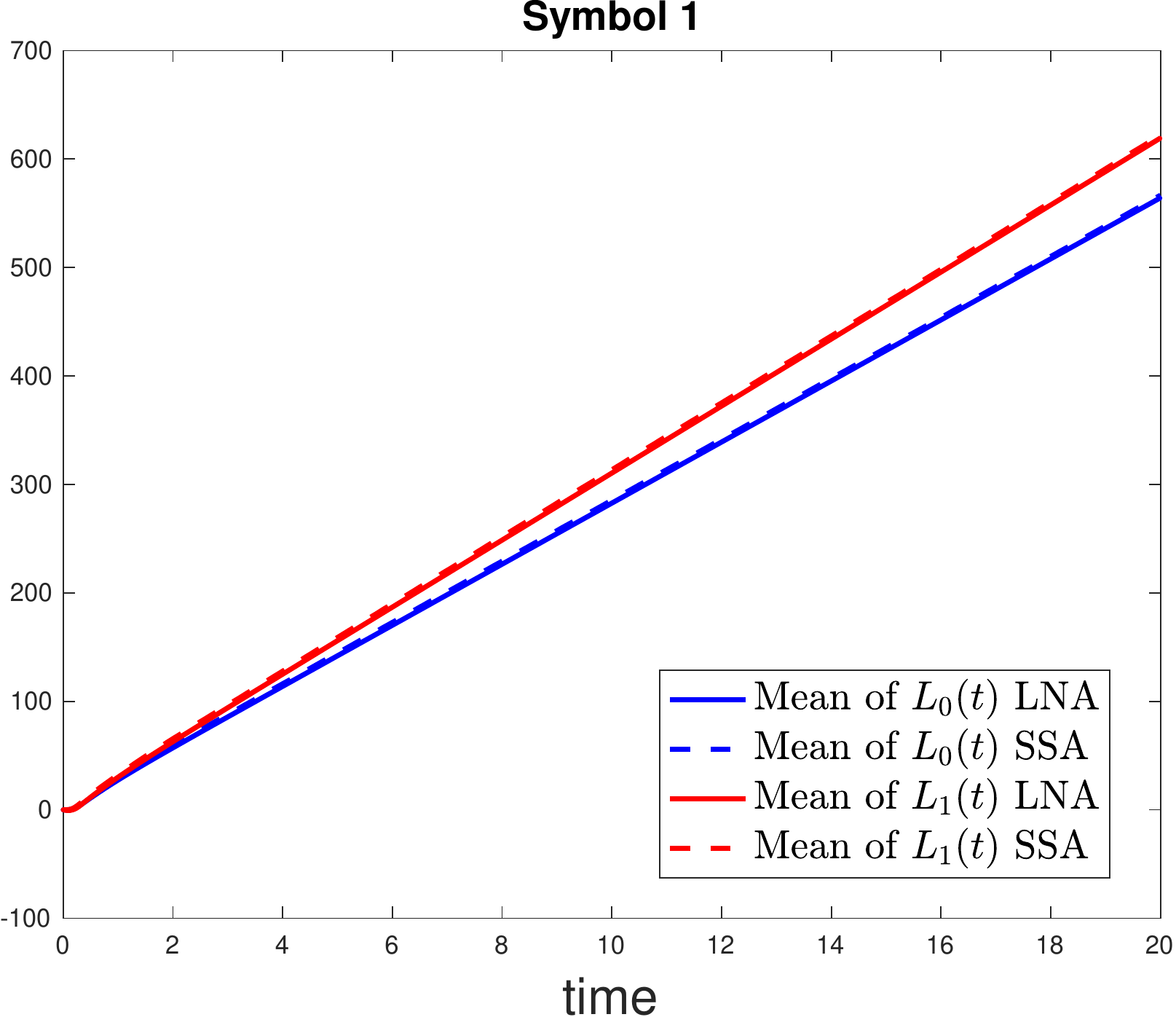}
        \caption{}
        \label{fig:lna_u1_mean}
    \end{subfigure}
     \begin{subfigure}[t]{0.45\textwidth}
        \centering        \includegraphics[scale=0.35]{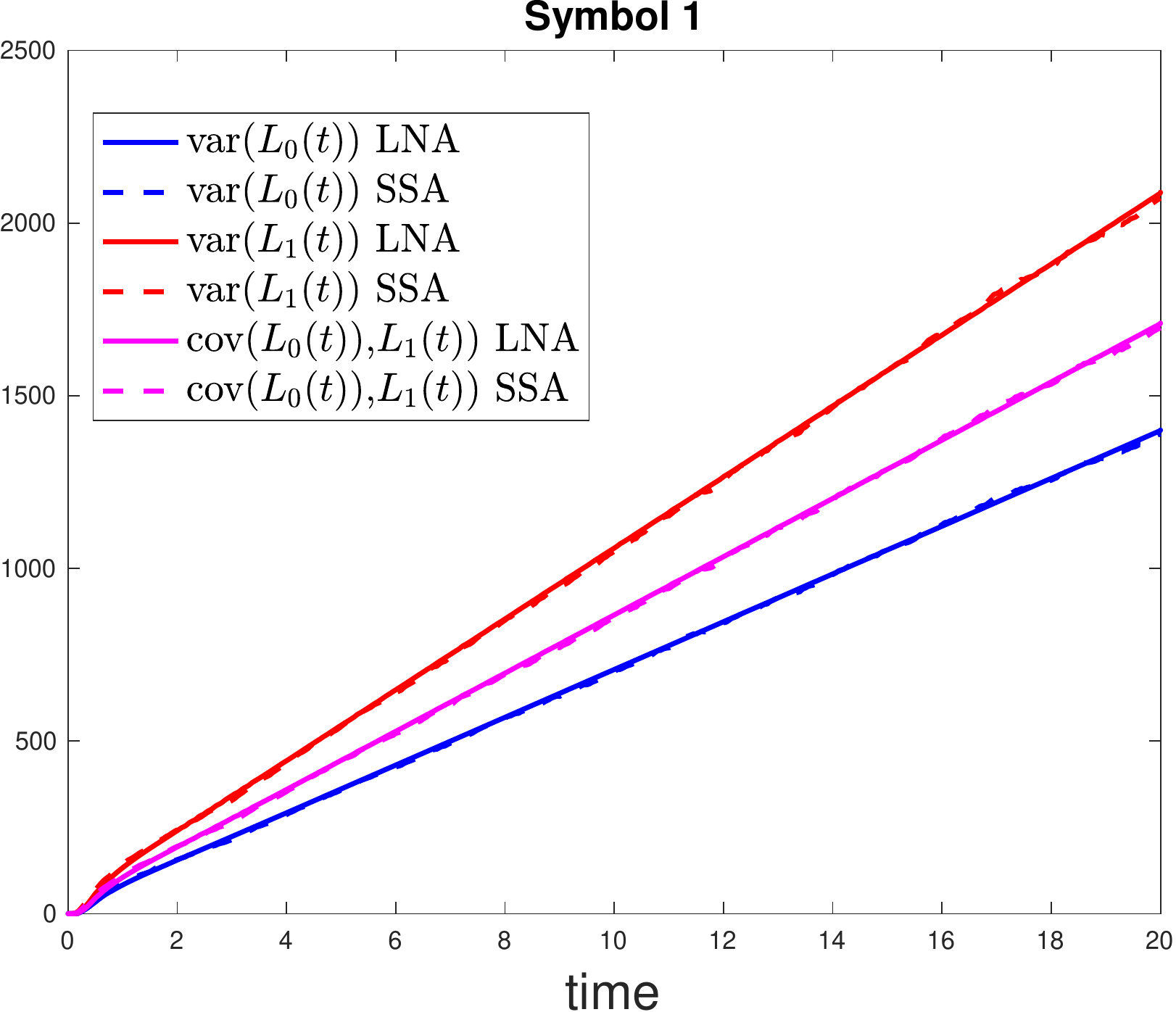}
        \caption{}
        \label{fig:lna_u1_var}
    \end{subfigure}
    \caption{Mean and covariance of $Z_1(t)$ and $Z_1(t)$ by LNA. Symbol 1.}  
    \label{fig:lna_u1}
\end{figure}
\fi

\subsection{What if it is possible to compute $A_p(t)$?}
In the approximate demodulator \eqref{eqn:demod:mixed}, we used $\left[ \frac{dX^{*}_{p}(t)}{dt} \right]_+$ instead of $A_p(t)$ because there does not appear to be a way to distinguish between activation events from diffusion events. In this numerical study, we ask what if it is possible to compute $A_p(t)$ and use it in \eqref{eqn:demod:mixed} instead of $\left[ \frac{dX^{*}_{p}(t)}{dt} \right]_+$. 

We use the same set up in Section \ref{sec:num:ber} and use SSA simulations to estimate the average BER for $d_r = 0.1$ and $d_r = 0.2$ if $A_p(t)$ is used. We compare these BERs against that of the partitioned configuration ($d_r = 0$) in Fig.~\ref{fig:possible}. It shows that if $A_p(t)$ can be obtained, then the mixed configuration has the same BER as the partitioned configuration. 
This shows that partitioning is a useful method to reduce BER because it takes away hard task of identifying the activation events in the mixed configuration. 

\section{Conclusions and Future Work}
\label{sec:SECTINO_7_Conclusions}

This paper considers the demodulation in a molecular communication system where the receiver can be in the partitioned or mixed configuration. We derive the MAP demodulator for both configurations. Our numerical experiment shows that partitioning, or a small diffusion coefficient for the receiver species, can lead to a lower BER. The use of partitioning does not seem to have been studied before so this leads to a new degree of freedom to improve the performance of molecular communication. 

This paper models the shape of a receiver by using multiple voxels. This can be considered as using the finite difference method to model the shape of a receiver. The finite difference method is a very basic method to model the shape of 3-dimensional objects. The modern approach is based on finite element or similar methods. We see this as an interesting future direction. The diffusion of molecules in the finite element setting can possibly be handled by the method in \cite{isaacson2006incorporating}. However, reaction-diffusion setting is a lot more complicated than the pure diffusion case. This is because RDME is fundamentally a spatially discrete method to approximate the behaviour of the fine grained Smoluchowski equation \cite{smoluchowski1918versuch}. This approximation is only accurate if the discretisation length scale is chosen correctly \cite{hellander2016reaction}. Fortunately, there is some recent work on using RDME on an unstructured mesh \cite{hellander2016reaction,isaacson2018unstructured}. Since our demodulator design is based on RDME, these recent work will allow us to extend our work to unstructured mesh to better model the 3-dimensional shape of the receiver.


\bibliographystyle{IEEEtran}
\bibliography{nano,book}

\begin{thebibliography}{10}
\providecommand{\url}[1]{#1}
\csname url@samestyle\endcsname
\providecommand{\newblock}{\relax}
\providecommand{\bibinfo}[2]{#2}
\providecommand{\BIBentrySTDinterwordspacing}{\spaceskip=0pt\relax}
\providecommand{\BIBentryALTinterwordstretchfactor}{4}
\providecommand{\BIBentryALTinterwordspacing}{\spaceskip=\fontdimen2\font plus
\BIBentryALTinterwordstretchfactor\fontdimen3\font minus
  \fontdimen4\font\relax}
\providecommand{\BIBforeignlanguage}[2]{{%
\expandafter\ifx\csname l@#1\endcsname\relax
\typeout{** WARNING: IEEEtran.bst: No hyphenation pattern has been}%
\typeout{** loaded for the language `#1'. Using the pattern for}%
\typeout{** the default language instead.}%
\else
\language=\csname l@#1\endcsname
\fi
#2}}
\providecommand{\BIBdecl}{\relax}
\BIBdecl

\bibitem{Akyildiz:2008vt}
I.~Akyildiz, F.~Brunetti, and C.~Bl{\'a}zquez, ``{Nanonetworks: A new
  communication paradigm},'' \emph{Computer Networks}, vol.~52, pp. 2260--2279,
  2008.

\bibitem{Nakano:2014fq}
T.~Nakano, T.~Suda, Y.~Okaie, M.~J. Moore, and A.~V. Vasilakos, ``{Molecular
  Communication Among Biological Nanomachines: A Layered Architecture and
  Research Issues},'' \emph{NanoBioscience, IEEE Transactions on}, vol.~13,
  no.~3, pp. 169--197, 2014.

\bibitem{akyildiz2015internet}
I.~F. Akyildiz, M.~Pierobon, S.~Balasubramaniam, and Y.~Koucheryavy, ``The
  internet of bio-nano things,'' \emph{IEEE Communications Magazine}, vol.~53,
  no.~3, pp. 32--40, 2015.

\bibitem{farsad2016comprehensive}
N.~Farsad, H.~B. Yilmaz, A.~Eckford, C.-B. Chae, and W.~Guo, ``A comprehensive
  survey of recent advancements in molecular communication,'' \emph{IEEE
  Communications Surveys \& Tutorials}, vol.~18, no.~3, pp. 1887--1919, 2016.

\bibitem{Pierobon:2010kz}
M.~Pierobon and I.~Akyildiz, ``{A physical end-to-end model for molecular
  communication in nanonetworks},'' \emph{IEEE JOURNAL ON SELECTED AREAS IN
  COMMUNICATIONS}, vol.~28, no.~4, pp. 602--611, 2010.

\bibitem{kilinc2013receiver}
D.~Kilinc and O.~B. Akan, ``Receiver design for molecular communication,''
  \emph{IEEE Journal on Selected Areas in Communications}, vol.~31, no.~12, pp.
  705--714, 2013.

\bibitem{mahfuz2015comprehensive}
M.~U. Mahfuz, D.~Makrakis, and H.~T. Mouftah, ``A comprehensive analysis of
  strength-based optimum signal detection in concentration-encoded molecular
  communication with spike transmission,'' \emph{IEEE transactions on
  nanobioscience}, vol.~14, no.~1, pp. 67--83, 2015.

\bibitem{jamali2017design}
V.~Jamali, A.~Ahmadzadeh, and R.~Schober, ``On the design of matched filters
  for molecule counting receivers,'' \emph{IEEE Communications Letters},
  vol.~21, no.~8, pp. 1711--1714, 2017.

\bibitem{tiwari2017estimate}
S.~K. Tiwari and P.~K. Upadhyay, ``Estimate-and-forward relaying in
  diffusion-based molecular communication networks: Performance evaluation and
  threshold optimization,'' \emph{IEEE Transactions on Molecular, Biological
  and Multi-Scale Communications}, vol.~3, no.~3, pp. 183--193, 2017.

\bibitem{fang2017convex}
Y.~Fang, A.~Noel, N.~Yang, A.~W. Eckford, and R.~A. Kennedy, ``Convex
  optimization of distributed cooperative detection in multi-receiver molecular
  communication,'' \emph{IEEE Transactions on Molecular, Biological and
  Multi-Scale Communications}, vol.~3, no.~3, pp. 166--182, 2017.

\bibitem{noel2014improving}
A.~Noel, K.~C. Cheung, and R.~Schober, ``Improving receiver performance of
  diffusive molecular communication with enzymes,'' \emph{IEEE Transactions on
  NanoBioscience}, vol.~13, no.~1, pp. 31--43, 2014.

\bibitem{noel2014optimal}
------, ``Optimal receiver design for diffusive molecular communication with
  flow and additive noise,'' \emph{IEEE transactions on nanobioscience},
  vol.~13, no.~3, pp. 350--362, 2014.

\bibitem{farahnak2019medium}
M.~Farahnak-Ghazani, G.~Aminian, M.~Mirmohseni, A.~Gohari, and
  M.~Nasiri-Kenari, ``On medium chemical reaction in diffusion-based molecular
  communication: a two-way relaying example,'' \emph{IEEE Transactions on
  Communications}, vol.~67, no.~2, pp. 1117--1132, 2019.

\bibitem{mugler2013spatial}
A.~Mugler, F.~Tostevin, and P.~R. Ten~Wolde, ``Spatial partitioning improves
  the reliability of biochemical signaling,'' \emph{Proceedings of the National
  Academy of Sciences}, p. 201218301, 2013.

\bibitem{chou2015markovian}
C.~T. Chou, ``A markovian approach to the optimal demodulation of
  diffusion-based molecular communication networks,'' \emph{IEEE Transactions
  on Communications}, vol.~63, no.~10, pp. 3728--3743, 2015.

\bibitem{awan2017generalized}
H.~Awan and C.~T. Chou, ``Generalized solution for the demodulation of reaction
  shift keying signals in molecular communication networks,'' \emph{IEEE
  Transactions on Communications}, vol.~65, no.~2, pp. 715--727, 2017.

\bibitem{Hiyama:2010jf}
S.~Hiyama and Y.~Moritani, ``{Molecular communication: Harnessing biochemical
  materials to engineer biomimetic communication systems},'' \emph{Nano
  Communication Networks}, vol.~1, no.~1, pp. 20--30, May 2010.

\bibitem{Nakano:2012dv}
T.~Nakano, M.~J. Moore, F.~Wei, A.~V. Vasilakos, and J.~Shuai, ``{Molecular
  Communication and Networking: Opportunities and Challenges},'' \emph{IEEE
  Transactions on Nanobioscience}, vol.~11, no.~2, pp. 135--148, 2012.

\bibitem{ShahMohammadian:2012iu}
H.~ShahMohammadian, G.~G. Messier, and S.~Magierowski, ``{Optimum receiver for
  molecule shift keying modulation in diffusion-based molecular communication
  channels},'' \emph{Nano Communication Networks}, vol.~3, no.~3, pp. 183--195,
  Sep. 2012.

\bibitem{Kuran:2011tg}
M.~Kuran, H.~Yilmaz, T.~Tugcu, and I.~Akyildiz, ``{Modulation Techniques for
  Communication via Diffusion in Nanonetworks},'' in \emph{Communications
  (ICC), 2011 IEEE International Conference on}, 2011, pp. 1--5.

\bibitem{Mahfuz:2011te}
M.~Mahfuz, D.~Makrakis, and H.~Mouftah, ``{On the characterization of binary
  concentration-encoded molecular communication in nanonetworks},'' \emph{Nano
  Communication Networks}, vol.~1, pp. 289--300, 2010.

\bibitem{awan2017improving}
H.~Awan and C.~T. Chou, ``Improving the capacity of molecular communication
  using enzymatic reaction cycles,'' \emph{IEEE transactions on
  nanobioscience}, vol.~16, no.~8, pp. 744--754, 2017.

\bibitem{Awan:2015:IRM:2800795.2800798}
\BIBentryALTinterwordspacing
------, ``Impact of receiver molecular circuits on the performance of reaction
  shift keying,'' in \emph{Proceedings of the Second Annual International
  Conference on Nanoscale Computing and Communication}, ser. NANOCOM' 15.\hskip
  1em plus 0.5em minus 0.4em\relax New York, NY, USA: ACM, 2015, pp. 2:1--2:6.
  [Online]. Available: \url{http://doi.acm.org/10.1145/2800795.2800798}
\BIBentrySTDinterwordspacing

\bibitem{Mahfuz:2014vs}
M.~U. Mahfuz, D.~Makrakis, and H.~T. Mouftah, ``{Strength-based optimum signal
  detection in concentration-encoded pulse-transmitted OOK molecular
  communication with stochastic ligand-receptor binding},'' \emph{Simulation
  Modelling Practice and Theory}, vol.~42, pp. 189--209, 2014.

\bibitem{Pierobon:2011ve}
M.~Pierobon and I.~F. Akyildiz, ``{Noise Analysis in Ligand-Binding Reception
  for Molecular Communication in Nanonetworks},'' \emph{IEEE Transactions on
  Signal Processing}, vol.~59, no.~9, pp. 4168--4182, 2011.

\bibitem{Pierobon:2011vr}
------, ``{Diffusion-based Noise Analysis for Molecular Communication in
  Nanonetworks},'' \emph{IEEE Transactions on Signal Processing}, vol.~59,
  no.~6, pp. 2532--2547, 2011.

\bibitem{Chou:rdmex_tnb}
C.~T. Chou, ``Extended master equation models for molecular communication
  networks,'' \emph{IEEE Transactions on Nanobioscience}, vol.~12, no.~2, pp.
  79--92, 2013, \url{doi:10.1109/TNB.2013.2237785}.

\bibitem{awan2016reducing}
H.~Awan, ``Reducing the effect of reaction rate constants on the performance of
  molecular communication networks,'' in \emph{Proceedings of the 3rd ACM
  International Conference on Nanoscale Computing and Communication}.\hskip 1em
  plus 0.5em minus 0.4em\relax ACM, 2016, p.~8.

\bibitem{Chou:gc}
C.~T. Chou, ``{Maximum A-Posteriori Decoding for Diffusion-Based Molecular
  Communication Using Analog Filters},'' \emph{Nanotechnology, IEEE
  Transactions on}, vol.~14, no.~6, pp. 1054--1067, Nov. 2015.

\bibitem{Gardiner}
C.~Gardiner, \emph{Stochastic methods}.\hskip 1em plus 0.5em minus 0.4em\relax
  Springer, 2010.

\bibitem{thomas2016capacity}
P.~J. Thomas and A.~W. Eckford, ``Capacity of a simple intercellular signal
  transduction channel,'' \emph{IEEE Transactions on information Theory},
  vol.~62, no.~12, pp. 7358--7382, 2016.

\bibitem{aminian2015capacity}
G.~Aminian, M.~Farahnak-Ghazani, M.~Mirmohseni, M.~Nasiri-Kenari, and F.~Fekri,
  ``On the capacity of point-to-point and multiple-access molecular
  communications with ligand-receptors,'' \emph{IEEE Transactions on Molecular,
  Biological and Multi-Scale Communications}, vol.~1, no.~4, pp. 331--346,
  2015.

\bibitem{Chou:2014jca}
C.~T. Chou, ``{Molecular communication networks with general molecular circuit
  receivers},'' in \emph{ACM The First Annual International Conference on
  Nanoscale Computing and Communication}.\hskip 1em plus 0.5em minus
  0.4em\relax New York, New York, USA: ACM Press, 2014, pp. 1--9.

\bibitem{kuscu2018maximum}
M.~Kuscu and O.~B. Akan, ``Maximum likelihood detection with ligand receptors
  for diffusion-based molecular communications in internet of bio-nano
  things,'' \emph{IEEE transactions on nanobioscience}, vol.~17, no.~1, pp.
  44--54, 2018.

\bibitem{Chou:hf}
C.~T. Chou, ``{Impact of Receiver Reaction Mechanisms on the Performance of
  Molecular Communication Networks},'' \emph{IEEE Transactions on
  Nanotechnology}, vol.~14, no.~2, pp. 304--317, Mar. 2015.

\bibitem{Atakan:2010bj}
B.~Atakan and O.~B. Akan, ``{Deterministic capacity of information flow in
  molecular nanonetworks},'' \emph{Nano Communication Networks}, vol.~1, no.~1,
  pp. 31--42, May 2010.

\bibitem{Einolghozati:2011cj}
A.~Einolghozati, M.~Sardari, A.~Beirami, and F.~Fekri, ``{Capacity of discrete
  molecular diffusion channels.}'' \emph{ISIT}, pp. 723--727, 2011.

\bibitem{Pierobon:2013cl}
M.~Pierobon and I.~Akyildiz, ``{Capacity of a Diffusion-Based Molecular
  Communication System With Channel Memory and Molecular Noise},''
  \emph{Information Theory, IEEE Transactions on}, vol.~59, no.~2, pp.
  942--954, 2013.

\bibitem{aquino2011optimal}
G.~Aquino, D.~Clausznitzer, S.~Tollis, and R.~G. Endres, ``Optimal
  receptor-cluster size determined by intrinsic and extrinsic noise,''
  \emph{Physical Review E}, vol.~83, no.~2, p. 021914, 2011.

\bibitem{chatterjee2017spatially}
G.~Chatterjee, N.~Dalchau, R.~A. Muscat, A.~Phillips, and G.~Seelig, ``A
  spatially localized architecture for fast and modular dna computing,''
  \emph{Nature nanotechnology}, vol.~12, no.~9, p. 920, 2017.

\bibitem{riaz2018using}
M.~U. Riaz, H.~Awan, and C.~T. Chou, ``Using spatial partitioning to reduce
  receiver signal variance in diffusion-based molecular communication,'' in
  \emph{Proceedings of the 5th ACM International Conference on Nanoscale
  Computing and Communication}.\hskip 1em plus 0.5em minus 0.4em\relax ACM,
  2018, p.~12.

\bibitem{chou2018designing}
C.~T. Chou, ``Designing molecular circuit for approximate maximum a posteriori
  demodulation of concentration modulated signals,'' \emph{IEEE Transactions on
  Communications}, 2019.

\bibitem{smoluchowski1918versuch}
M.~v. Smoluchowski, ``Versuch einer mathematischen theorie der
  koagulationskinetik kolloider l{\"o}sungen,'' \emph{Zeitschrift f{\"u}r
  physikalische Chemie}, vol.~92, no.~1, pp. 129--168, 1918.

\bibitem{Erban:2007we}
R.~Erban, J.~Chapman, and P.~Maini, ``{A practical guide to stochastic
  simulations of reaction-diffusion processes},'' \emph{arXiv preprint
  arXiv:0704.1908}, 2007.

\bibitem{DelVecchio:book}
D.~Del~Vecchio and R.~M. Murray, \emph{Biomolecular Feedback Systems}.\hskip
  1em plus 0.5em minus 0.4em\relax Princeton University Press, 2014.

\bibitem{noel2017simulating}
A.~Noel, K.~C. Cheung, R.~Schober, D.~Makrakis, and A.~Hafid, ``Simulating with
  accord: Actor-based communication via reaction--diffusion,'' \emph{Nano
  Communication Networks}, vol.~11, 2017.

\bibitem{hellander2017mesoscopic}
S.~Hellander, A.~Hellander, and L.~Petzold, ``Mesoscopic-microscopic spatial
  stochastic simulation with automatic system partitioning,'' \emph{The Journal
  of chemical physics}, vol. 147, no.~23, p. 234101, 2017.

\bibitem{Ke:2018jq}
Y.~Ke, C.~Castro, and J.~H. Choi, ``{Structural DNA Nanotechnology: Artificial
  Nanostructures for Biomedical Research},'' \emph{Annual Review of Biomedical
  Engineering}, vol.~20, no.~1, pp. 375--401, Jun. 2018.

\bibitem{Xavier:2018gv}
P.~L. Xavier and A.~R. Chandrasekaran, ``{DNA-based construction at the
  nanoscale: emerging trends and applications},'' \emph{Nanotechnology},
  vol.~29, no.~6, pp. 062\,001--29, Jan. 2018.

\bibitem{Han:2013ju}
D.~Han, S.~Pal, Y.~Yang, S.~Jiang, J.~Nangreave, Y.~Liu, and H.~Yan, ``{DNA
  gridiron nanostructures based on four-arm junctions.}'' \emph{Science}, vol.
  339, no. 6126, pp. 1412--1415, Mar. 2013.

\bibitem{Milo:BioNumbers}
R.~Milo and R.~Phillips, \emph{Cell Biology by the Numbers}.\hskip 1em plus
  0.5em minus 0.4em\relax Garland Science, 2015.

\bibitem{riaz2019nanocom}
M.~U. Riaz, H.~Awan, and C.~T. Chou, ``Maximum a posteriori-based molecular
  circuit demodulators for spatially partitioned molecular communication
  receivers,'' in \emph{Proceedings of 6th ACM International Conference on
  Nanoscale Computing and Communication.}\hskip 1em plus 0.5em minus
  0.4em\relax ACM, 2019.

\bibitem{Zechner:2016is}
C.~Zechner and M.~Khammash, ``{A molecular implementation of the least mean
  squares estimator.}'' \emph{CDC}, 2016.

\bibitem{Briat:2016ha}
C.~Briat, C.~Zechner, and M.~Khammash, ``{Design of a Synthetic Integral
  Feedback Circuit: Dynamic Analysis and DNA Implementation},'' \emph{ACS
  Synthetic Biology}, vol.~5, no.~10, pp. 1108--1116, Jul. 2016.

\bibitem{Munsky:2006es}
B.~Munsky and M.~Khammash, ``{The finite state projection algorithm for the
  solution of the chemical master equation},'' \emph{The Journal of Chemical
  Physics}, vol. 124, p. 044104, 2006.

\bibitem{Cardelli:2016bp}
L.~Cardelli, M.~Kwiatkowska, and L.~Laurenti, ``{Stochastic analysis of
  Chemical Reaction Networks using Linear Noise Approximation},''
  \emph{Biosystems}, vol. 149, pp. 26--33, Nov. 2016.

\bibitem{Erban:2009us}
R.~Erban and S.~J. Chapman, ``{Stochastic modelling of reaction-diffusion
  processes: algorithms for bimolecular reactions},'' \emph{Physical Biology},
  Mar. 2009.

\bibitem{Gillespie:1977ww}
D.~Gillespie, ``{Exact stochastic simulation of coupled chemical reactions},''
  \emph{The journal of physical chemistry}, 1977.

\bibitem{Hellander:2017ea}
S.~Hellander, A.~Hellander, and L.~Petzold, ``{Mesoscopic-microscopic spatial
  stochastic simulation with automatic system partitioning},'' \emph{The
  Journal of Chemical Physics}, vol. 147, no.~23, pp. 234\,101--14, Dec. 2017.

\bibitem{hellander2016reaction}
S.~Hellander and L.~Petzold, ``Reaction rates for a generalized
  reaction-diffusion master equation,'' \emph{Physical Review E}, vol.~93,
  no.~1, p. 013307, 2016.

\bibitem{isaacson2006incorporating}
S.~A. Isaacson and C.~S. Peskin, ``Incorporating diffusion in complex
  geometries into stochastic chemical kinetics simulations,'' \emph{SIAM
  Journal on Scientific Computing}, vol.~28, no.~1, pp. 47--74, 2006.

\bibitem{isaacson2018unstructured}
S.~A. Isaacson and Y.~Zhang, ``An unstructured mesh convergent
  reaction--diffusion master equation for reversible reactions,'' \emph{Journal
  of Computational Physics}, vol. 374, pp. 954--983, 2018.

\end{thebibliography}

\iftcom
\else
\appendices
\section{Partitioned Configuration}
\label{app:part}
In this appendix we show how the demodulation filter Eq.\eqref{MAP_generalized_partitioned} can be derived. We first explain how the case for $P = 2$ can be derived and then explain how it can be generalised. 

We follow the method in \cite{awan2017generalized}. The first step of the derivation is determine the probability $\mathbf{P}[X^{*}_{1}(t+\Delta t),X^{*}_{2}(t+\Delta t) | k, {\cal X}^{*}_{1}(t),{\cal X}^{*}_{2}(t)]$. In \cite{awan2017generalized} we present an algorithm to write down the expression of this probability by identifying all the reactions that can change the count of the output species, i.e. \cee{X^{*}_{1}} and \cee{X^{*}_{2}}. By using the algorithm in \cite{awan2017generalized}, we can show that:
\begin{align}
  &  \mathbf{P}[X^{*}_{1}(t+\Delta t),X^{*}_{2}(t+\Delta t) | k, {\cal X}^{*}_{1}(t),{\cal X}^{*}_{2}(t)]  =  \nonumber  \\ 
 &  \delta(X^{*}_{1}(t+\Delta t)  = X^{*}_{1}(t) + 1,  X^{*}_{2}(t+\Delta t)  = X^{*}_{2}(t))  Q_{1,a}+ \nonumber  \\
 &  \delta(X^{*}_{1}(t+\Delta t)  = X^{*}_{1}(t)-1  , X^{*}_{2}(t+\Delta t) = X^{*}_{2}(t) )  Q_{1,d}+   \nonumber  \\
&   \delta(X^{*}_{1}(t+\Delta t)  = X^{*}_{1}(t) ,  X^{*}_{2}(t+\Delta t)  = X^{*}_{2}(t)+1) Q_{2,a}+ \nonumber  \\ 
&   \delta(X^{*}_{1}(t+\Delta t)  = X^{*}_{1}(t), X^{*}_{2}(t+\Delta t) = X^{*}_{2}(t) - 1)  Q_{2,d}+ \nonumber  \\
&   \delta(X^{*}_{1}(t+\Delta t)  = X^{*}_{1}(t), X^{*}_{2}(t+\Delta t) = X^{*}_{2}(t))  Q_{0}
 \label{app:part:pred}
\end{align}

where $\delta()$ is an indicator function which takes the value of 1 if all the conditions within $()$ are true, otherwise its value is 0. In addition, we have: 
\begin{align} & Q_{1,a} = g_+(M_1-X^{*}_{1}(t))E[N_{R,1}(t) | k, {\cal X}^{*}_{1}(t),{\cal X}^{*}_{2}(t)] \Delta t  \nonumber \\ 
  &  Q_{1,d} = g_-X^{*}_{1}(t) \Delta t  \nonumber \\ 
  &  Q_{2,a} =g_+(M_1-X^{*}_{1}(t)) E[N_{R,2}(t) | k, {\cal X}^{*}_{1}(t),{\cal X}^{*}_{2}(t)] \Delta t  \nonumber \\ 
  &  Q_{2,d} = g_-X^{*}_{2}(t) \Delta t  \nonumber \\ 
  &  Q_{0} =  1-(Q_{1,a}+Q_{1,d}+Q_{2,a}+Q_{2,d}) 
  \label{app:part:q0}   
  \end{align}

Note that the term $Q_{1,a}$ in Eq.~\eqref{app:part:pred} corresponds to the case where the activation reaction \eqref{cr:on} takes place in receiver voxel 1 because $\delta(X^{*}_{1}(t+\Delta t)$ is one greater than $X^{*}_{1}(t)$. The subscripts 1 and $a$ in $Q_{1,a}$ refer to receiver voxel 1 and activation reaction. Similarly, $Q_{1,d}$ refers to deactivation reaction in voxel 1. The terms $Q_{2,a}$ and $Q_{2,d}$ are for voxel 2. Lastly, the term $Q_{0}$ corresponds to no reactions taking place. 

The next step is to derive the ODE which shows how the log-posteriori probability $L_k(t)$ evolves over time. From \cite{awan2017generalized}, we have:
\begin{align}
\frac{dL_k(t)}{dt} =& \lim_{\Delta t \to 0} \frac {\log((\mathbf{P}[X^{*}_{1}(t+\Delta t),X^{*}_{2}(t+\Delta t) | k, {\cal X}^{*}_{R}(t)])}{\Delta t} + L'(t)
\end{align}
where $L'(t)$ is a term independent of symbol $k$. Since $L_k(t)$ does not appear on the RHS of the above equation and $L'(t)$ adds the same contribution to all $L_k(t)$ for all $k = 0,...,K -1$, we can therefore ignore $L'(t)$ for the purpose of demodulation since it is the relative (rather than the absolute) magnitude of $L_k(t)$ which is needed for demodulation. 

By dropping $L'(t)$, we can compute the shifted version of the log-posteriori probability $L_k(t)$. For conciseness, we use $L_k(t)$ to denote the shifted version of the log-posteriori probability. We therefore have: 
\begin{align}
&\frac{dL_k(t)}{dt} = \lim_{\Delta t \to 0} \frac {\log(\mathbf{P}[X^{*}_{P}(t+\Delta t) | k, {\cal X}^{*}_{P}(t)])}{\Delta t} 
\label{app:part:Lt1} 
\end{align}

The next step is to substitute Eq.~\eqref{app:part:pred} into Eq.~\eqref{app:part:Lt1}. After some lengthy manipulations, we arrive at:  
\begin{align}
\frac{dL_k(t)}{dt} =& 
\sum_{p=1}^{2} ( \left[ \frac{dX^{*}_{P}(t)}{dt} \right]_+  \log( {\mathbf E}[N_{R,p}(t) | k, {\cal X}^{*}_{R}(t)] ) - \nonumber \\ 
& g_+  (M_p - X^{*}_{P}(t)){\mathbf E}[N_{p}(t) | k, {\cal X}^{*}_{R}(t)] ) 
\label{app:part:Lt2}
\end{align}
which is Eq.~\eqref{MAP_generalized_partitioned} for the case of $P = 2$. 

For general $P$, there will be $(2P+1)$ terms in the counterpart of Eq.~\eqref{app:part:pred}. Out of these $(2P+1)$ terms, $2P$ of them are $Q_{p,a}$ and $Q_{p,d}$ for $p = 1, \ldots, P$. The last term is $Q_{0}$, which equals to $1 - \sum_{p=1}^P (Q_{p,a} + Q_{p,d})$. After writing down the counterpart of Eq.~\eqref{app:part:pred} for $P$ voxels, we can follow the above procedure to derive Eq.~\eqref{MAP_generalized_partitioned}. 


\section{Mixed Configuration}
\label{app:mixed}
In this appendix we derive the demodulation filter for the mixed configuration for $P = 2$. The derivation in this appendix is similar to that for the partitioned configuration in \ref{app:part}. 

The first step of the derivation is to determine the probability $\mathbf{P}[X^{*}_{1}(t+\Delta t),X^{*}_{2}(t+\Delta t) | k, {\cal X}^{*}_{1}(t),{\cal X}^{*}_{2}(t)]$. As mentioned in Appendix \ref{app:part}, we can use the method in \cite{awan2017generalized} which is to identify all the reactions that can change the counts of the output species, i.e. \cee{X^{*}_{1}} and \cee{X^{*}_{2}}. It is important to point out here that the term {\sl reactions} here takes on a generalized meaning. In the modelling framework of RDME, the diffusion of a species from one voxel to another voxel is considered as a first order chemical reactions \cite{Erban:2007we}. Therefore, when we consider the reactions that can change the counts of \cee{X^{*}_{1}} and \cee{X^{*}_{2}}, we will also need to include the diffusion of the \cee{X^*} species between the voxels. Recall that $D_r$ is the diffusion coefficient of \cee{X^*}. Let $d_r = \frac{D}{w^2}$ where $w$ is the lenght of a voxel edge. By using the algorithm in \cite{awan2017generalized}, we have: 
\begin{align}
  &  \mathbf{P}[X^{*}_{1}(t+\Delta t),X^{*}_{2}(t+\Delta t) | k, {\cal X}^{*}_{1}(t),{\cal X}^{*}_{2}(t)]  =  \nonumber  \\ 
 &  \delta(X^{*}_{1}(t+\Delta t) = X^{*}_{1}(t) + 1,  X^{*}_{2}(t+\Delta t)  = X^{*}_{2}(t))  Q_{1,a}+ \nonumber  \\
 &  \delta(X^{*}_{1}(t+\Delta t) = X^{*}_{1}(t)-1  , X^{*}_{2}(t+\Delta t) = X^{*}_{2}(t) )  Q_{1,d}+   \nonumber  \\
&   \delta(X^{*}_{1}(t+\Delta t) = X^{*}_{1}(t) ,  X^{*}_{2}(t+\Delta t)  = X^{*}_{2}(t)+1) Q_{2,a}+ \nonumber  \\ 
&   \delta(X^{*}_{1}(t+\Delta t) = X^{*}_{1}(t), X^{*}_{2}(t+\Delta t) = X^{*}_{2}(t) - 1)  Q_{2,d}+   \nonumber  \\ 
&  \delta(X^{*}_{1}(t+\Delta t) = X^{*}_{1}(t)-1, X^{*}_{2}(t+\Delta t) = X^{*}_{2}(t)+1) \nonumber  \\
&   Q_{1 \rightarrow 2} +\delta(X^{*}_{1}(t+\Delta t)=X^{*}_{1}(t)+1, X^{*}_{2}(t+\Delta t)=\nonumber  \\ 
&    X^{*}_{2}(t)-1)Q_{2 \rightarrow 1}+ \delta(X^{*}_{1}(t+\Delta t) = X^{*}_{1}(t), X^{*}_{2}(t+\Delta t) =\nonumber  \\
& X^{*}_{2}(t)) Q_{0}
 \label{eqn:predictbx1sq_mixed}
\end{align}
where 
\begin{align}
  &  Q_{1,a} = g_+ E[X_1(t)N_{R1}(t) | k, {\cal X}^{*}_{1},{\cal X}^{*}_{2}(t)] \Delta t  \nonumber \\ 
  &  Q_{1,d} = g_-X^{*}_{1}(t) \Delta t  \nonumber \\ 
  &  Q_{2,a} =g_+ E[X_2(t)N_{R2}(t) | k, {\cal X}^{*}_{1},{\cal X}^{*}_{2}(t)] \Delta t  \nonumber \\ 
  &  Q_{2,d} = g_-X^{*}_{2}(t) \Delta t  \nonumber \\ 
  &  Q_{1 \rightarrow 2} = d_rX^{*}_{1}(t)  \Delta t  \nonumber  \\ 
  &  Q_{2 \rightarrow 1} =  d_rX^{*}_{2}(t)  \Delta t  \nonumber  \\ 
  &  Q_{0} =  1-(Q_{1,a}+Q_{1,d}+Q_{2,a}+Q_{2,d}+Q_{1 \rightarrow 2}+Q_{2 \rightarrow 1})    
 \label{qvalues_mixed}  \nonumber 
  \end{align} 

The meanings of the terms $Q_{1,a}$, $Q_{1,d}$ etc. are the same as those in Appendix \ref{app:part}. The term $Q_{1 \rightarrow 2}$ corresponds to the diffusion of an \cee{X^*} molecule from receive voxel 1 to receiver voxel 2. Starting with Eq.~\eqref{eqn:predictbx1sq_mixed}, we can now follow the same procedure mentioned in Appendix \ref{app:part} to obtain the demodulation filter \eqref{eqn:demod:mixed:opt} for $P = 2$. 

Similarly, we can generalise Eq.~\eqref{eqn:predictbx1sq_mixed} to general $P$. We need to include terms $Q_{p,a}$ and $Q_{p,b}$, as well as diffusion terms $Q_{p_1 \rightarrow p_2}$ and $Q_{p_2 \rightarrow p_1}$ between any pairs of receiver voxels $p_1$ and $p_2$ which are neighbours. After forming the counterpart of Eq.\eqref{eqn:predictbx1sq_mixed} for $P$ receiver voxels, we can obtain the demodulation filter \eqref{eqn:demod:mixed:opt}. 

\section{Example for Section \ref{sec:LNA}}
\label{app:LNA} 

The evolution of the mean concentration of the species in the system is governed by: 

\begin{align}
\left[
\begin{array}{c}
\frac{d \bar{n}_{R,1}(t)}{dt} \\ 
\frac{d \bar{n}_{R,2}(t)}{dt} \\ 
\frac{d \bar{n}_{3}(t)}{dt} \\ 
\frac{d \bar{x}_1(t)}{dt} \\
\frac{d \bar{x}^*_1(t)}{dt} \\
\frac{d \bar{x}_2(t)}{dt} \\
\frac{d \bar{x}^*_2(t)}{dt} 
\end{array}
\right]
&=
\underbrace{
\left[
\begin{array}{rr  rr rr rr rr rr}
-1  &  1  &  0   &  0  &   0  &   0  & 0   &  0  &   0  &   0  & 0   &  0     \\ 
 1  & -1  & -1  &    1  & 0   &  0  &   0  &   0  & 0   &  0  &   0  &   0  \\
0   &  0  &   1  &  -1  &  0   &  0  &   0  &   0  & 0   &  0  &   0  &   0  \\
0   &  0  &   0  &   0  &  -1  &   1  &   0   &    0  &  -1 &  1 &   0  &   0 \\
0   &  0  &   0  &   0  &   1  &  -1  &   0   &    0  &   0 &  0 &  -1  &   1 \\
0   &  0  &   0  &   0  &   0  &   0  &   -1  &    1  &   1 & -1 &   0  &   0 \\
0   &  0  &   0  &   0  &   0  &   0  &    1  &   -1  &   0 &  0 &   1  &  -1 
\end{array}
\right]
}_{S} 
\left[
\begin{array}{c}
d  \bar{n}_{R,1}(t) \\     
d  \bar{n}_{R,2}(t)  \\    
d  \bar{n}_{R,2}(t) \\     
d  \bar{n}_{3}(t)\\         
\hat{k}_+ \bar{n}_{R,1}(t) \bar{x}_1(t) \\ 
k_- \bar{x}^*_1(t) \\
\hat{k}_+ \bar{n}_{R,2}(t) \bar{x}_2(t) \\ 
k_- \bar{x}^*_2(t) \\
 d_r \bar{x}_1(t) \\
 d_r \bar{x}_2(t) \\
 d_r \bar{x}^*_1(t) \\
 d_r \bar{x}^*_2(t)
\end{array}
\right] + 
\left[
\begin{array}{rr  rr rr rr rr rr}
0    \\ 
0    \\ 
1    \\ 
0    \\ 
0    \\ 
0    \\ 
0    \ 
\end{array}
\right]
u_k(t) 
\end{align}
where $u_k(t)$ is the concentration of the transmission symbol $k$ at time $t$. 

\end{document}